\documentclass[conference]{IEEEtran}
\ifCLASSINFOpdf
\else
\fi
\usepackage{CJKutf8}
\usepackage{multirow}
\usepackage{graphicx}
\usepackage{amsmath}
\usepackage{amssymb}
\usepackage{setspace}
\usepackage{array}
\usepackage[caption=false]{xcolor}
\usepackage{float}
\usepackage{subcaption}
\usepackage{arydshln}
\usepackage{algorithm}
\usepackage[noend]{algpseudocode}

\let\labelindent\relax
\usepackage{enumitem}
\setlist{leftmargin=5.5mm}

\newcommand{\parlabel}[1]{\vspace{0mm}\noindent{\bf \em #1} }

\usepackage[normalem]{ulem}

\begin{document}
\title{\LARGE Fingerprinting Search Keywords over HTTPS at Scale \vspace{-10mm}}

\author{\IEEEauthorblockN{Junhua Yan, Hasan Faik Alan and Jasleen Kaur}
\IEEEauthorblockA{University of North Carolina at Chapel Hill\\
\{junhuayan, alan, jasleen\}@cs.unc.edu}
}
\maketitle

\begin{abstract}
The possibility of fingerprinting the search keywords issued by a user on popular web search engines is a significant threat to user privacy.
This threat has received surprisingly little attention in the network traffic analysis literature. In this work, we consider the problem of keyword fingerprinting of HTTPS traffic---we study the impact of several factors, including client platform diversity, choice of search engine, feature sets as well as classification frameworks.
We conduct both closed-world and open-world evaluations using nearly 4 million search queries collected over a period of three months. Our analysis reveals several insights into the threat of keyword fingerprinting in modern HTTPS traffic.
\end{abstract}

\section{Introduction}
\label{sec:intro}

Search engines are among the most popular websites worldwide \cite{Alexa-topsites-url}. In fact, a significant fraction of referrals to even other popular websites (for instance, 55.3\% of referrals to the fifth most popular website, wikipedia.org), come from search engines \cite{wikipediareferralcitation}.
% ---for instance, 55.3\% of referrals to {\em wikipedia.org} (the fifth most popular website) come from search engines \cite{wikipedia-referral-citation}. 
Indeed, search query keywords are invaluable for improving user search results, preventing click fraud, reducing irrelevant advertising, and even detecting the spread of epidemics like influenza \cite{ginsberg2009detecting}.
However, the same keywords---if leaked---may also reveal quite sensitive information about the user, including health issues, marital problems,  abuse, and controversial political stance \cite{AOLleakage}. Hence, the possibility of fingerprinting search queries, from the network traffic they generate, is a significant privacy concern.

Over the past decade, the traffic analysis literature has devoted significant attention to the topic of website and webpage fingerprinting---in which network traffic headers are used to fingerprint which website, or webpage within a website, a user visited \cite{herrmann2009website,panchenko2011website,lu2010website,miller2014know,panchenko2016website,hayes2016k,abe2016fingerprinting,rimmer2017automated,yan2018feature,alan2019client}. However, this large body of work has mostly ignored the 
% even-more concerning possibility 
more challenging problem of fingerprinting search queries issued by a user---in which instead of simply discovering that a user is visiting the search engine {\em google.com}, traffic analysis is used to fingerprint the actual keywords (such as {\em pregnancy} or {\em depression}) that she may be searching for.\footnote{Note that keyword fingerprinting is much more challenging than webpage fingerprinting, since the search results returned by a search engine for a given keyword can change within short time spans. Furthermore, compared to pages within most websites, the number of search result pages generated by a search engine is enormous, with most sharing the format of results returned, while often sharing content across different search terms---in short, it is more difficult to distinguish between search pages and fingerprint keywords reliably.}
% variations of a search term may yield search result pages that are not very distinguishable.}

The notable recent exception is \cite{oh2017fingerprinting}, which studies keyword fingerprinting in the context of the {\em Tor anonymization network}---and
% ---this work 
% uses search query datasets with up to 300 targeted keywords and 80,000 non-targeted keywords and applied prominent website fingerprinting models to 
shows that the presence of one among 300 targeted keywords can be flagged with 80\% recall and 91\% precision, and the specific keyword can be identified with up to 48\% accuracy.
To the best of our knowledge, however, there is no prior work that helps us understand how fingerprintable are search keywords in modern {\em HTTPS} traffic, which is the dominant transfer scenario used by the vast majority of Internet users \cite{https-traffic-usage}.
It is important to note that keyword fingerprinting of HTTPS traffic is quite different from that of Tor traffic, due to differences in structural information revealed in the traffic.

% \markrm{While it may be tempting to assume that} It may be tempting to assume that since HTTPS traffic reveals more information than Tor traffic, fingerprinting attacks shown to work well in the latter would work well in the former as well.
% the analysis framework of \cite{oh2017fingerprinting} can be applied to HTTP traffic to yield higher fingerprinting accuracy than demonstrated in \cite{oh2017fingerprinting}---however,
% the ability to fingerprint keywords in the former is likely to be higher than demonstrated in \cite{oh2017fingerprinting}---however, recent examples from the domain of website fingerprinting caution us against making such a presumption \cite{papers-that-show-that-tor-features-give-poor-performance-in-https}.
% about HTTPS versus Tor traffic.

% Several factors make the study of keyword fingerprinting challenging. 
In this paper, we focus on HTTPS traffic and explore the limits of fingerprinting search keywords at Internet scale, by considering the impact of several influencing factors. 
% Specifically, we ask:
We ask:

\begin{enumerate}
    
    \item How accurately can one of a targeted set of search queries be identified, when a user relies on prominent web search engines? 
    
    \item Does the use of some search engines make a user more vulnerable to privacy violations/attacks?
    
    \item 
    % Given that different client platforms may generate different network traffic \cite{sanders2015influence,alan2019client}, 
    How does diversity in client browser platforms impact the accuracy of fingerprinting search queries? 
    Do some browsers make a user more vulnerable to privacy attacks? 
    % \markhasan{merge with above item?}
    
    % \item How much will privacy enhancing technologies affect the performance of search query fingerprinting under ``closed-world" assumption, such as encrypted tunnel (e.g., VPN and OpenSSH \cite{venkatachalam2007openssh}) or padding-based countermeasures (e.g., PadToMTU and BuFLO \cite{dyer2012peek})  proposed to against fingerprinting website visits? \textcolor{blue}{Countermeasures in closed-world scenario may not be necessary if the accuracy achieved with the large dataset is low.}
    
%    \item Finally, what is the limit of search query fingerprinting over HTTPS in the ``open-world'', at Internet scale?

    \item What type of traffic features aid in keyword fingerprinting? What type of noise is an impediment?

    \item Given the volatility of search results, how often must a classifier be re-trained for keyword fingerprinting?
    
    \item Finally, how fingerprint-able are targeted keywords in large scale ``open-world'' scenarios, in which a large mix of non-targeted background keywords may also be encountered?
    
\end{enumerate}
To address these questions, we collect a large scale dataset over a period of three months, consisting of {\em nearly 4 million} targeted and non-targeted web search queries, using four popular web search engines (DuckDuckGo, Google, Bing and Yahoo), and four prominent browsers (Chrome, Firefox, Edge, Safari). Our analysis reveals:
% that:
%
\begin{itemize}
    \item {\em Vulnerability With Different Search Engines:}
    The ability to fingerprint search queries differs significantly across users of the four major search engines studied.
    Search traffic generated by DuckDuckGo yields the highest fingerprinting accuracy (up to 96\%), followed by Google, Yahoo, and Bing (up to 45\%). % Bing search traffic yields the least fingerprinting accuracy.
    % Our analysis shows that the traffic generated by different search engines can indeed differ fundamentally in how discriminatively it can be clustered on the basis of targeted keywords.
    We hypothesize that these differences are partially due to the different levels of tracking/advertisement/news 
    background traffic generated by different search engines, which is inherently dynamic in nature and may not reflect the search keyword. The presence of such noise hinders the ability to achieve high classification accuracy. 
    % \textcolor{blue}{For example, Yahoo search initiates a large number of TCP connections to third-party service providers, which do not offer unique signatures that can aid in fingerprinting.}
    % Furthermore, if traffic generated with different search queries  follows a specific pattern, 
    % DuckDuckGo, which generates the least amount of traffic among the four search engines, is the most vulnerable to keyword fingerprinting. .
    
    \item {\em Impact of Browser:}
    Browsers differ in how vulnerable their traffic is to keyword fingerprinting. Furthermore, a classifier that is trained on traffic samples from a single-browser, can not be used successfully to test samples from different browsers---the classification accuracy achieved may be significantly lower. Attackers will either need to incorporate a diverse set of browsers/devices during training, or will need to first determine the client browser before testing. 
    
    \item \emph{Impact of Feature Selection:} We evaluate prominent feature sets used previously for website fingerprinting---the two best performing feature sets consistently outperform in all evaluations. Furthermore, our analysis reveals the presence of noisy traffic components that generate significant amount of dynamic traffic that is not indicative of the search query keywords.
    When such noise is eliminated by considering only the primary domains that are directly related to a given search engine, classification accuracy improves significantly for the Yahoo search engine, even in cross-browser attacks.
    
    % Based on the above findings, we explore the idea of an attacker that reduces noise in the training and testing datasets by considering traffic generated only by a handful of domains directly related to a given serch engine. The hypothesis is that this allows the attacker to eliminate significant amount of dynamic traffic that is not indicative of the search query keywords. We find that \textcolor{blue}{the accuracy is significantly increased using samples collected with Yahoo in this case, even when training and testing with different browsers.}

    \item {\em Effectiveness of Countermeasures:} The presence of countermeasures that obfuscate packet sizes and packet ordering, can significantly lower the accuracy of keyword fingerprinting classifiers in HTTPS traffic.
    
    \item {\em Presence of Large Number of Non-targeted Keywords:} In an open-world scenario, a user may search for any keyword in the wild, including those not seen during training.
    % , which makes the fingerprinting task more challenging. 
    Based on the goal of attackers, we consider three different classification tasks: binary classification, multi-level classification, and multi-class classification. For each task, we consider up to 250k non-targeted search queries for training and testing.
    We find even though the classification performance gradually degrades as the number of non-monitored testing samples increases, it is still quite promising.
    For instance, an attacker may be able to fingerprint among 1,440 targeted keywords, in the presence of around 190k non-targeted search samples, with nearly 90\% average precision for DuckDuckGo/Chrome. 
    
\end{itemize}
The rest of paper is organized as follows: Section \ref{sec:attack} formulates the problem, Section \ref{sec:dataset} summarizes our data collection methodology. Section \ref{sec:feature-sets} discusses the candidate feature sets and the classification algorithm used.
Section \ref{sec:closed_experiment} presents ``closed-world" evaluations, and Section \ref{sec:open_world} presents ``open-world'' evaluations. 
% Section \ref{sec:countermeasures} considers prominent countermeasures. 
Section \ref{sec:conclusion} presents our conclusions.

\section{Problem Formulation}
\label{sec:attack}

\parlabel{Keyword Fingerprinting Threat Model}
In this paper, we consider the scenario in which a user uses a web search engine and attackers eavesdrop on the HTTPS traffic traversing the access link of the user and attempt to predict the search query keywords sent by the user.\footnote{``keyword" and ``search query" are used interchangeably in the paper.}  
% \textcolor{blue}{Although voice service such as Amazon Alexa, Google Home and Apple Siri has gained a lot of popularity in recent years, we only focus on browser-based text searches in this paper. Also, we do not consider mobile traffic for now. (Should we leave it to conclusion and limitation section)} \markred{Mention that we are restricting to browser-based text searches, and not spoken google voice searches, or with mobile web browsers.}
For every sensitive search keyword the attackers wants to identify, they train a machine-learning classifier to determine, given the network traffic generated by the web search, whether it is associated with the keyword. 
To collect labeled training data, attackers repeatedly conduct web searches for both the targeted keywords of interest to them, as well as other popular keywords (as negative examples), and capture the generated network traffic. 

\parlabel{State of the Art}
Prior to 2012, traffic analysis had successfully demonstrated that the deterministic packet sizes generated by the auto-complete feature of the Google search engine can be used to successfully fingerprint the keywords being typed by a user \cite{chen2010side,sharma2012implementing}. Since then, Google has adopted the use of variable-length packets for a given search query with payload randomization and compression. In  \cite{schaub2014attacking}, a stochastic algorithm is adopted to infer the keywords---for a given word length, a prefix tree is created to represent the set of all possible words based on a chosen dictionary and hierarchical matching is conducted based on the observed length of subsequent response packets.

% \markred{ISN'T \cite{schaub2014attacking} PRIOR WORK FOR HTTPS? HOW CAN WE CLAIM WE ARE THE FIRST ONES AFTER 2012?}

Oh et al. performed the first study that extends  website fingerprinting attacks to fingerprint individual web search queries in Tor networks~\cite{oh2017fingerprinting}. They considered up to 300 targeted keywords and 80,000 background keywords with Google, Bing, and DuckDuckGo search engines. They showed that the presence of a targeted keyword can be flagged in Tor traffic with 80\% recall and 91\% precision when 10,000 non-targeted keywords are included---and the specific targeted keyword can be identified with up to 48\% accuracy. 

% \textcolor{blue}{\sout{Even though Tor offers more privacy, HTTPS remains the dominant transfer setting used by web users \cite{https-prevalence, SSL-Pulse, https-statistics}.}}
To the best of our knowledge, there is no prior work that extends website/webpage fingerprinting analysis to fingerprinting web search queries in modern HTTPS traffic.\footnote{A broader discussion of related work is included in Appendix \ref{subsec:related}.} In this paper, we fill this gap.

\parlabel{HTTPS vs. Tor Traffic:  Information Available}
HTTPS encrypts the payload of each network packet using TLS, while providing complete access to TCP/IP headers. Furthermore, hostname of the visited website is often revealed by the SNI extension of TLS, or by reverse DNS of the server IP address.
Thus, 
% in the case of search query fingerprinting with HTTPS traffic, 
we assume that the attacker already knows which web search engine (such as {\em google.com}) is being used by the user.

Tor 
% protects the browsing activities of a user from eavesdropping by 
tunnels and routes traffic through several relay nodes via a TLS connection. It also segments packets into fixed-size cells (512 byte) \cite{Tor}. As a consequence, and in contrast to HTTPS traffic: (i) the packets across all TCP transfers initiated by a web search are interleaved into {\em one} tunneled connection; (ii) the server hostname is not visible; and (iii) the actual packet sizes are obfuscated. In a nutshell, Tor traffic hides a lot of the structural information that is visible in HTTPS traffic.

\parlabel{Why Study HTTPS Traffic?}
Given the above, it is natural to ask---\emph{given that keyword fingerprinting has already been demonstrated over a Tor network, why is it important to study HTTPS, which seems to represent a more relaxed threat model?}
We believe there are several important reasons for doing so: 
\vspace{-1mm}
\begin{itemize}
\item HTTPS remains the dominant transfer setting used by web users \cite{https-prevalence, SSL-Pulse, https-statistics}.\footnote{In fact, users in some countries do not even have access to advanced anonymization networks such as Tor \cite{tor-access}.} An attacker attempting keyword fingerprinting over HTTPS is potentially targeting a much larger fraction of Internet users. It is important to understand how feasible and mitigable is such an attack.

\item Features and classifiers that work well for a seemingly stronger attack, such as Tor-based website or keyword fingerprinting, may not work well in HTTPS \cite{yan2018feature}---a study of the former does not subsume the latter.

\item Several practical issues need to be addressed before Tor-based keyword or website fingerprinting become feasible in practice \cite{acritiqueofwebsitetraffic}.
A significant one is that of segmenting traffic corresponding to different webpages, which has not yet been realized in Tor, but is easier in HTTPS due to the access to per-connection headers and IP addresses \cite{wang2016realistically}.\footnote{It is important to note that Oh et. al. \cite{oh2017fingerprinting} attempt to show that search query traffic can be distinguished from webpage traffic in Tor---however, the webpage dataset used in \cite{oh2017fingerprinting} is provided by Panchenko et al. \cite{panchenko2016website} while the Tor search query dataset is collected by Oh et al. \cite{oh2017fingerprinting}---the use of two different datasets (collected at different times and using different platforms) raises the question of whether the high accuracy is due to the classifier's ability to differentiate between webpage/search query traces or differences in the the underlying data collection setups that may make the traffic differ. For instance, we run a simple experiment with two Tor datasets collected by Wang et al. \cite{wang2014effective, wang2017walkie}---the classifier we train is able to determine which dataset a testing sample comes from with 100\% accuracy!} 
The state of the art is, thus, in a better position to study the practical feasibility of HTTPS keyword fingerprinting.

\item Given the widespread usage of HTTPS, an attacker is likely to encounter a large volume of searches in practice. It is important for a study to use at-scale data sets that are representative of this volume (and are much larger than have been used in prior work).
% New, larger data sets needed --- represent what an attacker will encounter in practice.

% \item 
% More practical issues remain to be addressed in Tor before website/webpage/keyword fingerprinting become feasible in reality  \cite{a-critique-of-website-traffic}.    One of the most essential and challenging one is traffic segmentation. 
% Wang et al. \cite{wang2016realistically} demonstrates the in-feasibility of correctly splitting overlapping pages under Tor, while in HTTPS this issue is much easier due to the availability of IP addresses from TCP/IP headers. 
% Furthermore, although Oh et al. \cite{oh2017fingerprinting} attempts to show the possibility of identifying search query traffic from webpage traces in Tor, the result remains uncertain to us.\footnote{\textcolor{blue}{The Tor webpage dataset used in \cite{oh2017fingerprinting} is provided by Panchenko et al. \cite{panchenko2016website} while the Tor search query dataset is collected by Oh et al. \cite{oh2017fingerprinting} -- there is not clear evidence to show the setups for collecting the two datasets are exactly the same, which make us doubt whether the high accuracy is due to the classifier's ability to differentiate between webpage/search query traces or the underlining setups that may make the traffic differ, such as different platforms/visiting time. For example, when experiment with two Tor datasets collected by Wang et al. \cite{wang2014effective, wang2017walkie}, the classifier we trained is able to determine which dataset a testing sample comes from with 100\% accuracy.}}
\end{itemize}

\parlabel{Our Approach}
%
% Enabled by the availability of large compute power and advances in machine learning, 
While it is heartening to witness the tremendous growth in traffic analysis studies over the past decade, it is also important for this line of research to avoid the pitfalls that may lead to significant exaggeration of fingerprinting accuracy in the real world \cite{acritiqueofwebsitetraffic, juarez2014critical, panchenko2016website, alan2019client}. In this paper, we consider the impact of several factors, many of which have been shown in prior work to significantly influence the performance of traffic analysis techniques:
\begin{itemize}
    \item {\em Client Platform Diversity:}
%    There is significant diversity in the client platforms (browsers, devices, and operating systems) used to access the Internet. 
    Recent studies have shown that different client browsers may result in significant differences in the network traffic generated---so much so, that traffic classification accuracy is significantly impacted when different browsers are represented in the training and testing data sets \cite{alan2019client}. It is, therefore, important to address the questions: {\em how does the use of different browsers impact the accuracy of fingerprinting search keywords? And do some browsers make a user more vulnerable to keyword fingerprinting attacks?}

    \item {\em Feature Set Design:}
    The choice of input features can significantly influence the performance of a learning-based classifier \cite{yan2018feature,Li:2018:MIL:3243734.3243832}---distinguishing features that are stable even in the presence of noise are likely to yield better classification results in the real world. Hence, we ask the question: {\em which features yield unique signatures across classification labels, but are also robust to the presence of real-world noise?} 

    \item {\em Choice of Search Engine:}
    For a given keyword search, the network traffic pattern generated may differ across different web search engines. Since HTTPS traffic lets us identify which search engine is being used, it can also help us answer: {\em does the use of some search engines make a user more vulnerable to keyword fingerprinting attacks?}

    \item {\em Large Scale Data Collection:}
    There are two well-accepted observations with respect to the use of machine learning for traffic classification: (i) training data sets that are large as well as representative of the noise and diversity likely to be encountered during testing, will lead to better classifier performance in the real world; and (ii) it is important to test a classifier in large-scale ``open-world'' settings that include a large number of unseen data points---classification results can often be exaggerations if based on small and narrow-focused testing data. Hence, we ask in this paper: {\em how does keyword fingerprinting perform when larger number of unseen keywords are encountered?}
    
\end{itemize}
In what follows, we describe our methodology to collect and analyze data in order to address the above questions.

\section{DataSet}
\label{sec:dataset}

% For studying the threat model identified in Section \ref{sec:attack}, we need to collect and analyze network traffic generated by keyword searching in Internet search engines.

\subsection{Selecting Search Query Keywords}

\parlabel{Targeted Keywords}
For the purpose of this study, we consider the context in which the attacker is interested in tracking whether a given user is interested in blacklisted/sensitive content.
% \footnote{For instance, the attacker could be interested in getting access to sensitive private data about the user, such as medical conditions.} 
For creating the corresponding list of targeted search keywords, we use 447 keywords blacklisted from Google Instant\footnote{Google Instant has been deprecated since 2017 \cite{google-instant-deprecate} and not used in the paper.} \cite{google-instant} (as reported by 2600.com \cite{2600-google-blacklist-url}), as well as 1,000 keywords that are considered sensitive in China \cite{source-of-chinese-keywords}.
% ---we refer to these two lists as {\em Google} and {\em CHN}, respectively.
% It is important to note that many of these search queries consist of more than one word and varying number of characters (see \figurename{ \ref{fig:length_of_queries}}).
% \markred{What is the point of \figurename{ \ref{fig:length_of_queries}}? If we keep it, the x and y labels need to be changed. And add distribution for non-targeted keywords too.}

% \markred{Are the Chinese keywords ``blacklisted/banned'' or ``sensitive/private''?} \textcolor{blue}{I will prefer to say ``sensitive/private''}

\parlabel{Non-targeted Background Keywords}
When collecting traffic from a given population of users, an attacker is likely to encounter web search traffic for non-targeted keywords as well---the additional task for the attacker, then, is to sift through the web searches to identify those that include targeted keywords. In order to understand the problem of keyword fingerprinting in the real world, therefore, it is important to study the ``open-world'' setting, in which regular background search traffic is also incorporated.

We assume that background web search traffic is modeled well by including popular search queries. For this, we consider the most popular queries reported by a commercial keyword tool \cite{keywordtool-io-citation}, for two region/language settings: \emph{Global/English} and \emph{Hong Kong/Chinese Simplified (China)}---given an input sequence of characters, this tool returns a list of suggested search keywords that either start with that sequence or contain that sequence as a substring. The tool also provides the average number of search per month for each suggested search keyword.
% 
% \footnote{\url{https://keywordtool.io}} 
% in the following two region/language settings as background search queries: \emph{Global/English} and \emph{Hong Kong/Chinese Simplified (China)}.
To harvest the most popular keywords for each region, we adopt a 3-step approach:
\begin{enumerate}

    \item First, we extract the complete list of all suggested search queries that start with each of the 26 alphabet characters a-z.
    
    \item From the above list, then, we select search queries that log more than 50,000 average search volume per month---this step yields 2,647 and 1,814 keywords for the \emph{Global/English} and \emph{Hong Kong/Chinese Simplified} settings, respectively.
    
    \item For each search query obtained in the second step above (e.g., ``apple''), we then input it back to the keyword tool~\cite{keywordtool-io-citation}---which returns a list of suggested search terms that contain that search query in them (e.g., ``apple sauce''). In this step, we select all query phrases that log more than 3,000 average search volume per month.
\end{enumerate}
Our final list of background keywords consist of all search terms harvested in steps two and three above---in total, we harvest around 235,767 background keywords, with a diverse span of topics.\footnote{Examples of non-targeted search keywords can be found in Appendix \ref{appendix:example_non_monotored}.}
The average search volume for these keywords during 2018 was around $2.89 \times 10^{11}$, which is about 1,052 hours of Google search volume according to \cite{0ne-second-stats}.

\begin{figure}[htbp]
\centering
\includegraphics[width=0.8\linewidth]{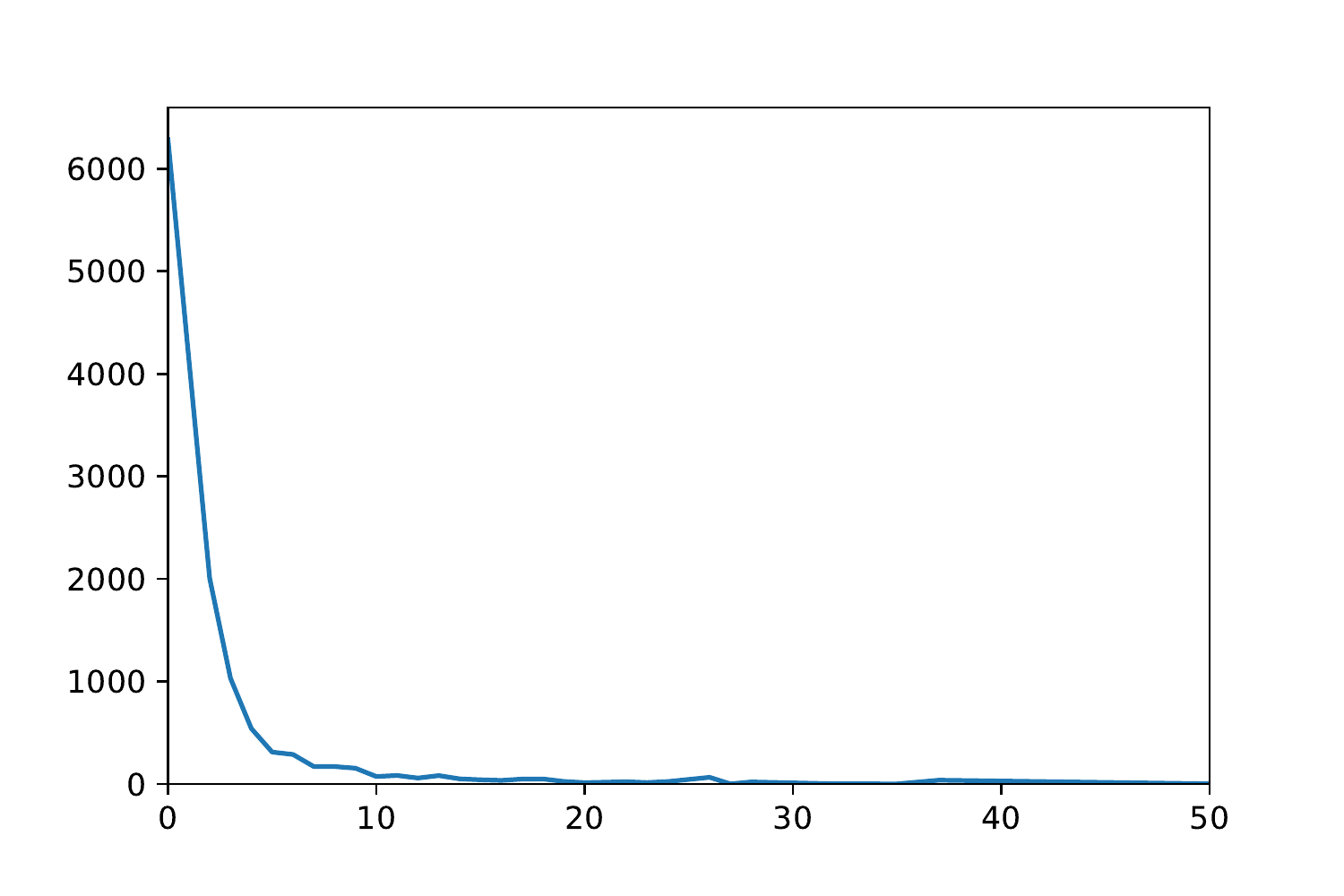}
\vspace{-3mm}
\caption{Distribution of non-targeted keywords with cosine similarity equal or larger than 0.8.}
\vspace{-3mm}
\label{fig:sim_distribution}
\end{figure}

\parlabel{Semantic closeness of Non-targeted Keywords}
% \textcolor{blue}{
Users may differ in how they search for a given topic -- for instance, a user may prefer ``tv” while another user may consider “television” instead. 
To reflect the variations, non-targeted keyword list is supposed to contain queries that are semantically close to each other. 
Thus we measure the semantic closeness of non-targeted keywords with a commonly-used technique in natural language processing. Specifically, we first embed the keywords and then calculate the cosine similarity between their embedding. 
By manually investigating groups of keywords grouped based on cosine similarity, we found 0.8 to be a reasonable threshold to balance between the variation and semantic similarity.\footnote{An example list of keywords with a cosine similarity more than 0.8 are ``maps maroon 5 lyrics",  ``maps by maroon 5 lyrics", ``maps maroon 5" and ``maps lyrics".}
% are (``walmart rings", ``walmart wedding rings",  ``walmart engagement rings") and  
\figurename{ \ref{fig:sim_distribution}} shows the distribution of the number of non-targeted keywords with cosine similarity equal to or larger than 0.8. The results indicate that around 53\% of keywords are unique while 17\% of them have one semantic similar neighbor.
% }

% Our monitored search query list consists of 447 queries from Google Instant blacklist reported by www.2600.com and 1,000 sensitive search queries in China. Information about the length of monitored queries is shown in \figurename{ \ref{fig:length_of_queries}}.

% \parlabel{Which Non-targeted Keywords to Include?}
% We consider the most popular queries reported by a commercial keyword tool\footnote{\url{https://keywordtool.io}} in the following two region/language settings as background search queries: \emph{Global/English} and \emph{Hong Kong/Chinese Simplified (China)}.
% To harvest the most popular search queries in each region, we adopt a hierarchical approach:
% \begin{enumerate}
%     \item Firstly, we identify a list of suggested search queries for each alphabet character from a-z;
%     \item Then, we obtain another list of suggested keywords for every popular query identified in the previous step that with more than 50,000 average search volume per month.
% \end{enumerate}
% The final background query list is composed of all queries with more than \textcolor{blue}{3,000} search volume per month from the suggested lists obtained in the previous two steps. In total, we harvest around \textcolor{blue}{235,800} background queries.

\subsection{Data Collection Methodology}

\parlabel{Web Search Engines}
We focus on four web search engines for our analysis---Google (\emph{www.google.com)}, Yahoo (\emph{www.yahoo.com}), Bing (\emph{www.bing.com}) and DuckDuckGo (\emph{duckduckgo.com}). The first three of these are the most popular web search engines worldwide \cite{top-search-engines}, while DuckDuckGo is the default search engine used in the Tor browser---a prominent feature of DuckDuckGo is that it claims to enhance user privacy by not tracking users as well as by blocking hidden trackers from Google \cite{duckduckgo}.

\parlabel{Client Browsers}
In order to study the impact of diverse client browser platforms on keyword fingerprinting, we consider four different browsers---Firefox, Chrome, Edge and Safari. 
A majority of the data is collected using Firefox and Chrome, which are run on 12 virtual machines and 3 desktop machines with Ubuntu 17.10. Edge instances are run on 8 Windows 10 virtual machines on Microsoft Azure \cite{azure-citation}, 
% \footnote{https://azure.microsoft.com/en-us/}, 
while Safari instances are running on a Mac Mini device with macOS High Sierra. We use Docker containers on the Linux virtual machines in order to scale data collection with Firefox and Chrome.

Our closed-world dataset is collected using all 4 browsers, while our larger open-world dataset is collected using Chrome.

\parlabel{Automated Traffic Capture}
We use \emph{Selenium} for web browser automation \cite{selenium-citation}. We define a {\em search session} as the process of issuing a search using a given combination of keyword, browser, and web search engine.
% , and capturing the network traffic generated.
% 
% targeted keyword is searched for with the four web search engines.
% \footnote{https://www.seleniumhq.org}.
% 
Within each search session, after successfully instantiating the browser, we capture the network traffic using either tcpdump\footnote{https://www.tcpdump.org} on macOS/Ubuntu or WinDump\footnote{https://www.winpcap.org/windump/} on Windows 10.

Furthermore, a search session is divided into two consecutive phases. 
The first phase opens the homepage of the web search engine (\emph{e.g., duckduckgo.com}), while the second phase simulates the typing of the search query by the user and pressing ENTER after the last character is typed. In order to capture all the information conveyed by auto-complete during typing, we impose a 1 second delay before typing the next character.\footnote{In practice, the interval is likely to be smaller than 1s. In Appendix \ref{subsec:type_speed}, we study how typing speeds affect the traffic trace and the vulnerability of being fingerprinted---we do not find any obvious correlation.} After loading the search page, we wait for an additional 5 seconds before stopping tcpdump/WinDump and closing the browser. Consequently, along with the actual search results, traffic related to loading the search engine homepage as well as auto-complete is also captured.

\parlabel{Simulating User Searching Modes}
There are two prominent web search modes used by Internet users---\emph{homepage searching}, in which a user first visits the homepage of the web search engine and then types in a query in the search box, and \emph{addressbar searching}, in which a user directly searches from the browser address-bar (with the configured default search engine) without visiting the homepage of the search engine. In order to study keyword fingerprinting under both modes, we consider network traffic as follows. For {\em homepage searching}, we consider all network traffic collected for a given search session. For {\em addressbar searching}, we filter out all packets captured before typing the first character in the search box.\footnote{For TCP connections that are established before typing the first character in the search box, we filter out only the packets observed before typing---since these may be persistent connections that are used to later carry search traffic.}
% {We only filter out individual packets instead of all packets in a TCP connection that established before typing the first letter in the search box considering persistent connections used in HTTP/1.1 and HTTP/2.0.} 

% Thus all packets transmitted  before typing the first letter in the search box are filtered out in \emph{addressbar searching} mode.\footnote{We only filter out individual packets instead of all packets in a TCP connection that established before typing the first letter in the search box considering persistent connections used in HTTP/1.1 and HTTP/2.0.} 

% \parlabel{Scaling Data Collection}
% Dockers, VMs, Selenium

\subsection{Dataset Summary}
\label{subsec:dataset_summary}
% \textcolor{blue}{(Remove the previous open-set dataset since the scale of current one seems to be enough.)}

% Due to the better availability of Linux platforms, Firefox and Chrome samples are collected over 4 days in February 2019, while Edge and Safari samples are collected over 6 weeks in January to February 2019.

% In total we succeeded to search \textcolor{blue}{add number} non-targeted search queries each at least \textcolor{blue}{three times} and \textcolor{blue}{add number} targeted search queries each at least \textcolor{blue}{times} with Google/DuckDuckGo on Chrome.

\tablename{ \ref{tab:dataset_overview}} summarizes the total number of targeted and non-targeted search sessions captured with different web search engines and client browsers. 
This dataset is collected over a three month period, in two phases. The first phase lasts seven weeks and focuses on  studying the influence of several factors in a closed-world scenario. 
% For each targeted keyword, we conduct multiple search sessions using different search engines and different browsers. 
For a given browser platform, we conduct search sessions with the four search engines in a round-robin manner, iterating multiple times over the different targeted keywords.\footnote{Due to access to several Linux platforms, data collection with Firefox and Chrome on Ubuntu 17.10 took significantly less time.}
% \textcolor{blue}{the first is a seven-week period in which search sessions with targeted keywords are captured using all four browsers;
% \markred{For each targeted keyword, we conduct search sessions with the four search engines in a round-robin manner using different browsers.}
Within the first phase, we successfully searched 1,440 targeted keywords each at least 54 times using Chrome and Firefox, 40 times using Edge, and 4 times using Safari.\footnote{Due to limited availability of Apple devices, search sessions conducted using Safari are used for testing only (not for training the machine learning classifiers).}

% ------------

The second phase lasted six weeks and focused on open-world data collection, using only the Chrome browser and with the DuckDuckGo and Google search engines---this phase conducted search sessions with both targeted and non-targeted keywords.
We randomly iterate over the non-targeted keywords, and randomly insert targeted keyword queries in between.
% For each non-targeted keyword, we conduct search sessions with only the Chrome browser and using only DuckDuckGo and Google search engines. \textcolor{blue}{In the meantime, we randomly search monitored queries between non-monitored query visits and consider them as monitored samples for testing in open-world scenario in order to eliminate potential time effects on fingerprinting performance.}
Overall, each non-targeted keyword is queried at least 5 times each with DuckDuckGo and Google, and each targeted keyword is queried at least 106 times.

In total, our dataset contains nearly 4 million search query visits with four search engines and browsers, including both targeted and non-targeted search queries.\footnote{The dataset can be shared to assist future studies per request.}

\begin{table}[t]
    \centering
    \caption{Total Number of Search Sessions Captured}
    \vspace{-2mm}
    \resizebox{1.0\linewidth}{!}{
    \begin{tabular}{|l|cccc|l|}
    \hline
        & Google& Bing &Yahoo & DuckDuckGo & Time Span \\ \hline
        \multicolumn{6}{c}{\bf \em Closed World Scenario} \\
        \hline
        Edge & 81,020 & 80,832& 84,662 & 88,246 & {Jan-Feb, 2019}\\
        Safari  & 5,784 & 5,784& 5,784& 5,784 & {Jan-Feb, 2019} \\
        Chrome-cw & 98,296  & 98,316 & 97,610 & 98,327 & {Feb, 2019} \\
        Firefox & 94,757  &  94,848 & 93,462  & 94,919 & {Feb, 2019}\\
        \hline
        \multicolumn{6}{c}{\bf \em Open World Scenario} \\
        \hline
        % \bf{Open-World} & {Google} & & &  {DuckDuckGo}  & Time Span \\ \hline
        Chrome-ow & {{1,223,727}} & &  &{{1,223,696}} & {Feb-Mar, 2019} \\
        % (open world)  & & & & & \\
        % Chrome-ow2  &\multicolumn{2}{c}{\multirow{2}{*}{220,024}} & \multicolumn{2}{c}{\multirow{2}{*}{220,018}}  & \multirow{2}{*}{Jan-Feb} \\
        % (open world)  & & & & & \\
        Chrome-targeted 
        & {{150,011}} & & 
        & {{150,023}}
        & {Feb-Mar, 2019} \\
        % (monitored)  & & & & & \\
    \hline
    \end{tabular}}
    \vspace{-3mm}
    \label{tab:dataset_overview}
\end{table}

\section{Feature Sets Considered}
\label{sec:feature-sets}

The performance of learning-based classification may be significantly influenced by the choice of input features. Indeed, in the traffic analysis literature, several different types of features have been derived from the headers of TCP/IP network traffic for the purpose of website and webpage fingerprinting, especially in the context of Tor traffic \cite{wang2013improved,juarez2014critical,wang2014effective,hayes2016k,panchenko2016website,yan2018feature,Li:2018:MIL:3243734.3243832}.
In this paper, we consider and evaluate the following five feature sets---none of these have been considered in the context of keyword fingerprinting over HTTPS traffic, but most have been shown to yield good classification performance in their respective application domains:

% \markred{USE IN DISCUSSION OF FEATURE SELECTION. This process is repeated multiple time and the headers from each packet in the resulting traffic are being captured, along with the time at which the packet is observed at the vantage point. 
% Five key fields in packet headers are source/destination IP address, source/destination port number and TCP segment length, from which a group of features can be extracted and fed to a supervised machine learning framework to fingerprint search queries. }

% \textcolor{blue}{Should we remove Tor feature sets? Why do we need to consider Tor feature sets under HTTPS?}

\begin{itemize}
    \item \emph{k-FP} (2016): K-fingerprinting is devised for fingerprinting web page visits over Tor using features based on packet number and ordering---including number of packets, ratio of incoming/outgoing packets, packet ordering, number of packets per second, concentration of outgoing packets, packet inter-arrival time, and the overall transmission time \cite{hayes2016k}.
    % Prior work shows that the performances of \emph{k-FP} is second to \emph{Wfin} but significantly better compared with other classifiers with Tor traffic \cite{hayes2016k, yan2018feature}.
   
   \item \emph{SvmResp}/\emph{EtResp} (2017): Targeted search keywords are identified in \cite{oh2017fingerprinting} by combining informative features for website fingerprinting, as well as additional novel features in Tor. 
  The features considered include number of total/incoming/outgoing packets, number of incoming bursts,\footnote{A burst is defined as a sequence of back-to-back packets sent in one direction between two packets sent from the opposite direction \cite{panchenko2011website}.} and the cumulative size of TLS records. We use the code provided by Oh et al. \cite{OhGit} for feature extraction.

    % \item \emph{InfLeak} (2018): \cite{Li:2018:MIL:3243734.3243832}. \markred{Add this.}
    
    \item \emph{Wfin} (2018): Recent work on website fingerprinting has focused on extracting many more features than considered previously \cite{yan2018feature,Li:2018:MIL:3243734.3243832}. Yan and Kaur \cite{yan2018feature} extract and analyze the importance of more than 36,000 fine-grained features (grouped into more than 100 feature categories) 
    % from
    % \textcolor{blue}{Yan and Kaur summarized 109 types of feature categories extracted from 
    % TCP/IP headers for website fingerprinting 
    in different communication scenarios.\footnote{Examples of features considered for HTTPS traffic are \emph{packet size count}, \emph{burst size count}, \emph{total No. of TCP connections}, \emph{transmitted bytes w.r.t. port 443}, and \emph{hostname count} (details in Section 11 of \cite{yan2018feature}).}
    % , and analyzes the importance of features.
    % with \emph{Wfin} \cite{yan2018feature}. 
    % For HTTPS-based communication, five levels of feature categories are defined and considered---packet-level, burst-level, TCP-level, port-level and IP address-level---yielding more than 36,000 fine-grained features.
    Feature selection is conducted based on the importance of each feature category for classification---the 
    % according to the \emph{mean decrease impurity} derived from decision tree-based ensemble methods \cite{louppe2013understanding}---the 
    selected feature categories are then used for website fingerprinting classification. Yan and Kaur \cite{yan2018feature} show that \emph{Wfin} achieves comparable or better website fingerprinting accuracy in several communication scenarios, including HTTPS.
    We use the code provided by the authors of \cite{yan2018feature} to select features for each combination of search engine, browser, browsing mode.
    % all HTTPS traffic features and perform feature selection with search session samples collected from each of four search engines.
    % respectively using the methodology in \cite{yan2018feature} to fingerprint different search queries. 
    % The intuition for considering \emph{Wfin} is because:
    % \begin{enumerate}
    %    \item it includes a comprehensive list of features extracted from TCP/IP headers with up to 109 types of feature and around 36,000 individual features;
    %    \item prior researches have shown that \emph{Wfin} is able to achieve comparable or better performance for website fingerprinting in a wide range of communication scenarios, including HTTPS \cite{ yan2018feature, alan2019client}.
    % \end{enumerate}
    % Thus our ultimate goal is to explore the limit of search query fingerprinting over HTTPS, we want to start with as many features as possible, from which to select a subset of features that hopefully yields the best performance.

    \item \emph{Wfin++}: 
    % \markred{move this to \ref{subsec:vul_search_engine}?}
    We also derive from \emph{Wfin} a feature set in two steps.
    % \begin{itemize}
      %  \item 
        First, inspired by features introduced in \cite{oh2017fingerprinting} for fingerprinting search keywords in Tor traffic, in addition to the 109 feature categories identified in \cite{yan2018feature}, we introduce features such as the sequence of reversed cumulative size of packets/bursts, total number of packets, maximum packet size, and the average packet size in the largest incoming burst.
        
        % \item 
        Second, in order to alleviate the impact of noise when too many features are considered \cite{jansen2018inside},
        % \markred{add reference suggested by last reviewer here}, 
        we reduce the final feature list returned by \emph{Wfin} by computing the validation accuracy yielded by the top-\emph{N} most informative features. We select the \emph{N} that yields the highest validation accuracy and use only the corresponding top-\emph{N} features for keyword fingerprinting---details are in Appendix \ref{appendix:wfin}.
        % similar to forward selection~\cite{guyon2003introduction,das2001filters}, after ranking features based on their importance, we iteratively add one type of features to the training feature sets at each step, starting from the most important one, and evaluate the classification performance. The feature set that achieves the highest accuracy is chosen as the final one for classification.\footnote{See Appendix \ref{appendix:wfin} for details.}
        % Instead of including all features selected by \emph{Wfin}, the additional forward selection procedure may further reduce the dimension of features, which may alleviate
        The resultant feature reduction also helps reduce 
        the need for intensive memory and computation resources with our large scale datasets.
        % while also improving classification performance.
    % \end{itemize}
    
    % We compare the performance of \emph{Wfin++} with \emph{Wfin}
    % , which does not include the additional features from \cite{oh2017fingerprinting} and the forward selection step, 
    % in Section \ref{subsec:vul_search_engine}.
   
    \item \emph{Packet Size Count} (\emph{PSC}): Packet size count, for a given traffic direction, is the frequency of each packet size encountered---it has been shown to be one of the most informative features for identifying individual web pages under both HTTPS and encrypted tunnels 
    % in prior works 
    \cite{liberatore2006inferring, herrmann2009website, miller2014know, yan2018feature, alan2019client}. Thus we also consider packet size count as the baseline feature set.
\end{itemize}
%
% We use the Extra-Trees classifier to perform keyword fingerprinting based on each of these feature sets. 
% The label predicted by each classifier is the one with the highest mean probability estimate across the trees in the Extra-Trees classifier.

% \parlabel{Why Extra-Trees?}
\parlabel{Machine Learning Algorithms}
Prior studies have used different types of machine learning algorithms for traffic analysis,
including multinomial Bayes \cite{herrmann2009website}, Support Vector Machine (SVM) \cite{scholkopf2001learning}, and decision tree-based ensemble methods, including Random Forests \cite{breiman2001random} and Extra-Trees \cite{geurts2006extremely}.
Among these, SVM and decision tree-based ensemble methods are able to achieve consistently high performance \cite{hayes2016k,panchenko2016website,yan2018feature}---\emph{SvmResp} uses SVM with 10-fold cross validation for parameters tuning, \emph{k-FP} directly/indirectly uses 
% \textcolor{blue}{classification results from} 
Random Forests,\footnote{In closed-world settings, \emph{k-FP} directly uses the classification output of the Random Forest; for open-world settings, the output of the forest is used as features to fit into \emph{k}-NN.} while \emph{Wfin} uses Extra-Trees. We observe in our evaluations that: (i) for large scale datasets, SVM has a prohibitively high training overhead during cross-validation; and (ii) with comparable performance, Extra-Trees is more computationally efficient than Random Forests, due to the randomization of cut-point choices and the use of whole learning samples for growing the trees  \cite{geurts2006extremely}---hence, we choose Extra-Trees with the above five features sets for both closed-world and open-world evaluations.\footnote{Performance of SVM and Extra-Trees is compared in Appendix \ref{subsec:https_tor}.}
% However, due to the training overhead incurred during cross-validation, SVM can not be easily extended to large scale datasets.\footnote{We trained \emph{SvmResp} with 50,400 samples, using \textcolor{blue}{LIBSVM \cite{libsvm}} and 10-fold cross validation, with \emph{C} ranging from $2^{-7}$ to $2^{7}$, and $\gamma$ from $2^{-5}$ to $2^{2}$, as described in \cite{oh2017fingerprinting}---we trained for more than 72 hours without finishing on a machine with 64 CPUs and 1TB memory.} 
% Furthermore, Extra-Trees is more computationally efficient compared to Random Forests, due to the randomization of cut-point choices and the use of whole learning samples for growing the trees with comparable performance \cite{geurts2006extremely}.\footnote{In one of our experiments on a machine with 64 CPUs and 1TB memory, the maximum number of trees that could be built with Random Forests was no more than 600, while it was around 1,000 with Extra-Trees, using the same number of features and training samples.} 
% Thus, we use Extra-Trees instead of SVM or Random Forests with the above candidate feature sets---\emph{SVMResp} is appropriately renamed as \emph{EtResp}\footnote{\textcolor{blue}{A comparison between the performance achieved with SVM and Extra-Trees using the same set of features is shown in Section \ref{subsec:https_tor}.}}.

\section{Closed World Evaluations}
\label{sec:closed_experiment}

We start with the ``closed-world" assumption (the attacker knows in advance that the search keyword to be fingerprinted belongs to a small known targeted set of keywords)---we use this constrained scenario to investigate the impact of five important factors on the performance of fingerprinting search keywords, including (1) the choice of search engines; (2) the traffic features used for classification; (3) the impact of client browser platforms; (4) the choice of searching modes;  and (5) the impact of the increasing time gap between training and testing. 
In this section, we primarily use the closed-world dataset summarized in \tablename{ \ref{tab:dataset_overview}}.

\begin{figure*}[t]
\centering
\includegraphics[width=1.0\textwidth]{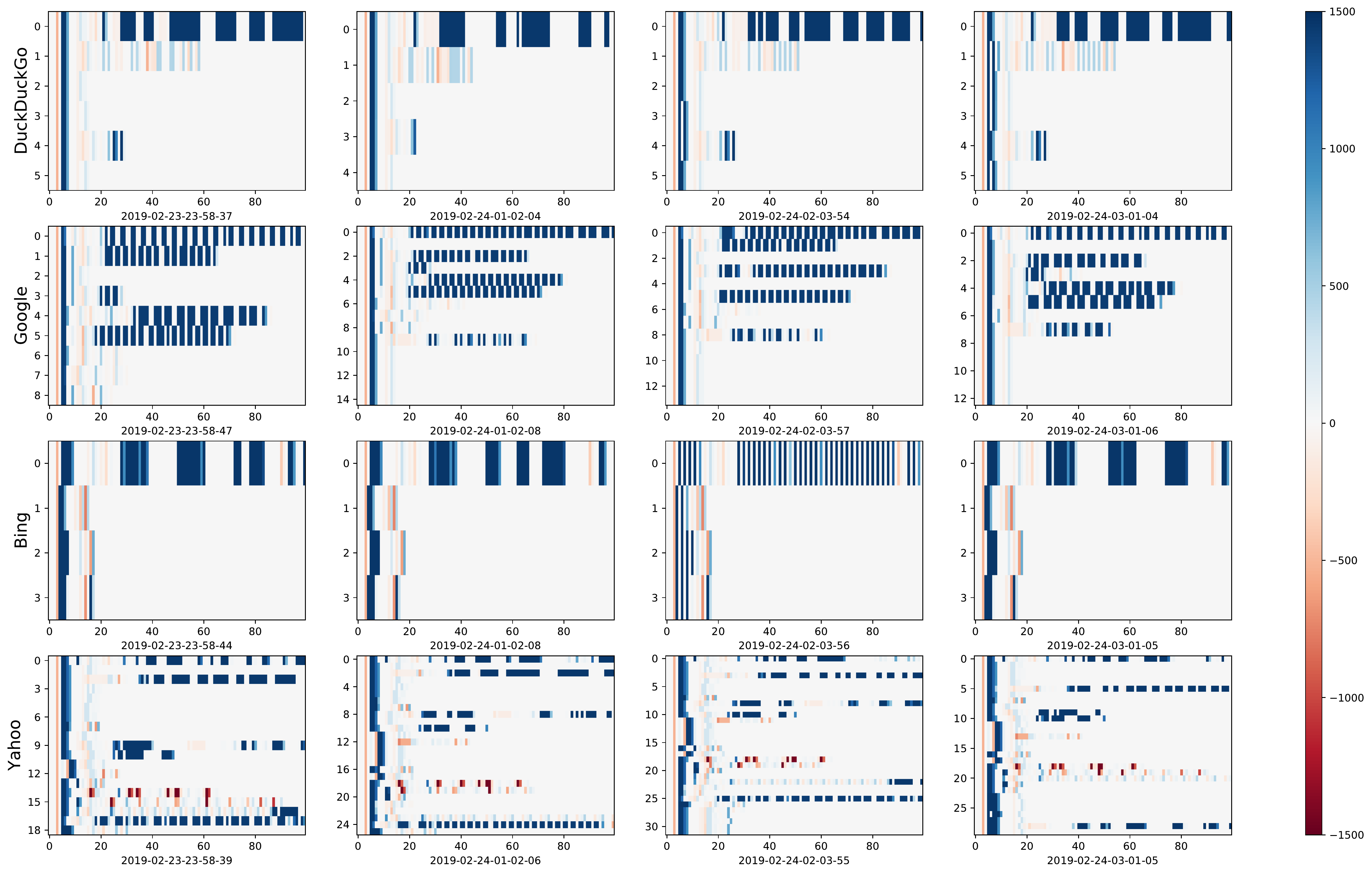}
\vspace{-6mm}
\caption{Network traffic generated using Chrome (caption below each sub-figure indicates the time of search) 
% \emph{Year-Month-Day-Hour-Minute-Second}. 
% \markhasan{we need browser labels on y axis}
}
\vspace{-6mm}
\label{fig:search_engine}
\end{figure*}

\subsection{Vulnerability of Different Search Engines}
\label{subsec:vul_search_engine}

One of the key differences between HTTPS and Tor traffic is that the attacker may be able to learn which search engine is used by the user in the former case, based on the IP address of the server as well as the server name in the SNI extension of TLS \cite{Nygren2017,Gonzalez:2016:UPT:2987443.2987451}. That is, the attacker can be assumed to know that a user is visiting \emph{google.com}---but the attacker wants to fingerprint the keywords that the user is using for their search query. Thus the first question we ask is: \emph{among Google, Bing, Yahoo and DuckDuckGo, which web search engine makes a user more vulnerable/robust to keyword fingerprinting?}

To answer this question, we use our searches for the 1,440 targeted keywords 
% \textcolor{blue}{collected for closed world scenario shown in \tablename{ \ref{tab:dataset_overview}}}
on Yahoo, Bing, Google and DuckDuckGo, using Firefox, Chrome, and Edge.\footnote{Results with the Edge browser are included in Appendix  \ref{appendix:edge_search_engine}.} 
For Chrome and Firefox, the targeted dataset is split into three disjoint subsets for training, validation and testing, with the ratio of \emph{4:1:1}---for every 6 consecutive visits of a search query, the first 4 samples are used for training, the 5th for validation, and the 6th for testing.\footnote{The reason for adopting this sub-sampling scheme is to minimize the  time gap between training and testing samples (the impact of time will be explicitly studied in Section \ref{subsec:time_effect}). In practice, it is indeed possible for the  attacker to capture training and testing traces close in time, if it is an offline testing model.} 
% \footnote{For edge samples, the ratio is 8:1:1 due to different number of samples available.}
Since each keyword is queried at least 54 times in the closed-world datasets for Chrome and Firefox, we consider 36 training samples, 9 validation samples, and 9 testing samples during evaluation of each targeted keyword.
The validation dataset is used for tuning parameters in Extra-Trees and for deriving the feature sets for \emph{Wfin} and \emph{Wfin++}. 
To obtain test accuracy, we include both training and validation samples for building the machine learning model, and use test samples for the fingerprinting evaluation (45:9 samples).\footnote{In Extra-Trees classifier, the split criteria is Gini Index with bootstrap disabled and the number of trees is set to 700. Details about parameter tuning are included in Appendix \ref{appendix:parameter_tuning}.}

% \vspace{-1mm} 
\begin{table}[htbp]
    \centering
    % \vspace{-5mm}
%    \caption{Classification Accuracy (\%) Achieved with four web search engines using Chrome and Firefox browsers.}
     \caption{Classification Accuracy Achieved (\%)}
   \vspace{-2mm}
   \resizebox{1\linewidth}{!}{
    \begin{tabular}{|l|cccc|}
    \hline
        \textbf{Chrome} & Bing & Yahoo & Google & DuckDuckGo  \\ \hline
         \emph{Wfin++} 
         & 44.86 $\pm$ 0.07 
         & \fbox{64.63 $\pm$ 0.24} 
         & \fbox{60.32 $\pm$ 0.07} 
         & 96.15 $\pm$ 0.01 \\
          \emph{PSC} 
          & \fbox{44.98 $\pm$ 0.12}
          & 57.66 $\pm$ 0.09 
          & 57.72 $\pm$ 0.05 
          & \fbox{96.33 $\pm$ 0.04} \\
          \emph{Wfin} 
         & 41.63 $\pm$ 0.02 
         & 52.57 $\pm$ 0.05 
         & 58.25 $\pm$ 0.14 
         & 94.06 $\pm$ 0.04 \\
         
         \emph{EtResp} 
         & 15.54 $\pm$ 0.06
         & 0.57 $\pm$ 0.03 
         & 27.39 $\pm$ 0.07 
         & 42.38 $\pm$ 0.08 \\
         
        %  \emph{EtResp} 
        %  & 6.86 $\pm$ 0.06
        %  & 0.70 $\pm$ 0.03 
        %  & 27.73 $\pm$ 0.07 
        %  & 34.26 $\pm$ 0.08 \\
         \emph{k-FP} 
         & 8.95 $\pm$ 0.08
         & 1.48 $\pm$ 0.06 
         & 20.70 $\pm$ 0.14 
         & 33.45 $\pm$ 0.08 \\
       
        \hline
        % \multicolumn{5}{c}{} \\ 
        \hline
         
        \textbf{Firefox} & Bing & Yahoo & Google & DuckDuckGo  \\ \hline
         \emph{Wfin++} 
         & 44.73 $\pm$ 0.13 
         & \fbox{58.14 $\pm$ 0.23} 
         & 75.56 $\pm$ 0.07 
         & 91.95 $\pm$ 0.10\\
         \emph{PSC} 
         & \fbox{44.83 $\pm$ 0.08} 
         & 49.90 $\pm$ 0.11 
         & \fbox{76.75 $\pm$ 0.02} 
         & \fbox{92.23 $\pm$ 0.02} \\ 
         \emph{Wfin} 
         & 41.22 $\pm$ 0.03 
         & 43.06 $\pm$ 0.09
         & 73.55 $\pm$ 0.21 
         & 89.70 $\pm$ 0.03 \\ 
         
         \emph{EtResp} 
         & 8.07 $\pm$ 0.10
         & 0.12 $\pm$ 0.01 
         & 24.07 $\pm$ 0.02 
         & 28.08 $\pm$ 0.06 \\
         
        %  \emph{EtResp} 
        %  & 5.02 $\pm$ 0.10
        %  & 0.55 $\pm$ 0.01 
        %  & 10.93 $\pm$ 0.02 
        %  & 10.27 $\pm$ 0.06 \\
         \emph{k-FP} 
         & 5.34 $\pm$ 0.06 
         & 0.81 $\pm$ 0.03 
         & 9.69 $\pm$ 0.06 
         & 15.82 $\pm$ 0.17 \\
    \hline
    \end{tabular}
    }
    \label{tab:search_engine}
    % \vspace{-5mm}
\end{table}
% \vspace{-1mm} 

\tablename{ \ref{tab:search_engine}} summarizes the test accuracy (mean and standard deviation) for fingerprinting among 1,440 targeted keywords, obtained with different feature sets and search engines.
% \footnote{The standard deviation less than 0.01\% is omitted in the table.} 
We find that:
\begin{itemize}

    \item \emph{Feature Sets:}
    As reported for website fingerprinting \cite{yan2018feature}, feature sets that work well for Tor may not work well for HTTPS---the performance of \emph{EtResp} and \emph{k-FP} are not comparable to  \emph{Wfin}, \emph{Wfin++}, or \emph{PSC}. 
    For instance, the accuracy obtained for DuckDuckGo/Chrome is around 33-42\% with \emph{EtResp} and \emph{k-FP}, but ranges around 94-96\% for the others.
    % is around 34\% while with the other three classifiers is between 94\% to 96\%.
    
    Second, the \emph{Wfin++} feature set outperforms \emph{Wfin} in all cases---the performance gains are as large as 15\% for Yahoo/Firefox.
    In the rest of this paper, we only consider the derived \emph{Wfin++}.
    
    Finally, \emph{PSC} and \emph{Wfin++} outperform each other in different cases. However, \emph{Wfin++} performs at least comparably well in all cases, while \emph{PSC} may sometimes yield a significantly lower accuracy (Yahoo).

    \item \emph{Search Engines:} Among the four web search engines, traffic generated by targeted keywords with DuckDuckGo are most vulnerable to be fingerprinted---the highest accuracy achieved (boxed in Table \ref{tab:search_engine}) with DuckDuckGo is above 96\%, while it ranges from 44\% (Bing) to 76\% (Google) for the rest. Below, we further investigate differences across search engines.
    
    \item \emph{Client Browsers:} The vulnerability of fingerprinting searches on a given search engine also depends on the client browser used---this is most notable for Google, in which targeted keywords can be identified with 76\% accuracy when using Firefox, but only 60\% accuracy when using Chrome. We further investigate the impact of browsers in Section \ref{subsec:client_platform}.

    \item \emph{HTTPS vs. Tor:} For Google, the highest keyword fingerprinting accuracy achieved with 1,400 keywords in the closed-world setting of \tablename{ \ref{tab:search_engine}} ranges from 60\% to 76\% with different client browsers. The closed-world fingerprinting accuracy achieved for Google with 100 keywords in the Tor browser in \cite{oh2017fingerprinting} was around 64\%. However, it is important to resist the urge to compare these numbers---the dataset used in \cite{oh2017fingerprinting} is very different from the one used here (most notably, in the time period of data collection,  the number of targeted keywords, and the number of training/testing samples). In Appendix \ref{subsec:https_tor}, we provide a brief comparison of HTTPS and Tor by carefully controlling factors during data collection. In Appendix \ref{subsec:num_of_keywords} we study the impact of smaller number of keywords on the classification accuracy (we find that the accuracy of fingerprinting Google/Chrome keywords increases from 60\% to 80\% as the number of targeted keywords is reduced to 100).  \\
    
\vspace{-0mm}
\end{itemize}
%
%
% The fingerprinting accuracy observed with the search engines follows the trend DuckDuckGo$>$Google$>$Yahoo$>$Bing.
To understand what makes the search engines differently vulnerable to keyword fingerprinting, we examine the network traffic.
\figurename{ \ref{fig:search_engine}} illustrates the traffic generated using Chrome with four search engines, when searching for a specific targeted keyword at four different times within five hours on Feb 24-25, 2019. The \emph{x-axis} represents the packet index within each TCP transfer, and the \emph{y-axis} represents the index of the TCP connection in the network trace.\footnote{Only the first 100 packets in each TCP connection are depicted for ease of visualization. TCP connections are sorted according to the timestamp of the first packet observed in each TCP connection. The time gap between the start of different TCP connection is not shown in the figure.} Different colors are used to describe the packet length together with the direction (``-" sign indicates packets sent from the client to the server). We observe that the network traffic  differs significantly, when searching the exact same query at approximately the same time across different search engines (any given column in \figurename{ \ref{fig:search_engine}}).
% The search engines from top to bottom are DuckDuckGo, Google, Bing, and Yahoo. 
% We observe several differences:
% 

In \tablename{ \ref{tab:packet_num}}, we summarize statistics with respect to the number of packets generated and number of TCP connections initiated by the four search engines, for \emph{all} search sessions collected using Chrome. We observe that:
\begin{figure*}[htbp]
\centering
\begin{subfigure}{0.3\textwidth}
    \centering
    \smallskip
    \includegraphics[width=1\textwidth]{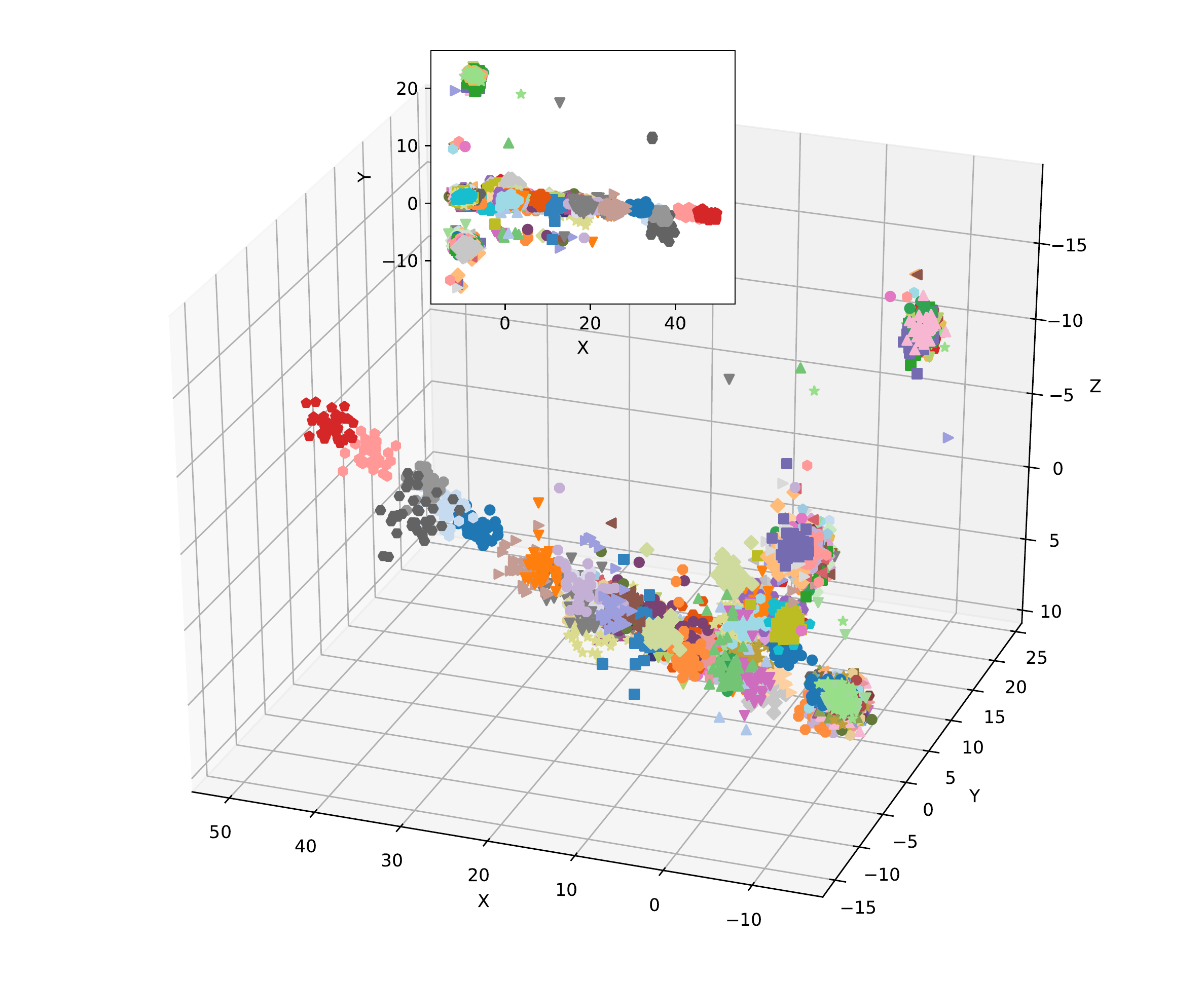}
    \vspace{-5mm}
    \caption{Bing} %{Light Unit}
    \label{fig:lda_bing}
\end{subfigure}
\begin{subfigure}{0.3\textwidth}
    \centering
    \smallskip
    \includegraphics[width=1\textwidth]{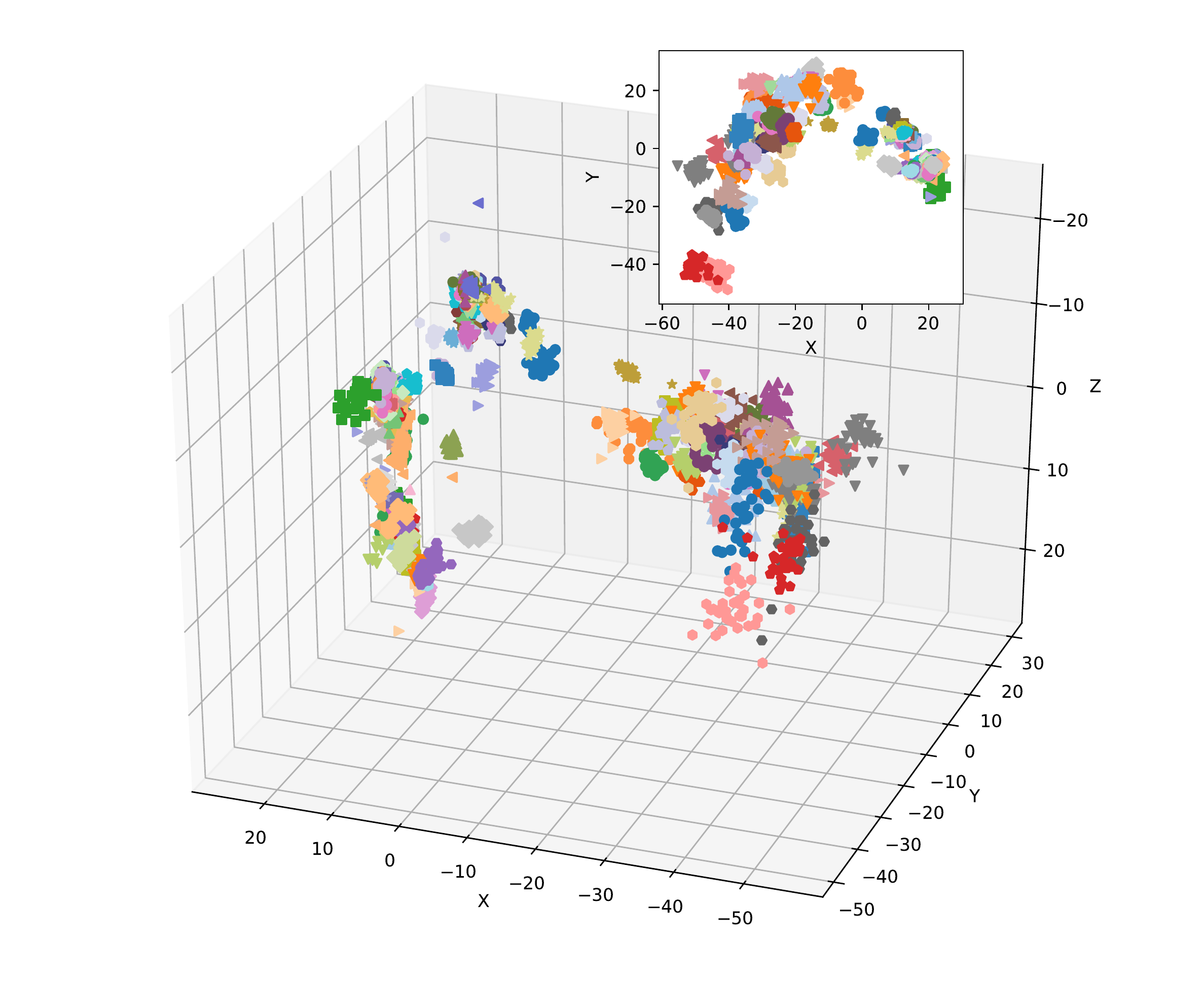}
    \vspace{-5mm}
    \caption{DuckDuckGo} %{Light Unit}
    \label{fig:lda_ddg}
\end{subfigure}
\begin{subfigure}{0.3\textwidth}
    \centering
    \smallskip
    \includegraphics[width=1\textwidth]{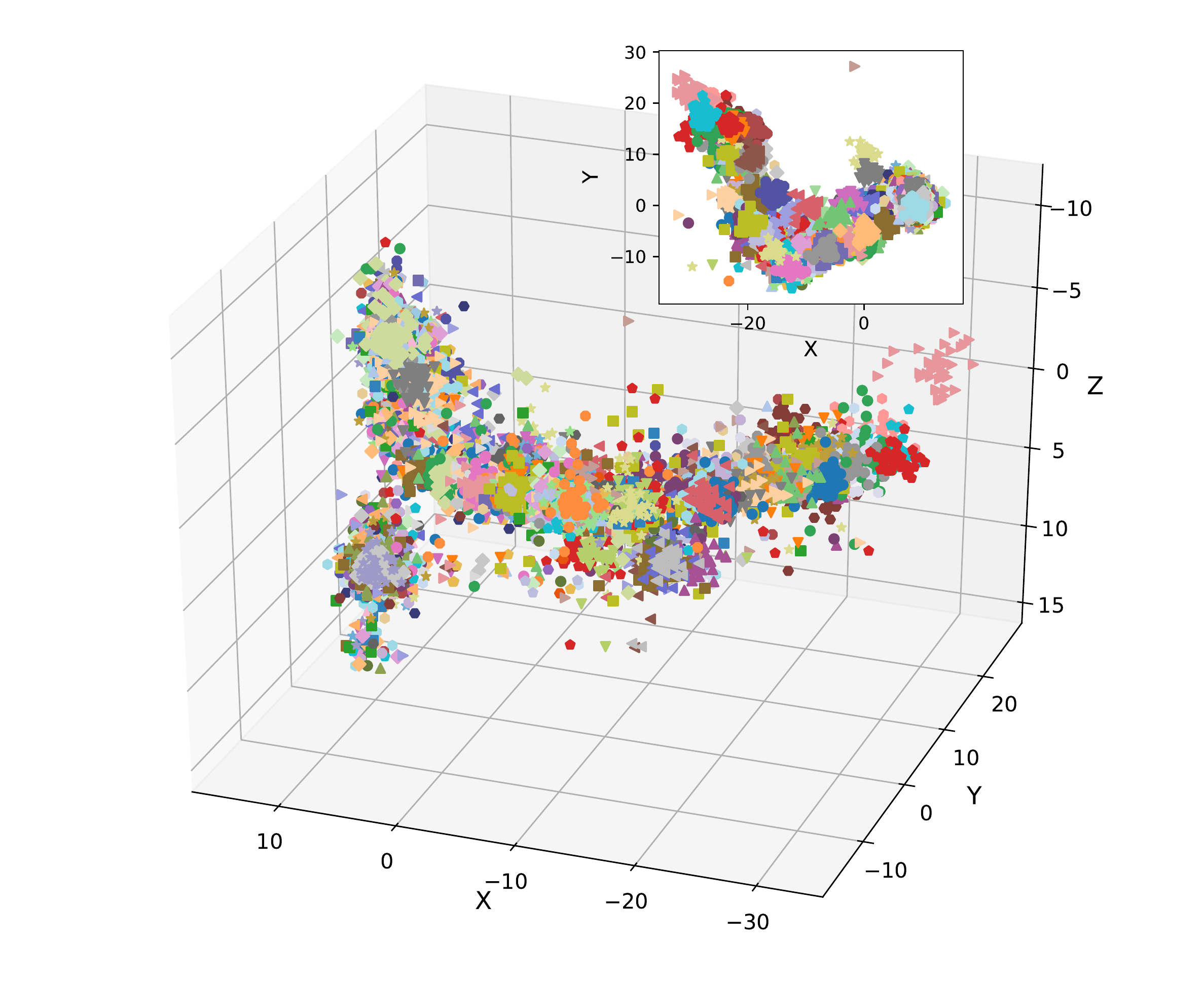}
    \vspace{-5mm}
    \caption{Yahoo} %{Light Unit}
    \label{fig:lda_yahoo}
\end{subfigure}
\vspace{-4mm}
\caption{Linear Discriminant Analysis (200 targeted search queries, using \emph{PSC} as feature)
% collected with Bing (a) and DuckDuckGo (b) using packet size count features. 
% The percentage of variance explained by each component is (,,) for Bing samples and (,,) for DuckDuckGo samples.
}
\vspace{-7mm}
\label{fig:lda}
\end{figure*}

% \begin{figure*}[!htb]
%     \centering
%     \begin{tabular}[t]{ccc}
%     % \hline
% \begin{subfigure}{0.33\textwidth}
%     \centering
%     \smallskip
%     \includegraphics[width=1\textwidth]{figures/movie_1_bing.pdf}
%     \caption{Cat 1} %{Light Unit}
% \end{subfigure}
%     &
%         \begin{tabular}{c}% if you add [t], than sub images are pushed down
%         \smallskip
%             \begin{subfigure}[t]{0.3\textwidth}
%                 \centering
%                 \includegraphics[width=0.8\textwidth]{figures/movie_2_bing.pdf}
%                 \caption{Cat 2}
%             \end{subfigure}\\
%             \begin{subfigure}[t]{0.3\textwidth}
%                 \centering
%                 \includegraphics[width=0.8\textwidth]{figures/movie_2_ddg.pdf}
%                 \caption{Cat 3}
%             \end{subfigure}
%         \end{tabular}
        
%         &
% \begin{subfigure}{0.33\textwidth}
%     \centering
%     \smallskip
%     \includegraphics[width=1\textwidth]{figures/movie_1_ddg.pdf}
%     \caption{Cat 1} %{Light Unit}
% \end{subfigure}
%     \end{tabular}
%     \caption{Cats}
% \end{figure*}
\vspace{-2mm}
\begin{table}[H]
    \centering
    \caption{Network Traffic Statistics (Chrome)}
    \vspace{-3mm}
    \resizebox{1.0\linewidth}{!}{
    \begin{tabular}{|l|cccc|}
    \hline
    & Yahoo &  Bing & Google & DuckDuckGo \\ \hline
        
    % Mean no. of packets & 3018 & 1511 & 1223 & 732  \\ 
    Median no. of packets &  2989 & 1476 & 1195 & 723  \\ 
    Std Dev of no. of packets & 343.3 & 192.3 & 128.1 & 66.1 \\ 
    % \hline
    
    % Mean no. of TCP conn. & 28 & 4 & 9.7 & 6.1 \\ 
    Median no. of TCP conn. & 27 & 4 & 8 & 6  \\ 
    Std Dev of no. of TCP conn. & 6.4 & 0.7 & 3.2 & 1.1 \\ 
    % \hline
    
    Median homepage load time (s) & 3.05 & 0.52 & 1.00 & 1.05 \\
    \hline
    \end{tabular}
    }
    \label{tab:packet_num}
\end{table}
\vspace{-3mm}
\begin{table}[ht]
    \centering
    \caption{Second-level Server Names (Chrome)}
    \vspace{-2mm}
    \resizebox{0.75\linewidth}{!}{
    \begin{tabular}{lr|lr}
        \hline
        \multicolumn{2}{l|}{\emph{2019-02-23-23-58-37}} & \multicolumn{2}{l}{\emph{2019-02-24-01-02-04}} \\ \hline
        duckduckgo.com & 6 & duckduckgo.com & 5 \\ \hline
        % \vspace{-2mm}
        % \multicolumn{4}{c}{} \\ \hline
        \multicolumn{2}{l|}{\emph{2019-02-24-02-03-54}} &
        \multicolumn{2}{l}{\emph{2019-02-24-03-01-04}} \\ \hline
        duckduckgo.com & 6 & duckduckgo.com & 6 \\ \hline
        
        \multicolumn{4}{c}{\multirow{2}{*}{\textbf{(a) DuckDuckGo}}} \\
        \\
        \hline
        \multicolumn{2}{l|}{\emph{2019-02-23-23-58-39}} & \multicolumn{2}{l}{\emph{2019-02-24-01-02-06}} \\ \hline
        yimg.com & 9 & yimg.com & 9 \\ 
        yahoo.com & 5 & yahoo.com & 6 \\
        adtechus.com & 1 & atwola.com & 5 \\
        atwola.com & 1 & scorecardresearch.com & 1 \\
        scorecardresearch.com & 1 & adtechus.com & 1 \\
        google.com & 1 & nexac.com & 1 \\
        bing.com & 1 & addthis.com & 1 \\
         &  & google.com & 1  \\
         &  & bing.com & 1 \\
        \hline
%        \vspace{-2mm}
%        \multicolumn{4}{c}{} \\ \hline
        \multicolumn{2}{l|}{\emph{2019-02-24-02-03-55}} &
        \multicolumn{2}{l}{\emph{2019-02-24-03-01-05}} \\ \hline
        yimg.com & 9 & yimg.com & 9 \\
        yahoo.com & 6 & yahoo.com & 6 \\
        atwola.com & 5 & bing.net & 6 \\
        yahoodns.net & 5 & atwola.com & 5 \\
        scorecardresearch.com & 1 & scorecardresearch.com & 1 \\
        adtechus.com & 1 & adtechus.com & 1 \\
        krxd.net & 1 & krxd.net & 1 \\
        nexac.com & 1 & bing.com & 1 \\
        addthis.com & 1 &  & \\
        google.com & 1 &  & \\
        bing.com & 1 &  & \\
        \hline
        \multicolumn{4}{c}{\multirow{2}{*}{\textbf{(b) Yahoo}}} \\
    \end{tabular}}
    \vspace{-3mm}
    \label{tab:hostname_yahoo_ddg}
\end{table}
\begin{itemize}
    \item \emph{Amount of Traffic Generated:} 
    % The network traffic generated when searching the exact same query at approximately the same time with four search engines (any given column in \figurename{ \ref{fig:search_engine}}) differs significantly.
    % In \tablename{ \ref{tab:packet_num}}, we summarize statistics with respect to the number of packets generated and number of TCP connections initiated by the four search engines, for \emph{all} search sessions collected using Chrome.
    % \markred{in the traces illustrated in Figure \ref{fig:search_engine}}\textcolor{blue}{(not for the example in figure 1 but the overview for all visits)}.
    % \item 
    The median number of packets (as well as standard deviation) is largest for Yahoo, followed by Bing, Google, and DuckDuckGo (\tablename{ \ref{tab:packet_num}}). 
    A smaller number of packets may result from simpler content in the responses returned for a search query. For example, most of the responses returned by DuckDuckGo are text-based, in contrast to a significant presence of image-based responses from the other three search engines.
    % \sout{and Yahoo searches initiate connections to many third-party servers}.
        
    \item \emph{Third-party Connections:} Yahoo initiates a much larger number of TCP connections (and with largest deviation). To understand who a user communicates with when using different search engines, we study the second-level server names extracted from the SNI extension field of TLS in the network traces of \figurename{ \ref{fig:search_engine}}. \tablename{ \ref{tab:hostname_yahoo_ddg}} lists the number of TCP connections initiated to different servers when using DuckDuckGo and Yahoo.
    While DuckDuckGo searches contact only \emph{duckduckgo.com} for serving query results, Yahoo searches involve communicating with several third-party servers. Among those servers, some are owned by other web search engines, such as \emph{google.com} and \emph{bing.com},\footnote{By further inspecting the traffic, we observe contents retrieved under \emph{google.com} are mostly from \emph{ade.googlesyndication.com}, \emph{www.googletagservices.com} and \emph{googleads.g.doubleclick.net}. On the other hand, Yahoo retrieved some images on the result page from \emph{bing.com} \cite{bing-yahoo}. }
    while some are other marketing/advertisement service providers, such as \emph{adtechus.com} \cite{adtechus}, \emph{scorecardresearch.com} \cite{scorecardresearch} and \emph{atwola.com}
     \cite{atwola}. 
    Content served from marketing and advertisement servers can be quite dynamic---we believe this adds significant noise to the traffic that contains the actual search query results, and may affect the classification accuracy. Hence the accuracy achieved with Yahoo is relatively lower.

    \item \emph{Stability of Response Over Time:} Overall, the generated traffic pattern for the same search query---packet ordering and packet sequence within each TCP connection---is relatively more stable over time with DuckDuckGo and Bing, compared to the other two search engines (\figurename{ \ref{fig:search_engine}}).
    Such stability ensures greater similarity between training and test samples and helps attackers fingerprint with higher accuracy.
    % samples may help attackers to correctly classify testing samples by measuring similarities with training samples. 
    Thus, fingerprinting accuracy with DuckDuckGo is expected to be high (\tablename{ \ref{tab:search_engine}}).
    % However, Bing is the most robust in face of fingerprinting among those four . 
%    So time alone does not explain the performance differences.
    
    \item \emph{Identifiability of Keywords:} Finally, we perform linear discriminant analysis \cite{mika1999fisher} using Bing, DuckDuckGo, and Yahoo traffic samples, and using packet size count as features. 
    We consider 200 targeted search queries from the training dataset, each with 36 samples, and project them onto the three-dimensional subspace that optimizes both the variance of the data and the separation between different search keywords in \figurename{ \ref{fig:lda}}---different colors and markers represent different search keywords.
    As can be seen, samples associated with \emph{different} keywords are more densely clustered with Bing, while these are more separable with DuckDuckGo---which indicates that it is much easier to separate samples for different keywords collected by DuckDuckGo, as compared to Bing. Thus, even though Bing does not generate significant third party connections,\footnote{Most Bing/Chrome traffic traces consists of 4 TCP connections -- one connection from each of \emph{bing.com}, \emph{bingparachute.com}, \emph{live.com} and \emph{microsoftonline.com}.} the accuracy is still low due to similarity between traces associated with different keywords. 
    % Besides, the least variances observed with Bing samples in \tablename{ \ref{tab:packet_num}} in terms of the number of TCP connections further demonstrate the similarity.
    
    For Yahoo, samples for the \emph{same} keyword do not cluster as well as the other two, possibly due to the dynamic background and third-party traffic---which explains a lower classification accuracy.
    % \textcolor{blue}{Thus the main cause of the low classification accuracy with Bing and Yahoo differs: for Bing, the main reason }
    
\end{itemize}

\subsection{Homepage vs. Addressbar Searching}
\label{subsec:searching_mode}

Users may enter keywords either in the search box on the homepage of a search engine, or directly in the address bar of their browser (using the default search engine). We simulate two corresponding searching modes---\emph{homepage searching} and \emph{addressbar searching}---depending on whether or not we include the network traffic generated when loading the homepage of the search engine. \tablename{ \ref{tab:search_mode}} summarizes the accuracy obtained in \emph{addressbar searching} mode with 1,440 targeted keywords using the same setup as in Section \ref{subsec:vul_search_engine}.
Compared with the results obtained in \emph{homepage searching} mode (\tablename{ \ref{tab:search_engine}}), we observe:

\vspace{-2mm}
\begin{table}[htbp]
    \centering
    \caption{Classification Accuracy: \emph{Addressbar Searching}}
    \vspace{-2mm}
    \resizebox{1.0\linewidth}{!}{
    \begin{tabular}{|l|cccc|}
    \hline
        \textbf{Chrome} & Bing & Yahoo  & Google & DuckDuckGo  \\ \hline
         \emph{Wfin++} 
         & 45.18 $\pm$ 0.07
         & \fbox{86.62 $\pm$ 0.08} 
         & \fbox{64.05 $\pm$ 0.06}
         & 96.51 $\pm$ 0.02\\
         \emph{PSC} 
         & \fbox{45.69 $\pm$ 0.14} 
         & 84.22 $\pm$ 0.02 
         & 59.05 $\pm$ 0.05 
         & \fbox{96.61 $\pm$ 0.02} \\ 
         \emph{EtResp} 
         & 21.54 $\pm$ 0.04
         & 0.80  $\pm$   0.02
         & 28.61 $\pm$ 0.05 
         & 55.59 $\pm$ 0.06\\
        %  \emph{EtResp} 
        %  & 8.77 $\pm$ 0.04
        %  & 1.01	  $\pm$   0.02
        %  & 30.08 $\pm$ 0.05 
        %  & 43.85 $\pm$ 0.06\\
         \emph{k-FP} 
         & 8.97 $\pm$ 0.12 
         & 2.83	 $\pm$  0.03
         & 26.24 $\pm$ 0.13 
         & 41.12 $\pm$ 0.09\\
       
        \hline
        % \multicolumn{5}{c}{} \\ 
        \hline
         
        \textbf{Firefox} & Bing & Yahoo & Google & DuckDuckGo  \\ \hline
         \emph{Wfin++} 
         & \fbox{45.76 $\pm$ 0.05} 
         & \fbox{86.00 $\pm$ 0.04} 
         & 78.44 $\pm$ 0.13 
         & \fbox{92.87 $\pm$ 0.03} \\
         \emph{PSC} 
         & 45.69 $\pm$ 0.17 
         & 83.65 $\pm$ 0.13 
         & \fbox{79.27 $\pm$ 0.08} 
         & 92.44 $\pm$ 0.04\\
%          \emph{EtResp} &  7.55 $\pm$ 0.12 & 1.76 $\pm$ 0.03 & 22.22 $\pm$ 0.09 & 24.72 $\pm$ 0.03
% \\

         \emph{EtResp} &  13.06 $\pm$ 0.12 & 0.61 $\pm$ 0.03 & 25.58 $\pm$ 0.09 & 29.85 $\pm$ 0.03
\\

         \emph{k-FP} & 7.49 $\pm$ 0.06 & 3.27 $\pm$ 0.02 & 17.77 $\pm$ 0.06 & 25.98 $\pm$ 0.04
 \\
         \hline
    \end{tabular}
    }
    \vspace{-2mm}
    \label{tab:search_mode}
\end{table}

\begin{itemize}

\item For Bing, Google, DuckDuckGo, the fingerprinting accuracy achieved with Chrome increase significantly with \emph{EtResp} and \emph{k-FP}, but increases only slightly with \emph{PSC} and \emph{Wfin++}.
% while the improvement is more significant with \emph{EtResp} and \emph{k-FP}. 
Since \emph{EtResp} and \emph{k-FP} use more coarse-grained features such as statistical derivatives of number of packets (versus fine-grained features such as packet size count used in \emph{PSC} and \emph{Wfin++}), their performance is more likely to be impaired by the additional noise introduced in homepage searching.

\item For Yahoo, the performance is increased  significantly even with \emph{Wfin++}---from  60\% to 86\%.
This may be attributed to the elimination of dynamic traffic generated when loading the homepage of Yahoo, which contains news, weather, and ads. 
Indeed, when we repeat the analysis of \figurename{ \ref{fig:search_engine}} for the \emph{addressbar searching} mode with Yahoo, we observe a significant decrease in the number of packets as well as number of TCP connections (details in Appendix \ref{appendix:traffic_pattern_addressbar}). 

% \vspace{-1mm}
\end{itemize}
In what follows, we present evaluations with only \emph{Wfin++} and \emph{PSC}, and homepage searching results are included only in the appendices.

% \textcolor{blue}{When taking a closer look at the generated traffic pattern and the second-level server name when searching the same query used for \figurename{ \ref{fig:search_engine}}}, we have the following observations (Appendix \ref{appendix:traffic_pattern_addressbar}:  \figurename{ \ref{fig:search_engine_no_homepage}}):
% \input{tables/search_mode_acc.tex}
% \begin{itemize}
%     \item For DuckDuckGo and Google, less packets are observed and the traffic pattern become relative more stable after removing homepage traffic, 
%     which may lead to the increasing fingerprinting accuracy;
%     \item For Yahoo, we observe a significant decreasing number of packets transferred in the 
%     \emph{addressbar searching} mode as well as less TCP connections. 
%     However, the traffic patterns are still quite dynamic over time compared with the others and there are tracking/advertisement servers transferring contents in the background, such as \emph{addthis.com}\cite{addthis} and \emph{krxd.net}\cite{krux} ( displayed in Appendix \ref{appendix:traffic_pattern_addressbar}: \tablename{ \ref{tab:hostname_yahoo_no_homapge}}). 
% \end{itemize}

\subsection{Impact of Disregarding Client Platforms?}
\label{subsec:client_platform}

Alan and Kaur \cite{alan2019client} show that differences in client browser platforms represented in training and test data can significantly impair the accuracy of fingerprinting webpages. Our evaluations presented so far have trained and tested with data collected using the \emph{same} client browser.
% examine how client diversity impacts the performance of HTTPS webpage fingerprinting due to significant browser differences that resulted in the various size of packets carrying HTTP request, HTTP/2.0 implementations and client specific connections . 
Next, we investigate the performance of keyword fingerprinting in cross-browser attacks (that train with data from one type of browser and test with another), using \emph{Wfin++} and \emph{PSC}. 

To minimize the variations introduced by different time of data collection  on classification performance, we study the cross-browser impact with Firefox vs. Chrome samples (both collected in Feb 2019), and with Edge vs. Safari samples (both collected in Jan-Feb 2019).\footnote{Cross-browser attack results with Edge vs. Safari are included in Appendix \ref{appendix:cross_browser_edge}.}
For comparison, we also evaluate the classification accuracy when training with samples from \emph{both} Firefox and Chrome browsers.

\vspace{-2mm}
\begin{table}[htbp]
    \centering
    \caption{Classification Accuracy: Cross-browser Attacks in \emph{Addressbar Searching}}
    % \vspace{-4mm}
    \resizebox{1\linewidth}{!}{
    \begin{tabular}{|l|cc|cc|}
   
    % \multicolumn{5}{c}{\textbf{{Homepage Searching Mode}}} \\ 
     
    % \hline
    % \multirow{2}{*}{\textbf{DDG}} & \multicolumn{2}{c|}{Firefox} & \multicolumn{2}{c|}{Chrome} \\ \cline{2-5}
    % & \emph{Wfin++} & \emph{PSC}  & \emph{Wfin++} & \emph{PSC} \\ \hline
    % Firefox & 91.95 $\pm$ 0.10 & \fbox{92.23 $\pm$ 0.02} & 1.22 $\pm$ 0.09 & 1.82 $\pm$ 0.05 \\ 
    % Chrome & 0.22 $\pm$ 0.01&0.97 $\pm$ 0.05 & 96.15 $\pm$ 0.01 & \fbox{96.33 $\pm$ 0.04} \\
    % Fire/Chr & 91.40	$\pm$ 0.08& \fbox{92.14 $\pm$ 0.06}& \fbox{96.30 $\pm$ 0.04} & 96.04 $\pm$ 0.04 \\ 
    
    % \hline
    % % \multicolumn{5}{c}{} \\ 
    % \hline
    
    % \multirow{2}{*}{\textbf{Google}} & \multicolumn{2}{c|}{Firefox} & \multicolumn{2}{c|}{Chrome} \\ \cline{2-5}
    % & \emph{Wfin++} & \emph{PSC}  & \emph{Wfin++} & \emph{PSC} \\ \hline
    % Firefox & 75.56 $\pm$ 0.07 & \fbox{76.75 $\pm$ 0.02} & 0.38 $\pm$ 0.07 &1.42 $\pm$ 0.04  \\ 
    % Chrome &0.07& 0.98 $\pm$ 0.13 & \fbox{60.32 $\pm$ 0.07} & 57.72 $\pm$ 0.05 \\
    % Fire/Chr & 75.17 $\pm$ 0.13& \fbox{75.88 $\pm$ 0.13} & \fbox{61.91 $\pm$ 0.31} &57.12 $\pm$ 0.14\\ 
    % \hline  
    % \hline
    % \multirow{2}{*}{\textbf{Yahoo}} & \multicolumn{2}{c|}{Firefox} & \multicolumn{2}{c|}{Chrome} \\ \cline{2-5}
    % & \emph{Wfin++} & \emph{PSC}  & \emph{Wfin++} & \emph{PSC} \\ \hline
    % Firefox & 58.06 $\pm$ 0.25 & 49.87 $\pm$ 0.13 & 15.61 $\pm$ 0.35 & 12.42 $\pm$ 0.21 \\ 
    % Chrome & 4.01 $\pm$ 0.39& 2.47 $\pm$ 0.04& 64.60 $\pm$ 0.21 & 57.85 $\pm$ 0.26\\
    % Fire/Chr  & 
    % \\ 
    
    % \hline

    % % \multicolumn{5}{c}{} \\
    % \multicolumn{5}{c}{\textbf{{Addressbar Searching Mode}}} 
    % \\ 
    
    \hline
    \multirow{2}{*}{\textbf{DDG}} & \multicolumn{2}{c|}{Firefox} & \multicolumn{2}{c|}{Chrome} \\ \cline{2-5}
    & \emph{Wfin++} & \emph{PSC}  & \emph{Wfin++} & \emph{PSC} \\ \hline
    Firefox & \fbox{92.87 $\pm$ 0.03}&92.44 $\pm$ 0.04 &0.69 $\pm$ 0.06 &0.91 $\pm$ 0.03  \\ 
    Chrome & 0.80 $\pm$ 0.06 & 1.54 $\pm$ 0.07 & 96.51 $\pm$ 0.02 & \fbox{96.61 $\pm$ 0.02}\\
    Fire/Chr  & \fbox{92.26 $\pm$ 0.03} & 92.18 $\pm$ 0.03
    & 96.49 $\pm$ 0.02 & \fbox{96.52 $\pm$ 0.04}
    \\ 
    
    \hline
    % \multicolumn{5}{c}{} \\ 
    \hline
    
    \multirow{2}{*}{\textbf{Google}} & \multicolumn{2}{c|}{Firefox} & \multicolumn{2}{c|}{Chrome} \\ \cline{2-5}
    & \emph{Wfin++} & \emph{PSC}  & \emph{Wfin++} & \emph{PSC} \\ \hline
    Firefox & 78.44 $\pm$ 0.13 & \fbox{79.27 $\pm$ 0.08} & 1.41 $\pm$ 0.08 & 1.61 $\pm$ 0.04 \\ 
    
    Chrome &1.65 $\pm$ 0.08& 0.77 $\pm$ 0.01 & \fbox{64.05 $\pm$ 0.06} & 59.05 $\pm$ 0.05 \\ 
    
    Fire/Chr  & 77.57 $\pm$ 0.22 & \fbox{77.62 $\pm$ 0.05} &\fbox{64.94 $\pm$ 0.09} & 58.48 $\pm$ 0.17
    \\ 
    \hline
     % \multicolumn{5}{c}{} \\ 
    \hline
    
    \multirow{2}{*}{\textbf{Yahoo}} & \multicolumn{2}{c|}{Firefox} & \multicolumn{2}{c|}{Chrome} \\ \cline{2-5}
    & \emph{Wfin++} & \emph{PSC}  & \emph{Wfin++} & \emph{PSC} \\ \hline
    Firefox & \fbox{85.95 $\pm$ 0.09} & 83.63 $\pm$ 0.13 & 53.19 $\pm$ 0.52 & 48.68 $\pm$ 0.59 \\ 
    Chrome & 22.09 $\pm$ 0.38 & 25.39 $\pm$ 0.63& \fbox{86.66 $\pm$ 0.11} & 84.17 $\pm$ 0.08\\
    Fire/Chr  & \fbox{85.66 $\pm$ 0.17} & 83.71 $\pm$ 0.19 & \fbox{85.65 $\pm$ 0.07} & 83.88 $\pm$ 0.27 \\ 
    
    \hline
    
    \end{tabular}
    }
    \vspace{-2mm}
    \label{tab:cross_browser}
\end{table}

The classification results with Google, DuckDuckGo, and Yahoo are displayed in \tablename{ \ref{tab:cross_browser}}. In both \emph{addressbar searching} and \emph{homepage searching} modes, we find that when there is a mismatch between the training browser and testing browser, the classification accuracy is significantly lower. For instance, when the classifier is trained with DuckDuckGo/Firefox, the classification accuracy is around 92\% using test samples from DuckDuckGo/Firefox, but is less than 1\% while using test samples from DuckDuckGo/Chrome. An attacker that disregards the client browser platform, may fail in practice.

Furthermore, the performance achieved by including samples from \emph{both} Firefox and Chrome during training is comparable to (or even slightly better than) what is achieved when trained with a single matching browser. 
% compared with training with Firefox or Chrome samples alone. 
For instance, when testing with DuckDuckGo/Chrome, the accuracy obtained by \emph{PSC} and \emph{Wfin++} is around 96\% in both cases.
This observation stresses the importance of incorporating diverse browser platforms during training, in order to achieve good fingerprinting accuracy in practice. 

% Furthermore, we evaluate the performance of the cross-browser attack in \emph{addressbar searching} mode and show the result in \tablename{ \ref{tab:cross_browser}} with Google and DuckDuckGo respectively. 
% Consistent with the results obtained in Section \ref{subsec:searching_mode}, the accuracy is slightly improved  after eliminating homepage traffic compared with \tablename{ \ref{tab:cross_browser}}. 
% For instance, the highest accuracy is increased from 0.97\% to 1.54\% after removing homepage traffic when training with DuckDuckGo/Chrome and testing with DuckDuckGo/Firefox. 
% However, when training and testing samples are not generated using the same browser, the performance of classifier is degraded drastically. 
% Hence the impact of disregarding the client platform is not only limited to \emph{homepage searching} mode but also in \emph{addressbar searching} mode.

\parlabel{Browser Specific Communication} We  next examine browser-specific communication---\tablename{ \ref{tab:firefox_specfic}} lists the second-level server names observed uniquely in Firefox (not observed in Chrome), with DuckDuckGo and Yahoo. 
We find that Firefox generates a significant fraction of connections to its own servers---\emph{mozilla.com}, \emph{mozilla.net} and \emph{mozilla.org} (this was true across all four search engines---DuckDuckGo: 40.06\%, Google: 33.53\%, Bing: 39.93\%, and Yahoo: 9.73\%). 
Furthermore, almost all unique connections came from mozilla servers when Firefox used DuckDuckGo, Google, and Bing; however, many other unique server names were observed with Yahoo, such as \emph{alephd.com} and \emph{smartadserver.com}---this indicates the impact of different browsers may also differ across search engines.
This also suggests that the appearance of unique server names in a network trace may be used by an attacker to infer the browser (and then select the corresponding trained model for fingerprinting). An attacker may also train a multi-class classifier using features based on server names to help identify the browser \cite{yen2009browser}.

\vspace{-2mm}
\begin{table}[htbp]
    \centering
    \vspace{-2mm}
    \caption{Firefox-specific Server Names}
    \resizebox{0.95\linewidth}{!}{%
    \begin{tabular}{llllll}
    \hline
    
     \textbf{Rank} & \textbf{Server name} & \textbf{Popularity} & \multicolumn{3}{c}{\textbf{(Continued)}}
    \\ \hline
     % \multicolumn{3}{l}{} \\
     \multicolumn{3}{c}{\underline{\textbf{DuckDuckGo (40.06\%)}}}  
     & 9 & specificmedia.com & 0.08\% \\
     
    1 & mozilla.com & 22.89\% 
    & 10 & adroll.com & 0.06\% \\
    2 & mozilla.net & 11.45\% 
    & 11 & reson8.com & 0.04\%
    \\
    3 & mozilla.org & 5.72\% 
    & 12 & 3lift.com & 0.03\%
    \\
    
    % \multicolumn{3}{l}{\underline{\textbf{Google (33.53\%)}}}  \\
    % % Rank & Domain name & Popularity \\ \hline
    % 1 & mozilla.com & 19.16\% \\
    % 2 & mozilla.net & 9.59\% \\
    % 3 & mozilla.org & 4.79\% \\
    % 4 & 2mdn.net & $<$0.01\% \\
    
    % \multicolumn{3}{l}{} \\
    
    % \multicolumn{3}{l}{\underline{\textbf{Bing (39.93\%)}}}  \\ 
    % % Rank & Domain name & Popularity \\ \hline
    % 1 & mozilla.com & 22.86\% \\
    % 2 & mozilla.net & 11.35\% \\
    % 3 & mozilla.org & 5.72\% \\
    
   % \multicolumn{3}{l}{} \\
    
    \multicolumn{3}{c}{\underline{\textbf{Yahoo (10.76\%)}}} & 13 & marriott.com & 0.03\%
    \\
    1 & mozilla.com & 5.56\% 
    & 14 & impdesk.com & 0.03\%
    \\
    2 & mozilla.net & 2.78\% 
    & 15 & yieldmo.com & 0.03\%
    \\
    3 & mozilla.org & 1.39\% 
    & 16 & pswec.com & 0.03\%
    \\
    4 & alephd.com & 0.14\% 
    & 17 & media.net & 0.03\% 
    \\
    5 & smartadserver.com & 0.12\% 
    & 18 & solocpm.com & 0.02\%
    \\
    6 & vindicosuite.com & 0.12\% 
    & 19 & webmd.com & 0.02\% 
    \\
    7 & jivox.com & 0.11\% 
    & 20 & extend.tv & 0.02\% 
    \\
    8 & akamaihd.net & 0.1\% \\
    % 9 & specificmedia.com & 0.08\% \\
    % 10 & adroll.com & 0.06\% \\
    % 11 & reson8.com & 0.04\% \\
    % 12 & 3lift.com & 0.03\% \\
    % 13 & marriott.com & 0.03\% \\
    % 14 & impdesk.com & 0.03\% \\
    % 15 & yieldmo.com & 0.03\% \\
    % 16 & pswec.com & 0.03\% \\
    % 17 & media.net & 0.03\% \\
    % 18 & solocpm.com & 0.02\% \\
    % 19 & webmd.com & 0.02\% \\
    % 20 & extend.tv & 0.02\% \\
    % \hline
    % Rank & Domain name & Popularity \\ \hline
      
    \hline
    \end{tabular}%
    }
    \vspace{-2mm}
    \label{tab:firefox_specfic}
\end{table}

\subsection{Eliminate Noise From ``Other'' Domains?}
\label{subsec:feature_set}

Based on our observations so far, there are at least two sources of noisy \emph{background} traffic---tracking/advertisement connections, and browser specific connections---that (even in \emph{addressbar searching} mode) may hinder achieving high fingerprinting accuracy. 
To eliminate such noise, we next study keyword fingerprinting when only TCP connections that serve the actual search results are considered.
% by analyzing the traffic generated by different search engines. 
Specifically, based on manual analysis, we include connections from \emph{google.com} and \emph{gstatic.com} for Google, \emph{duckduckgo.com} for DuckDuckGo, \emph{bing.com} for Bing, and \emph{yahoo.com} for Yahoo.\footnote{It is natural to ask---\emph{can noisy (and irrelevant) background traffic be removed automatically by using feature selection techniques?} 
% \emph{Why can not feature selection help discover important domains?} 
For that, we would need to define existing features at an additional domain-level granularity---e.g., instead of just \emph{packet size count}, we would need to define \emph{packet size count w.r.t. domain X}, for all encountered domains X.
% \emph{packet size count w.r.t. google.com/gstatic.com}, \emph{number of incoming/outgoing packets w.r.t. google.com/gstatic.com}, which 
That will lead to an explosion of the feature space due to a large number of unique domains, especially when large feature sets are considered \cite{hayes2016k,yan2018feature,Li:2018:MIL:3243734.3243832}. It  also adds to the \emph{curse of dimensionality} problem \cite{donoho2000high}. Instead, we select the manual analysis approach. To the best of our knowledge, no prior work has attempted elimination of traffic from other domains.}

% \vspace{-2mm}

\tablename{ \ref{tab:top_domains_yahoo}} summarizes the classification accuracy achieved with Yahoo. We observe a significant increase in classification accuracy, from around 86\% (\tablename{ \ref{tab:search_mode}}) to 95\% (increase was less significant for the other search engines---see Appendix \ref{appendix:top_domains_others}). 
% \markred{Include DDG, Bing, Google in Appendix.}
% 
% As can be seen, the probability of correctly classifying Yahoo samples is further increased from around 86\% (\tablename{ \ref{tab:search_mode}}) to 95\%. 
More importantly, the accuracy is high for Yahoo even in cross-browser attacks---around 91\% when training (testing) with Chrome (Firefox), and 82\% when training (testing) with Firefox (Chrome).\footnote{We did not observe any significant improvement in Google, Bing, and DuckDuckGo by eliminating connections from other domains, since the traffic of these contains only a limited number of third-party connections.}
% As shown in Section \ref{subsec:searching_mode}, there are still traffic served from third-party servers when searching with Yahoo in \emph{addressbar searching} mode, which makes the same query searching at different times dynamic. 
% Through manually inspection, we found that nearly all search results are served from \emph{yahoo.com}.
Thus, we find that when significant amount of noise from other domains is eliminated, keyword fingerprinting 
% \textcolor{blue}{may become? since we did not observe much improvements in other three search engine} 
may become vulnerable to even cross-browser attacks.

\begin{table}[htbp]
    \centering
    \vspace{-3mm}
    \caption{Classification Accuracy (only \emph{yahoo.com} connections considered with Yahoo, \emph{addressbar searching} mode)}
    
    \resizebox{1\linewidth}{!}{
    \begin{tabular}{|l|cc|cc|}

    \hline
    \multirow{2}{*}{\textbf{Yahoo}} & \multicolumn{2}{c|}{Firefox} & \multicolumn{2}{c|}{Chrome} \\ \cline{2-5}
    & \emph{Wfin++} & \emph{PSC}  & \emph{Wfin++} & \emph{PSC} \\ \hline
    Firefox & \fbox{94.60 $\pm$ 0.03} & 94.37 $\pm$ 0.05 & 91.39 $\pm$ 0.09& \fbox{91.65 $\pm$ 0.12} \\ 
    Chrome & 76.91 $\pm$ 0.15 & \fbox{82.09 $\pm$ 0.18} & \fbox{96.27 $\pm$ 0.03} & 96.16 $\pm$ 0.06 \\
    \hline
    \end{tabular}
    }
    \vspace{-3mm}
    \label{tab:top_domains_yahoo}
\end{table}

% \vspace{-6mm}
% \begin{table}[htbp]
%     \centering
%     \caption{Classification Accuracy (all connections considered with Yahoo, \emph{addressbar searching} mode)}
%     \vspace{-3mm}
%     \resizebox{1\linewidth}{!}{
%     \begin{tabular}{|l|cc|cc|}

%     \hline
   
%     \hline
%     \end{tabular}
%     }
%     \vspace{-3mm}
%     \label{tab:all_yahoo}
% \end{table}

% after pruning traffic, we observe an obvious improvement.
% \markred{Why no Yahoo in \tablename{ \ref{tab:cross_browser}}?}

% \markred{You seem to be using DDG and Yahoo in some tables and DDG and Google in others---we should either include everyone, or be consistent in the ones included in main paper vs. Appendix. Homepage searching can be moved to Appendix.}

% \textcolor{blue}{for space consideration at that time I think. Should we move results for homepage searching mode after Section 5.3 to appendix?}

% \markred{Also add homepage searching results in Appendix.}

\subsection{How Often to Re-train Classifier?}
\label{subsec:time_effect}

The auto-suggestion list and search results returned by a search engine depend on the current social trends and can be quite dynamic over time---thus, an attacker may need to frequently re-train the machine learning model to keep up its performance.
In order to understand how frequently the model may need to be updated in a closed-world scenario, we next study how the classification accuracy changes as the time gap between training and test samples is increased.
For this, we focus on Chrome and use the first 36 samples (out of 54) for each of the 1,440 targeted keywords for training, next 9 samples (37-45) for validation, and test with five datasets (\emph{test-3/4/8/10/14}) collected in different time periods. Each test dataset is composed of 9 samples for each of the 1,440 targeted keywords---\emph{test-3} contains the last 9 samples (out of 54) from the Chrome-cw dataset (\tablename{ \ref{tab:dataset_overview}}); \emph{test-4/8/10/14} are composed of samples from the \emph{Chrome-targeted} dataset collected during the second phase. 
% To measure the time gap, we calculate the average searching time difference between the last/first sample in the testing dataset and the last sample in the training dataset for each monitored query.
% \begin{itemize}
%     \item \emph{test-3}: the last 9 samples from each search query visited in the first phase in February 2019. The average time gap between the training samples and \emph{test-3} samples is from 20 to 30 hours.
    
%     \item \emph{test-4/8/10/14}: each dataset contains 9 samples for each monitored query collected in the second stage between February and March 2019. 
%     The average time gap between the training samples and \emph{test-4/8/10/14} samples is within the range (38, 66), (164, 188), (219, 244) and (331, 352) hours respectively. 
    
% \end{itemize}
% \input{tables/time_effect_close_world.tex}

% \vspace{-4mm}
\begin{table}[htbp]
    \centering
   
    \caption{Classification Accuracy: Impact of Time (Chrome, \emph{addressbar searching} mode)}
     \vspace{-2mm}
    \resizebox{0.95\linewidth}{!}{
    \begin{tabular}{|l|ccccc|}
    
    \hline
    % & \emph{test-3}  & \emph{test-4}  & \emph{test-8} & \emph{test-10} & \emph{test-14}\\ \hline
    \emph{Gap (Hours)} & (20-30) & (38-66) & (164- 188)& (219-244) & (331-352)\\
        
    \hline   
    % \multicolumn{6}{c}{} \\ 
    \hline
    
    \bf{DDG} & \emph{test-3}  & \emph{test-4}  & \emph{test-8} & \emph{test-10} & \emph{test-14}\\ \hline
      \emph{Wfin++} & 92.20 & 90.86 & 81.84 & 72.90 & 72.46\\
       \emph{PSC} & 92.85 & 91.09 & 80.34 & 69.65 & 69.15\\ 
     % \emph{EtResp} & 31.22 & 22.09 & 11.07 & 7.94 & 7.29\\
      % \emph{k-FP} & 37.35 & 34.78 & 27.99 & 20.30 & 19.72\\
        
        \hline
    % \multicolumn{6}{c}{} \\ 
    \hline
        
       \bf{Google} & \emph{test-3}  & \emph{test-4}  & \emph{test-8} & \emph{test-10} & \emph{test-14}\\ \hline 
       \emph{Wfin++} & 55.14 &43.34 & 22.53 & 19.99 & 22.16\\
       \emph{PSC} & 47.74 & 34.67 & 9.75 & 10.14 & 14.72\\
      % \emph{EtResp} & 26.92 &24.05 & 18.78 & 12.91 & 12.34\\
      % \emph{k-FP}  & 23.96 & 24.05 & 22.19 & 17.29 & 15.54\\
        \hline
    \end{tabular}}
    \vspace{-2mm}
    \label{tab:time_effect_addressbar}
\end{table}
% \vspace{-2mm}

\tablename{ \ref{tab:time_effect_addressbar}} summarizes the classification accuracy (\emph{addressbar searching} mode) against the time gap between the last training sample and the testing samples (specified as a range in number of hours).
% for each testing dataset and the classification accuracy  for DuckDuckGo/Chrome in \emph{addressbar searching} mode.
As expected, the accuracy decreases as the time gap between training and testing samples increases. 
However, the rate of decline differs across search engines. 
For instance, the accuracy achieved with Google decreases from 55\% to 22\%, while with DuckDuckGo, it decreases from 92\% to 72\%.\footnote{Similar trend is observed in \emph{homepage searching} (Appx. \ref{appendix:time_effect_addressbar}:  \tablename{ \ref{tab:time_effect_addressbar}}).}
Thus, in order to achieve high fingerprinting accuracy, an attacker may need to re-train their classifier every 30 hours for Google and every 66 hours for DuckDuckGo. The lower retraining frequency needed for DuckDuckGo further strengthens our observation in Section \ref{subsec:vul_search_engine} about the stability of samples over time.

\section{Open World Evaluations}
% {---Impact of Background Query Set Size}
\label{sec:open_world}

\parlabel{Our Objective} 
We next go beyond the ``closed-world" assumption and consider the more realistic scenario in which the user may search for many keywords in the wild---the goal of the attacker is then to either: (i) determine whether the user is searching for one of the targeted keywords, or (ii) further identify the specific targeted keyword.
% establish which one. 
The first goal can be formalized as a binary classification problem. The second goal can be pursued by either applying an additional classifier trained with only targeted keywords, after differentiating targeted samples from non-targeted ones with the binary classifier (multi-level classification), or by training a new classifier using both targeted and non-targeted samples to directly determine which targeted keyword is being searched for by the user (multi-class classification).
Our aim is to study: (i) is keyword fingerprinting indeed a potential privacy concern in open-world scenarios? (ii) how does the number of non-targeted training and testing samples affect the classification performance? (iii) which classification scheme is likely to yield better performance for an attacker?
% and (iii) \markred{how often does the classifier need to be re-trained, especially given the time overhead involved in collecting a large scale non-targeted training dataset?}

\parlabel{How Many Non-targeted Keywords Can Be Encountered?}
% \textcolor{blue}{Move this after Limiting Time Effect?}
The number of non-targeted keywords represented in our dataset is 200K+. 
% Is this representative enough of the scale that an attacker might encounter in practice?
% testing samples in our experiments is around 203k (including both targeted and non-targeted ones), 
This corresponds to roughly 2-3 seconds worth of worldwide query traffic processed by Google \cite{0ne-second-stats}. Furthermore, according to \cite{pageload_time}, the average load time of a webpage was 3.21 seconds in 2017.
Given our threat model, in which an attacker is eavesdropping on the access link of a user, the volume of search queries he/she may encounter during a 3 second search session is likely to be significantly smaller than 200K.
Thus, we believe our dataset is representative enough of the scale that an attacker may encounter in practice.

\parlabel{Limiting the Time Effect} 
Our closed-world evaluations revealed that search traffic signatures can change with time, requiring frequent classifier re-training to achieve high performance. In order to reduce the impact of time in our open-world evaluations: 
% To reasonably compare the classification performance across different search engines in open-world scenario: 
(i) each keyword should be searched at approximately the same time with different search engines; and (ii) targeted and non-targeted keywords should be visited in similar time spans.
Given our scale of 200K+ keywords, visiting each non-targeted keyword once, with even {\em one} search engine, takes more than 3 days. 
Thus, for open-world evaluations,
we focus only on the DuckDuckGo and Google search engines, with Chrome browser in \emph{addressbar searching} mode---DuckDuckGo is the most vulnerable to keyword fingerprinting (Section \ref{sec:closed_experiment}), while Google is the most popular search engine worldwide.
For classifiers, we consider two best-performing feature sets in the closed-world scenario, \emph{Wfin++} and \emph{PSC}, with Extra-Trees (\emph{n\_estimators} = 700).\footnote{\emph{k-FP} and \emph{EtResp} do not perform as well even in closed-world evaluations. We do not expect them to perform well in open-world scenarios, which are more challenging due to large volumes of unseen non-targeted keywords.} We use the open-world dataset (\tablename{ \ref{tab:dataset_overview}})---45 samples are used for training and 9 for testing for each targeted keyword; and 1 sample is considered for either training or testing for each selected non-targeted search keyword (from a list of 235,767 keywords). 
The maximum number of testing samples in our experiments is around 203k (including both targeted and non-targeted ones).
% , which is approximately 2-3 seconds worth of worldwide query traffic processed by Google \cite{0ne-second-stats}. According to \cite{pageload_time}, the average load time of a webpage is 3.21 seconds in 2017.
% Since in the threat model, an attacker is residing in the same local network as the user, the volume of search queries he may encounter during a 3 seconds search session is likely to be order of magnitude less than 203k.
% Thus we believe the scale we considered is representative enough to illustrate keyword fingerprinting in reality.
% Recall that we have a list of 235,767 non-targeted keywords.
% The maximum number of non-targeted search samples considered \markred{in the network trace from a given user} is 235,767.

\begin{figure*}[htbp]
\centering

\begin{subfigure}{0.31\linewidth}
    \centering
    \smallskip
    \includegraphics[width=1\textwidth]{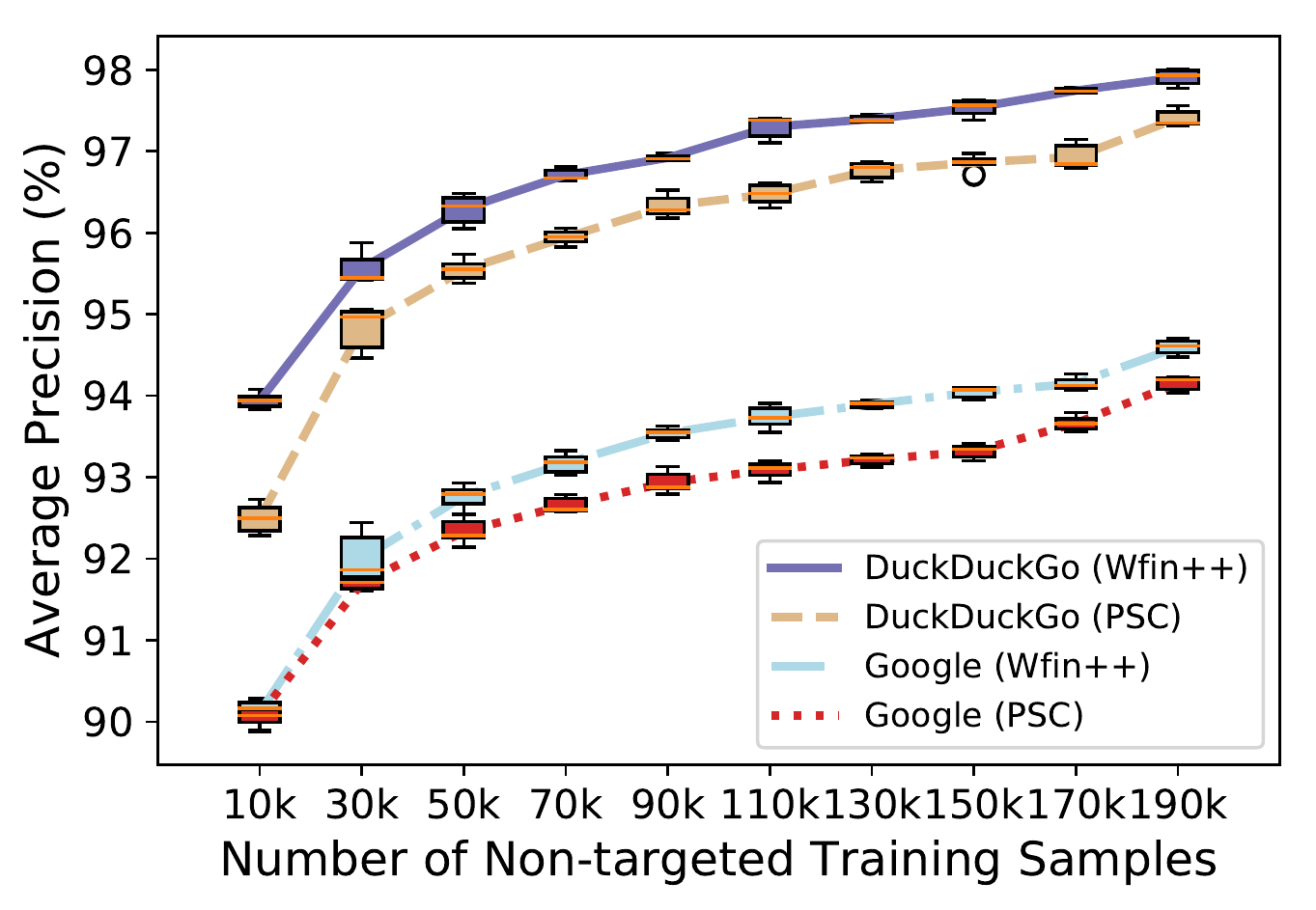}
        \vspace{-5mm}
\caption{AP: \# Non-targeted Training Samples} %{Light Unit}
    \label{fig:vary_binary_a}
\end{subfigure}
\begin{subfigure}{0.31\linewidth}
    \centering
    \smallskip
    \includegraphics[width=1\textwidth]{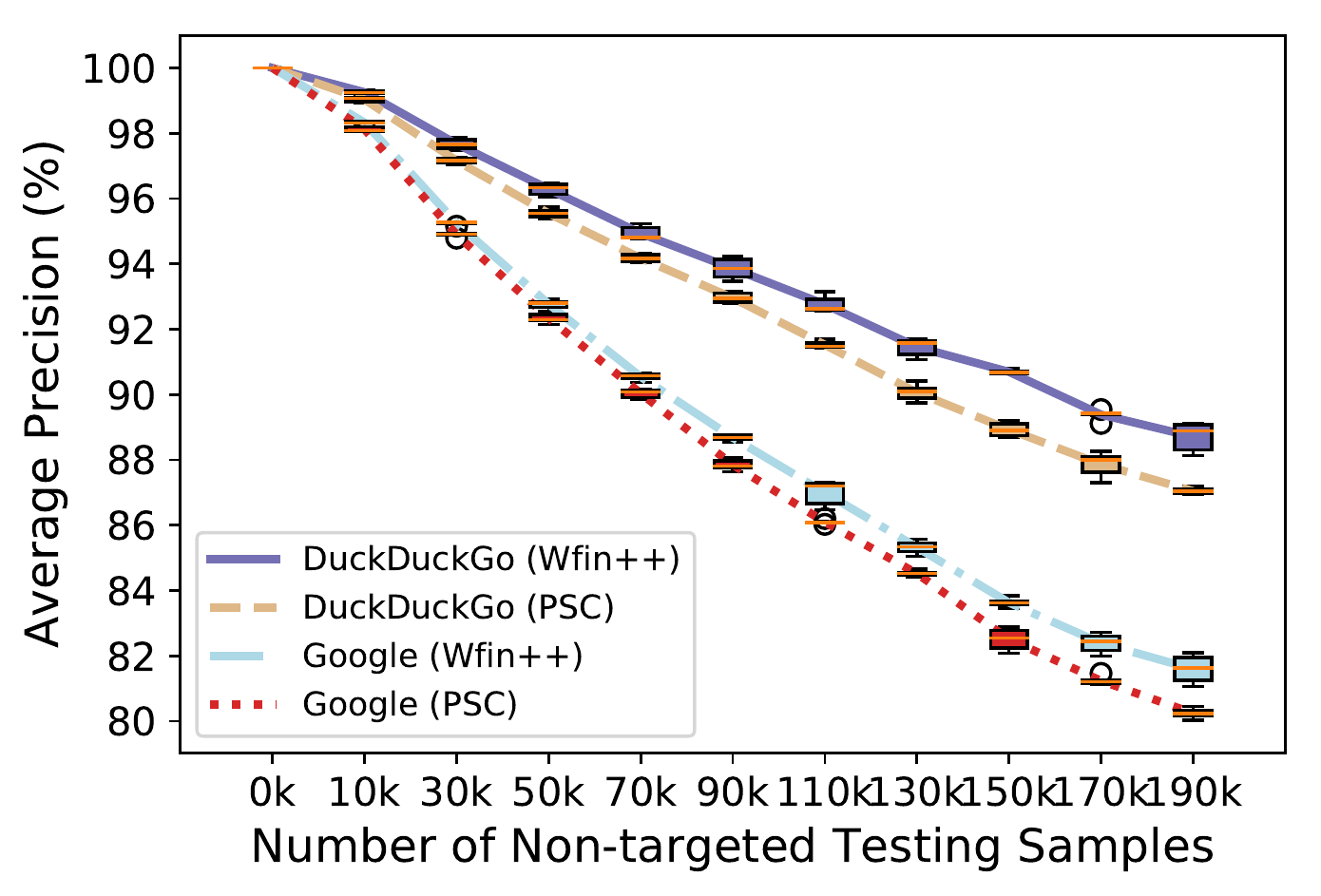}
        \vspace{-5mm}
\caption{AP: \# Non-targeted Testing Samples} %{Light Unit}
    \label{fig:vary_binary_b}
\end{subfigure}
\begin{subfigure}{0.31\linewidth}
    \centering
    \smallskip
    \includegraphics[width=1\textwidth]{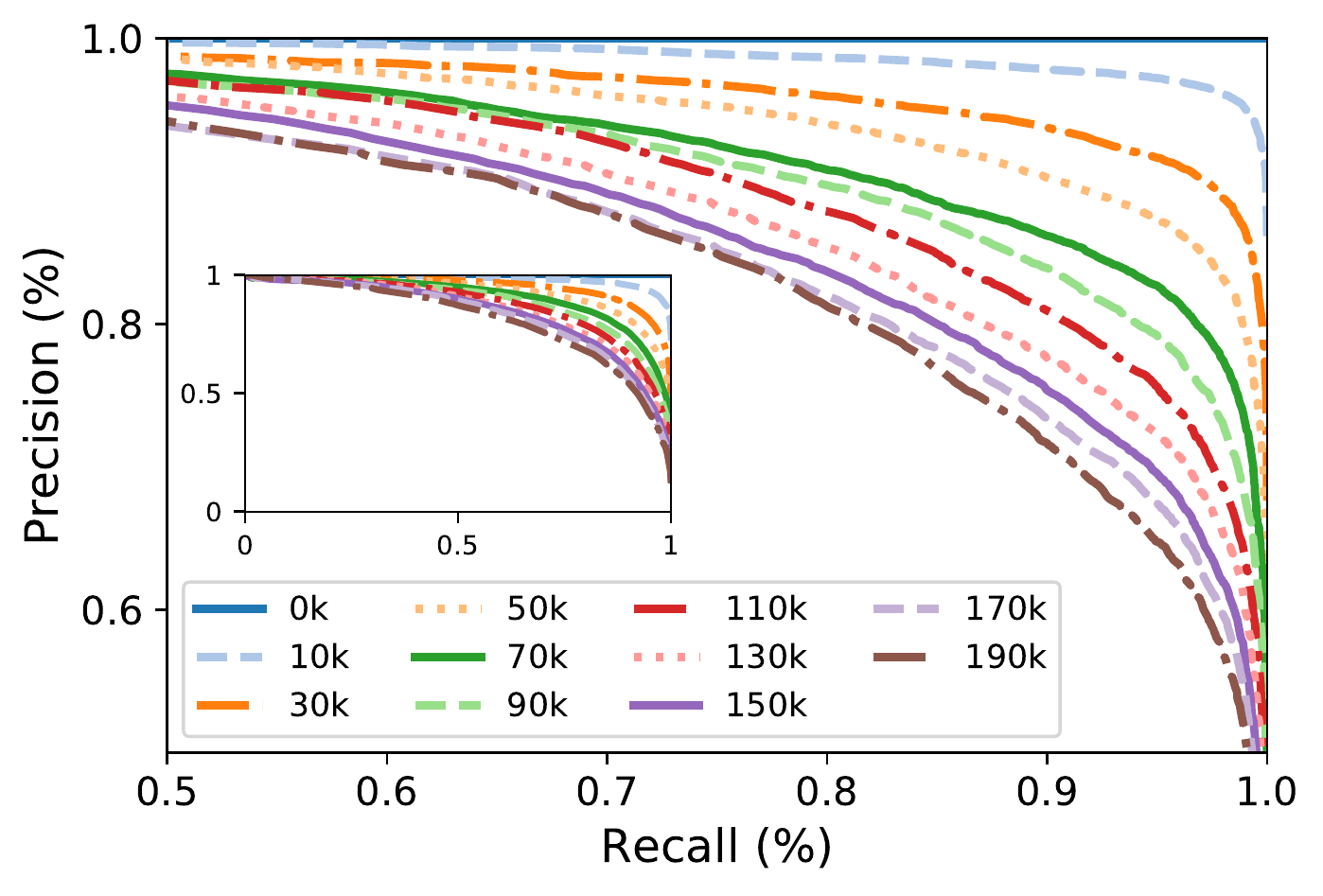}
    \caption{Precision recall curve (DDG, \emph{Wfin++})} %{Light Unit}
    \vspace{-5mm}
    \label{fig:vary_binary_c}
\end{subfigure}
\vspace{-4mm}
\caption{Binary Classification: Impact of Number of Non-targeted Training/Testing Samples (DuckDuckGo and Google)}
% performance achieved with DuckDuckGo and Google using \emph{Wfin++} and \emph{PSC} in binary classification scenario. \emph{Average precision} with varying number of non-monitored training (a) and testing (b) samples. \emph{Precision-recall-curve} achieved by \emph{Wfin++} with DuckDuckGo when varying number of non-monitored testing sample (c).}
\vspace{-3mm}
\label{fig:vary_binary}
\end{figure*}

\begin{figure*}[htbp]
\centering
\begin{subfigure}{0.31\textwidth}
    \centering
    \smallskip
    \includegraphics[width=1\textwidth]{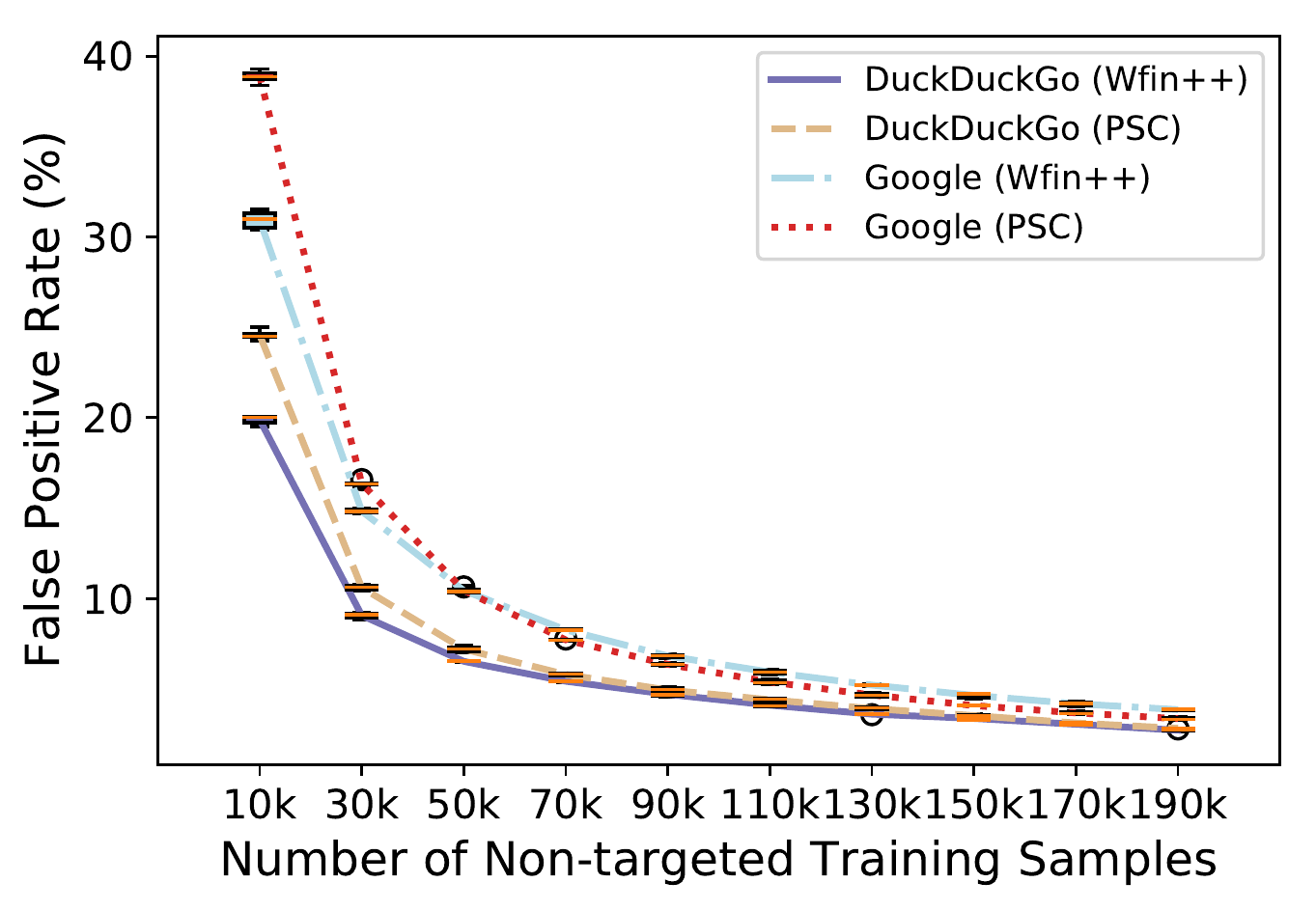}
    \vspace{-5mm}
    \caption{False Positive Rate} %{Light Unit}
    \label{fig:multilevel_openset_performance_a}
\end{subfigure}
\begin{subfigure}{0.31\textwidth}
    \centering
    \smallskip
    \includegraphics[width=1\textwidth]{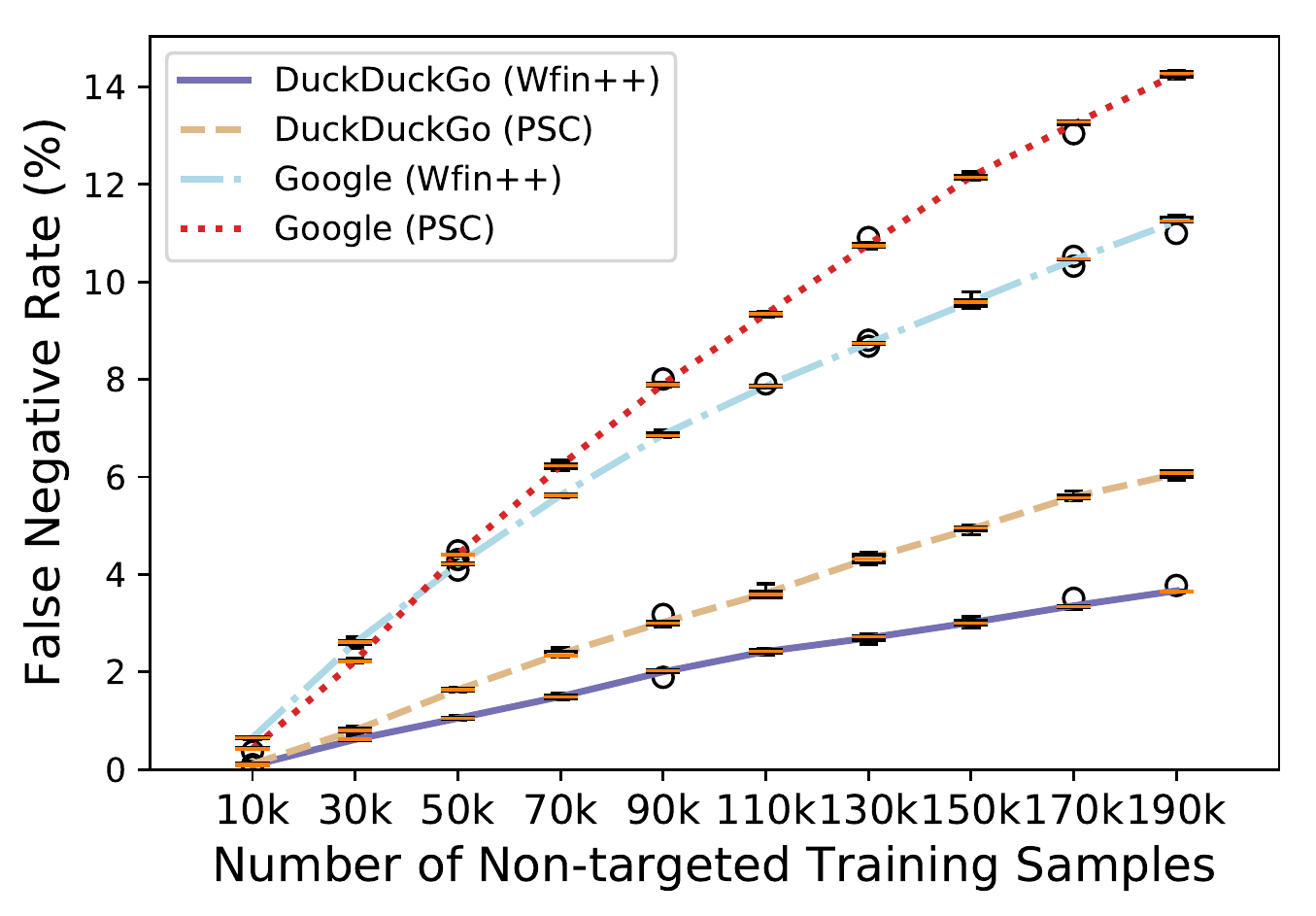}
    \vspace{-5mm}
    \caption{False Negative Rate} %{Light Unit}
    \label{fig:multilevel_openset_performance_b}
\end{subfigure}
\begin{subfigure}{0.31\textwidth}
    \centering
    \smallskip
    \includegraphics[width=1\textwidth]{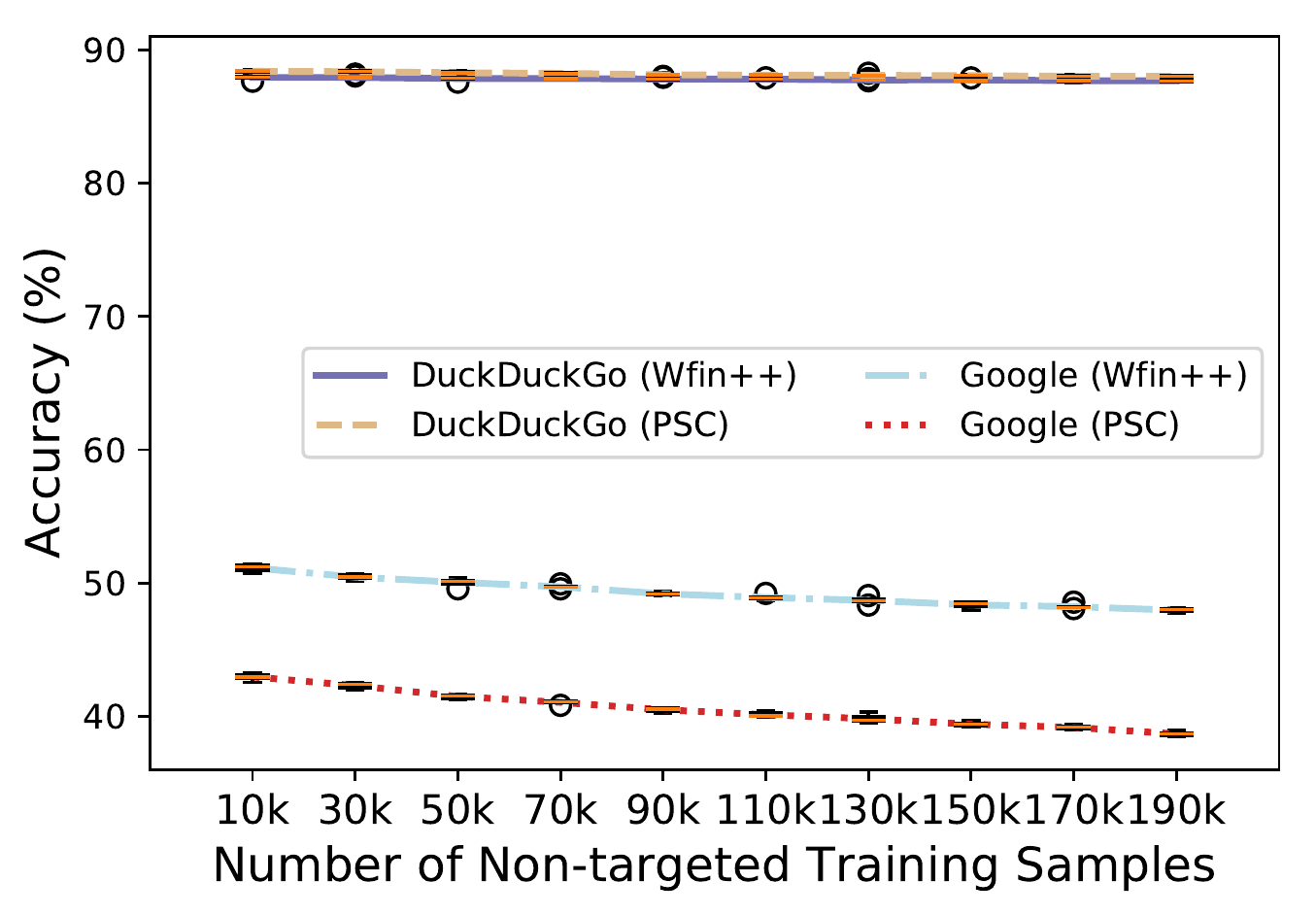}
    \vspace{-5mm}
    \caption{Accuracy} %{Light Unit}
    \label{fig:multilevel_openset_performance_c}
\end{subfigure}
\vspace{-4mm}
\caption{Multi-level Classification: Impact of Number of Non-targeted Training Samples (DuckDuckGo and Google)}
\vspace{-6mm}
\label{fig:multilevel_openset_performance}
\end{figure*}

\subsection{Binary Classification}
\label{subsec:binary}

We first consider the attacker task of determining whether or not a given user query is searching for a targeted keyword---this can be formulated as a binary classification problem with two labels: \emph{targeted} or \emph{non-targeted}. To evaluate classification performance, we consider:
\begin{itemize}
    \item \emph{Precision-Recall-Curve (PRC$_{bc}$)}: \emph{PRC$_{bc}$} is based on two evaluation measures---\emph{precision} and \emph{recall}.\footnote{\emph{PRC} is generated by sweeping over classifier score thresholds, and calculating the \emph{precision} and \emph{recall} at each threshold.} \emph{Precision}, a measure of exactness, is the fraction of samples that are correctly classified, when being classified as a targeted one ($\frac{TP}{TP+FP}$); \emph{Recall}, a measure of completeness, is the fraction of targeted samples that are classified correctly (true positive rate$=\frac{TP}{TP+FN}$).\footnote{True positives---TP (true negatives---TN) are number of the targeted (non-targeted) samples that are correctly classified. False positives---FP (false negatives---FN) are the number of non-targeted (targeted) samples that are incorrectly classified.} \emph{PRC$_{bc}$} reflects the trade-off between precision and recall for different decision thresholds.
    
    \item \emph{Average Precision (AP$_{bc}$)}: This summarizes a given \emph{precision-recall curve} as the weighted mean of precisions achieved at each classifier score threshold\footnote{$AP = \sum_n{(R_n-R_{n-1})*P_n}$, where $P_n$ and $R_n$ are precision and recall at the $n$th threshold \cite{wiki-ap}.}---it reflects how accurate the classifier is, when it labels a testing sample as targeted.
    
    % \markred{What do you mean by the ``weighted'' mean?} \textcolor{blue}{footnote 24 for equation}
    
\end{itemize}

\parlabel{Impact of Number of Non-targeted Training/Testing Samples:} 
We plot the \emph{AP$_{bc}$} achieved by \emph{Wfin++} and \emph{PSC} for different number of non-targeted training and testing samples in \figurename{ \ref{fig:vary_binary_a} and \ref{fig:vary_binary_b}}, respectively, and the \emph{PRC$_{bc}$} (\emph{Wfin++}, DuckDuckGo) for different number of non-targeted testing samples in \figurename{ \ref{fig:vary_binary_c}}. We observe: % \markred{(is the PRC only for DuckDuckGo?)} \textcolor{blue}{Yes. Footnote 22.}
% together with the .
% When comparing the results, we have the following observations:
\begin{itemize}
    \item Search queries using DuckDuckGo, with both higher precision and higher recall, are consistently more vulnerable to be fingerprinted than those using Google (similar to observations in closed-world evaluations). 
    % is observed in open-world scenario regarding to the vulnerability of Google and DuckDuckGo in face of search query fingerprinting: search queries collected by DuckDuckGo are more vulnerable to be fingerprinted compared with Google.
  
     \item Attackers can benefit from training on larger number of non-targeted training samples (\figurename{ \ref{fig:vary_binary_a}})---$AP_{bc}$ increases from 90-94\% to around 94-98\% as the number of non-targeted training samples increases from 10k to 190k (the number of non-targeted testing samples used is 50k).
     
     % \markred{What is number of testing samples in \figurename{ \ref{fig:vary_binary_a}}?}
     
     % \markred{What is number of training samples in \figurename{ \ref{fig:vary_binary_b}}?}
    %  However, it requires more time for attackers to collect a large scale non-monitored dataset -- e.g., it takes us more than 3 days to search each of the 230k non-monitored query once with 60 docker containers.
     
    \item As the number of non-targeted test samples increases (testing dataset becomes increasingly imbalanced), $AP_{bc}$ keeps decreasing (\figurename{ \ref{fig:vary_binary_b}})-- for example, \emph{AP$_{bc}$} drops from 100\% to around 82\%  as the number of non-targeted testing samples increases from 0 to around 190k with Google (the number of non-targeted training samples used is 50k). However, the attacker is still able to distinguish 12,960 targeted samples from 190k non-targeted ones with 80\% \emph{AP$_{bc}$} for Google and 90\% \emph{AP$_{bc}$} for DuckDuckGo.
    
    \item 
    Both \emph{precision} and \emph{recall} decrease when the number of non-targeted test samples increases (\figurename{ \ref{fig:vary_binary_c}}).\footnote{A similar trend is observed with Google using \emph{Wfin++} and \emph{PSC}---results are omitted due to space constraints.} 
    For instance, as number of non-targeted test samples increases from 70k to 190k, a \emph{precision} of 88\% can be maintained only at the expense of recall decreasing from 90\% to 70\%.
    % non-targeted testing samples, respectively.
    % \item 

\end{itemize}
We conclude that, in practice, it seems feasible for attackers to determine whether a query is targeted or not, even in the presence of large scale unseen samples.

    \begin{figure}
\centering
\includegraphics[width=0.68\linewidth]{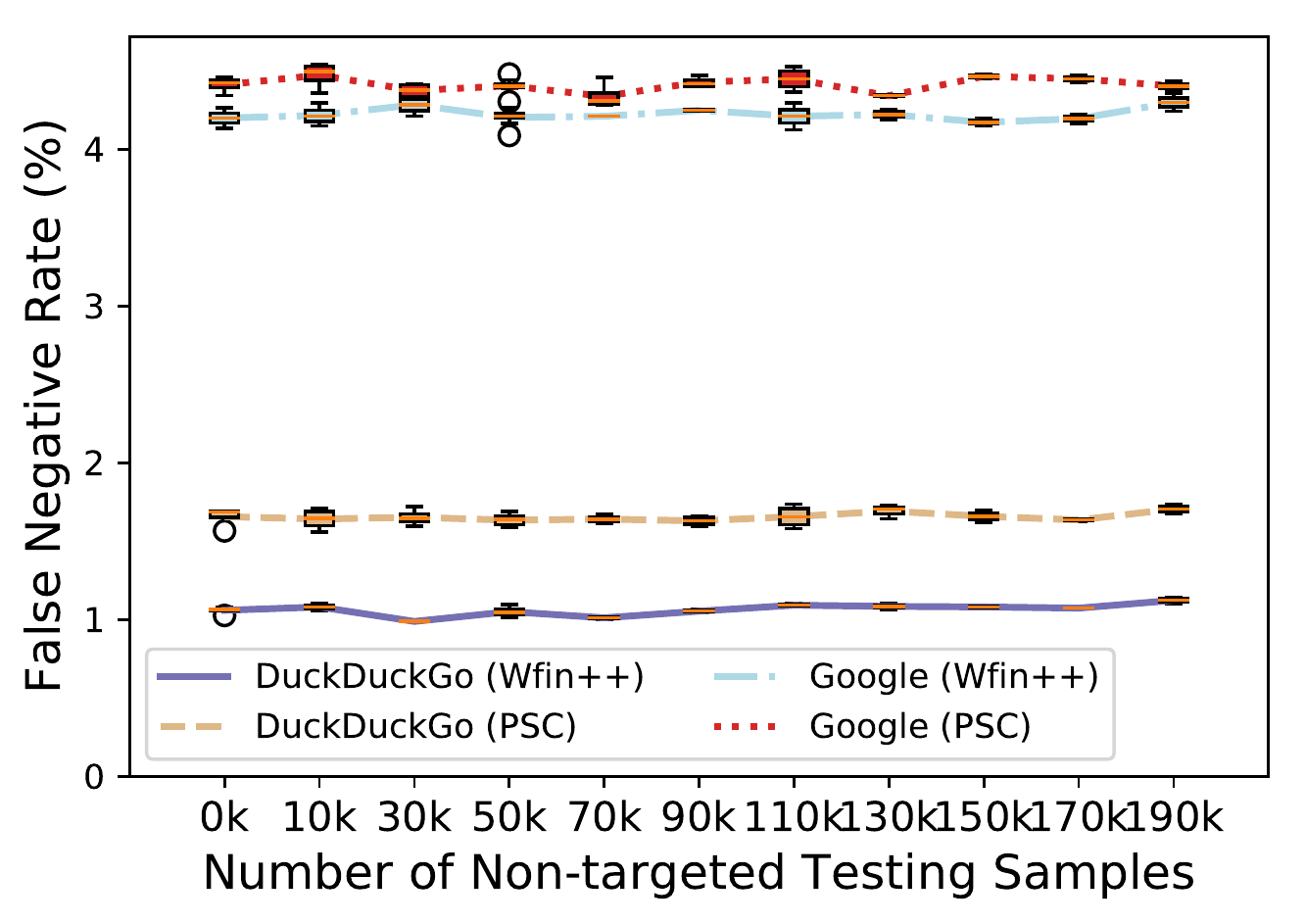}
% \vspace{-3mm}
\caption{Multi-level Classification: False negative rate with varying number of non-targeted testing samples (DuckDuckGo and Google).}
\vspace{-4mm}
\label{fig:fnr_vary_test_multilevel}
\end{figure}

\begin{figure*}[htbp]
\centering
\begin{subfigure}{0.31\linewidth}
    \centering
    \smallskip
    \includegraphics[width=1\textwidth]{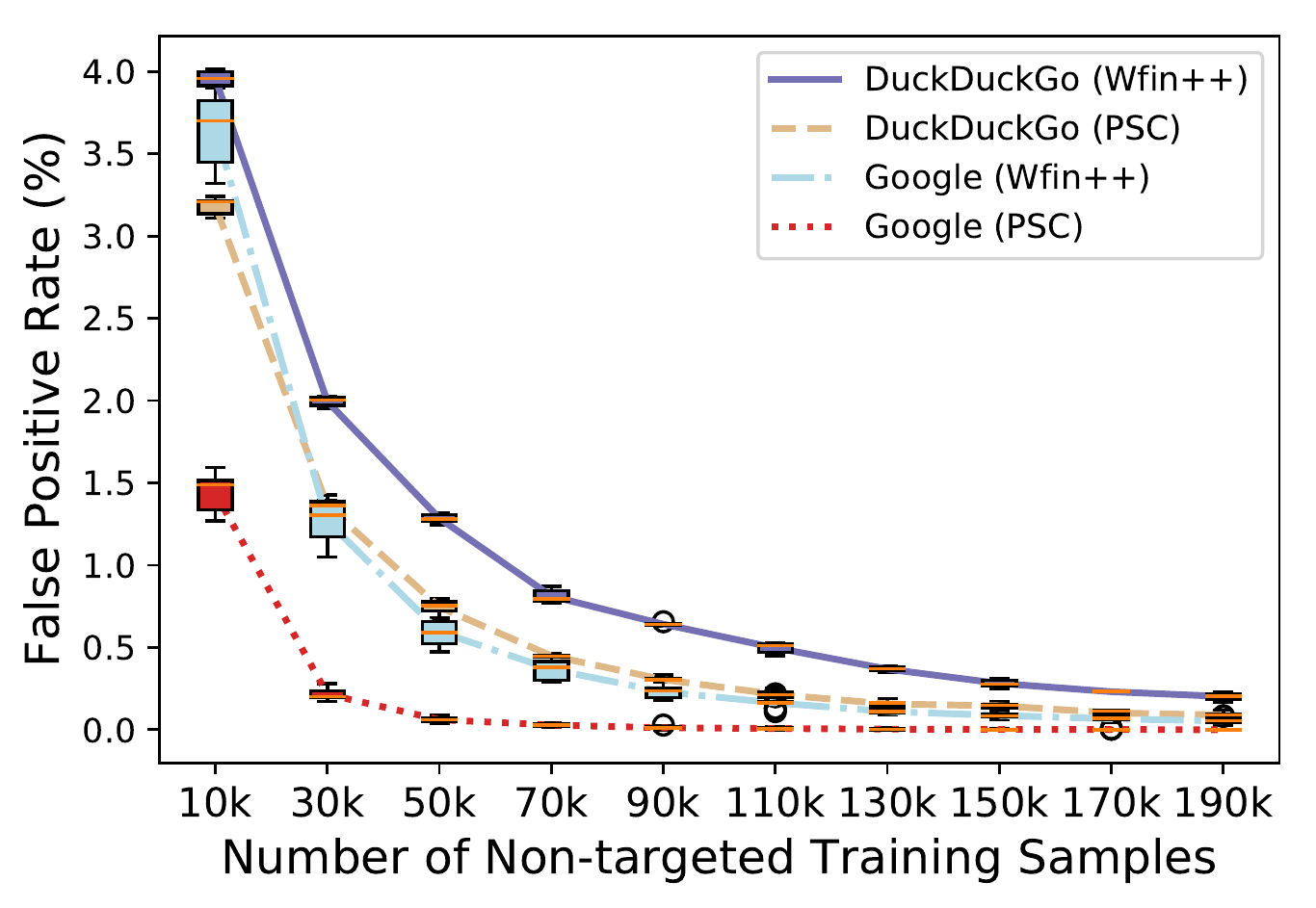}
    \vspace{-6mm}
    \caption{False Positive Rate} %{Light Unit}
\end{subfigure}
\begin{subfigure}{0.31\linewidth}
    \centering
    \smallskip
    \includegraphics[width=1\textwidth]{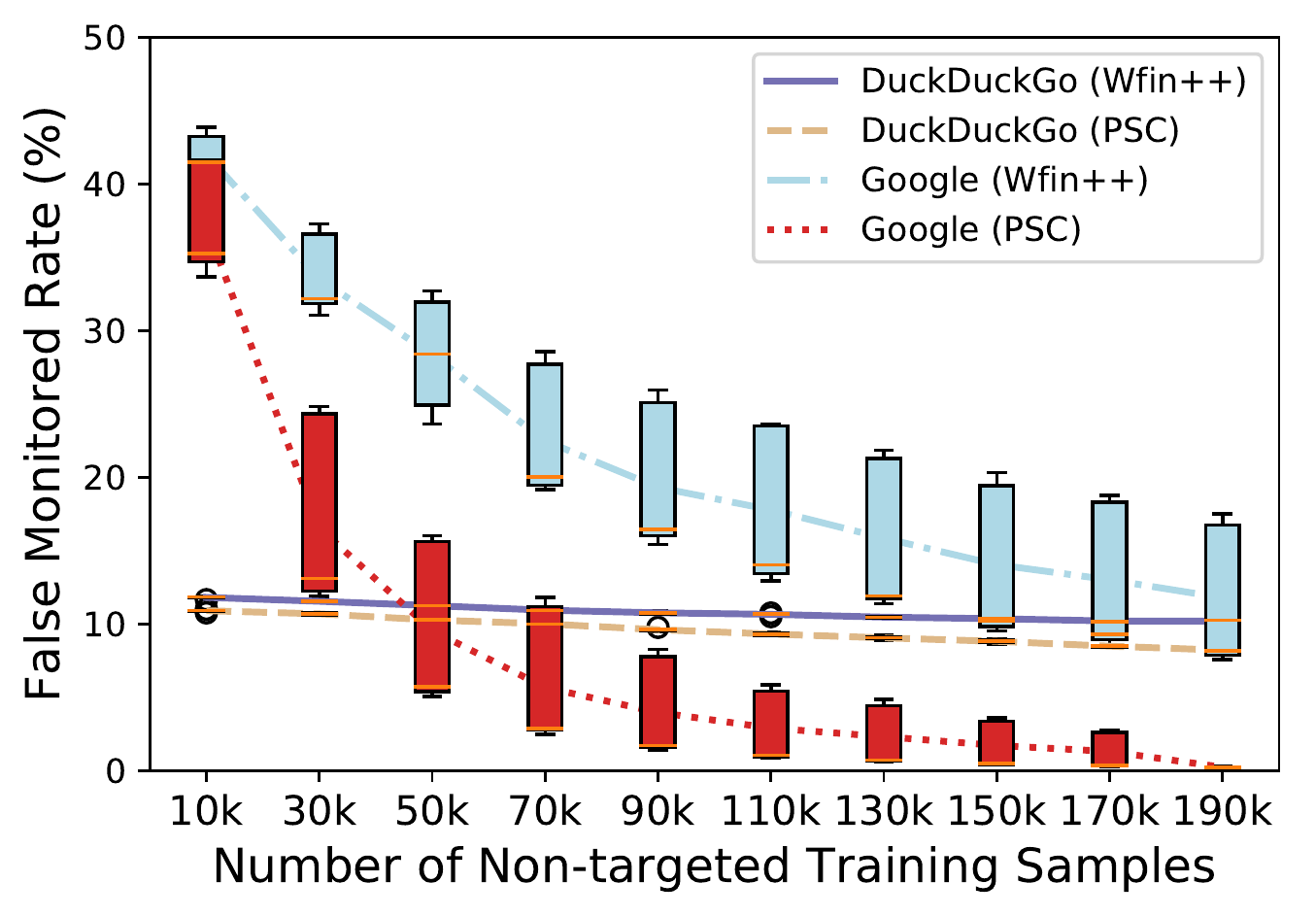}
        \vspace{-6mm}
    \caption{False Monitored Rate} %{Light Unit}
\end{subfigure}
\vspace{-3mm}
\begin{subfigure}{0.31\linewidth}
    \centering
    \smallskip
    \includegraphics[width=1\textwidth]{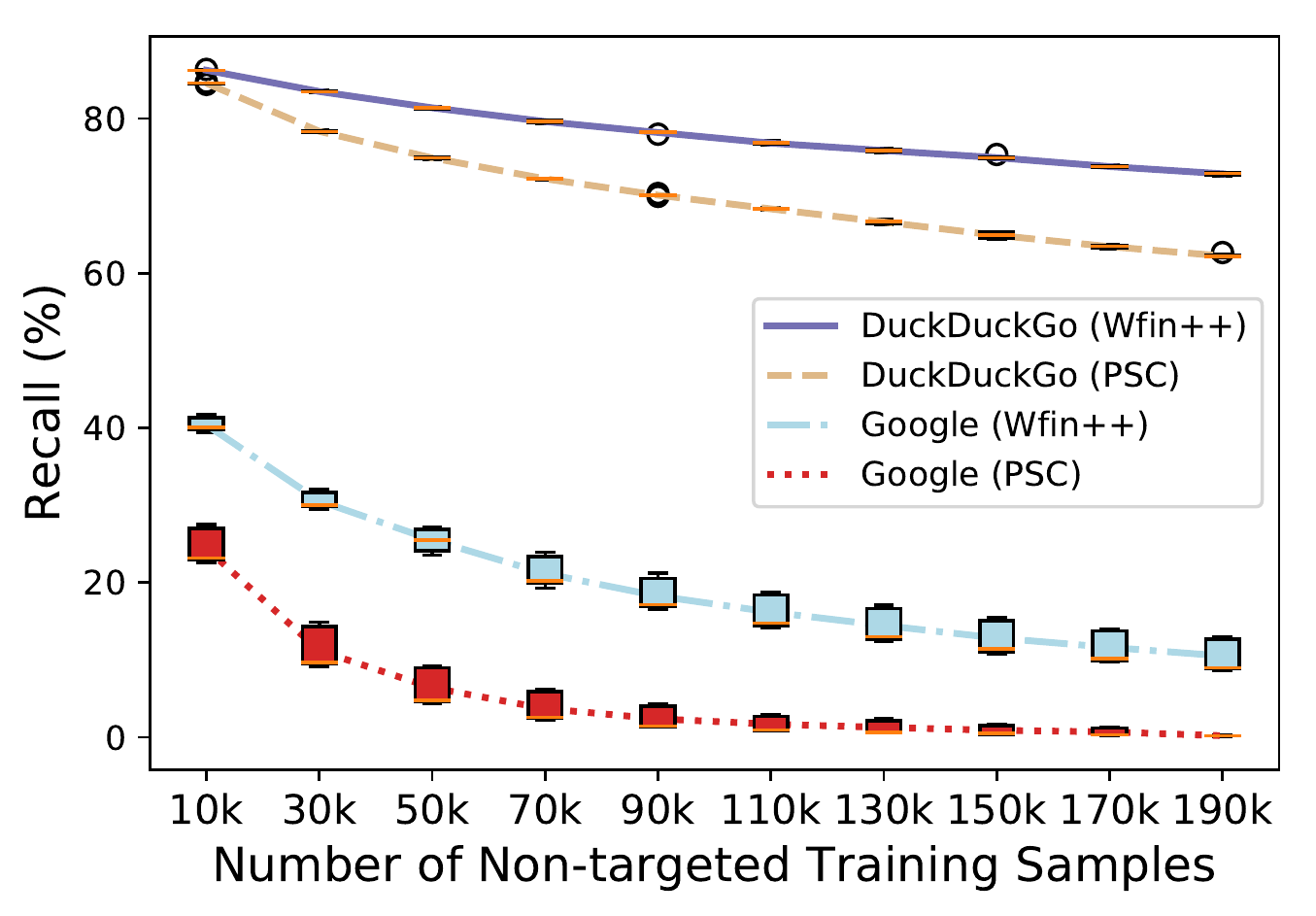}
        \vspace{-6mm}
    \caption{True Positive Rate (Recall)} 
    \label{fig:recall_vary_train_multiclass}
\end{subfigure}
\vspace{-1mm}
\caption{Multi-class Classification: Impact of Number of Non-targeted Training Samples}
\vspace{-6mm}
\label{fig:multiclass_performance}
\end{figure*}

\subsection{Multi-level Classification} 
\label{subsec:multilevel}

After determining that a given search query belongs to the targeted list (binary classification), an attacker can apply an additional classifier trained with only targeted samples (under the closed-world assumption), to further identify the {\em specific} targeted keyword. 
We evaluate this 2-level classification by feeding the samples that are classified as targeted by the binary classifier, to a second model that is trained with targeted samples (with 1,440 labels corresponding to the exact keyword). 
% The purpose of the first model is to decide whether or not a given trace is for a targeted keyword. 
% If it is, the trace is fit into the second model to further identify the specific targeted keyword. 
The label predicted by each classifier is the one with the highest mean probability estimate across the trees in the Extra-Trees classifier. % \textcolor{blue}{I would prefer to leave it here. The difference is how we measure binary classification problem and multiclass classification problem. In Section 6.1, we use precision-recall curve, not precision or recall or fpr or fnr, and average precision based on PRC to measure the performance. Since we only have two labels and the output of the classifier is an array consisting of two probabilities for two labels, and the sum is 1, we are able to use different define different score threshold and get the PRC. But in multi-class classification, the output probabilities is more than 3. In this case, defining a threshold does not work any more. In order for the second classifier to work, the binary classifier has to give a label to each sample to determine the testing samples for the second classifier, instead of a probability. The way how the binary classifier determines the final label is to use highest mean probability across trees. The same reason for why we do not measure the fpr and fnr in Section 6.1 but instead put it here.}
We measure:
\begin{itemize}
   \item \emph{False Positive Rate} (\emph{FPR$_{ml}$}): The fraction of non-targeted samples that are incorrectly classified, $\frac{FP}{FP+TN}$  (binary classification).
    % that are incorrectly misclassified as targeted, to the total number of targeted searches;

    % \markred{Is the above definition correct?}\textcolor{blue}{corrected}
    
    \item \emph{False Negative Rate} (\emph{FNR$_{ml}$}): The fraction of targeted samples that are incorrectly classified, $\frac{FN}{TP+FN}$  (binary classification).

     % \markred{FNR = 1 - Recall. Hasn't recall already been studied in Section \ref{subsec:binary}? Why are FPR and FNR being evaluated here, and not in Section \ref{subsec:binary}?}
     
%     \markred{Make sure you are computing all plots according to the definitions specified in this Section \ref{sec:open_world}.}
     % \textcolor{blue}{fnr = (1-recall), recall is true positive rate -- classify it as positive and actually positive,  here is false negative rate--classify it as negative but actually positive}

    % \markred{Definitions of precision, recall, FNR, FPR need to be precise.}
    
    \item \emph{Accuracy$_{ml}$}: The fraction of true positives 
    % \markred{true positives} \markred{(or targeted samples?)} \textcolor{blue}{targeted means positive. I would think in this case, targeted samples and true positives are the same.} \markred{Targeted samples = TP+FN} \textcolor{blue}{I understand targeted here as the ground-truth targeted... so in your case, if should be true positives}
    (identified by the first classifier) that are correctly classified at the second level.
    % of classifying them with correct label in the second level.
\end{itemize}
The first two metrics reflect the performance of the binary classifier in terms of sifting targeted samples from non-targeted ones, while the last metric reflects the ability of the second classifier to differentiate among targeted search queries. \figurename{ \ref{fig:multilevel_openset_performance}} shows the performance achieved in multi-level classification scenario with different number of non-targeted {\em training} samples using \emph{Wfin++} and \emph{PSC}.\footnote{Results with different number of non-targeted {\em testing} samples are in Appendix \ref{appendix:multi_level_vary_test}.} 
We observe:
% \vspace{-2mm}
\begin{itemize}
    \item Training the binary classifier on a larger number of non-targeted training samples reduces the \emph{FPR$_{ml}$}, but increases the \emph{FNR$_{ml}$} (\figurename{ \ref{fig:multilevel_openset_performance_a} and \ref{fig:multilevel_openset_performance_b}})---with 190k non-targeted training samples, the \emph{FNR$_{ml}$} of \emph{Wfin++} is around 4\% (out of 12,960) with DuckDuckGo and 11\% with Google, which shows classifiers' ability to identify targeted testing samples even in the presence of large number of non-targeted testing samples. Furthermore, \figurename{ \ref{fig:fnr_vary_test_multilevel}} shows how \emph{FNR$_{ml}$} changes as the number of non-targeted testing samples increases---we find that \emph{FNR$_{ml}$} is not significantly affected and remains to around 1\% for DuckDuckGo and 4\% for Googles with 50k non-targeted training samples.\footnote{Note that the attacker has the ability to achieve a different balance between the \emph{FPR$_{ml}$} and \emph{FNR$_{ml}$} of the binary classifier, by tuning classifier thresholds (similar to \figurename{ \ref{fig:vary_binary_c}}).}
    % \footnote{\figurename{ \ref{fig:multilevel_vary_test}} in the Appendix shows that \emph{FNR} is not affected by the number of non-targeted test samples.} 

    % When considering more non-monitored training samples, less non-monitored samples (approaching to 0) are being mistakenly classified as monitored ones (\figurename{ \ref{fig:multilevel_openset_performance_a}}), while 15\% of monitored Google samples and 6\% of DuckDuckGo samples are missed as non-monitored ones (\figurename{ \ref{fig:multilevel_openset_performance_b}}).
    
    % However, the \emph{FNR} are not affected by the non-monitored testing samples -- \emph{FNR} fluctuates around 1\% for DuckDuckGo and 4\% for Google as more  non-monitored testing samples are included (\figurename{ \ref{fig:multilevel_vary_test_b}}).
    
    \item The Accuracy$_{ml}$ of the second level classifier is not significantly impacted by the number of non-targeted training samples---it is around 50\% for Google and 90\% for DuckDuckGo (consistent with closed-world evaluations). 
    % That is to say, the second-level classifier is able to correctly classify monitored queries given the results from the first classifier.
    However, as a larger number of non-targeted training samples are used, the binary classifier labels more targeted samples as false negatives---hence, the overall performance of determining which targeted keyword a user is searching for, decreases.
    
    \item In all performance measures, \emph{Wfin++} consistently outperforms \emph{PSC}.
    
\end{itemize}

%
% \textcolor{blue}{\sout{It is important to note that the attacker has the ability to achieve a different balance between the \emph{FPR} and \emph{FNR} of the binary classifier, by tuning classifier thresholds (similar to \figurename{ \ref{fig:vary_binary_c}}).}}
% , depending on which metric is more important.

\vspace{-3mm}

\subsection{Multi-class Classification}
\label{subsec:multiclass}

The attacker may choose to simply train a single classifier to determine exactly which targeted query is being searched for. 
In this case, samples associated with different targeted keywords have different labels (as in closed-world scenario), while all non-targeted samples share the same label (e.g., -1). 

To evaluate performance, we define below an additional metric---the \emph{False Monitored Rate (FMR$_{mc}$)}. 
% \textcolor{blue}{Other than true positive ($TP_{mc}$) and false negative ($FN_{mc}$), an additional category of targeted testing samples in multi-class classification is false monitored (FM$_{mc}$)): the sample is labeled as the incorrect targeted keyword.}
Each targeted sample is categorized as either a true positive: $TP_{mc}$ (the keyword is correctly identified), a false negative: $FN_{mc}$ (the sample is labeled as non-targeted), or a false monitored: $FM_{mc}$ (the sample is labeled as an incorrect targeted keyword).
% \textcolor{blue}{I think we already gave the definition in footnote 23 for TP, FN etc.} : \textcolor{blue}{we use \emph{FM} to describe the number of samples in a special case in the open world multi-class scenario when the attacker may correctly identify a query as a targeted one, but fail to assign the correct targeted label. For example, \emph{target A} is classified as \emph{target B} (FM), instead of \emph{non-target} (FN) or \emph{target A} (TP).} 
% \textcolor{blue}{Thus use FM to describe this case -- number of the targeted samples A that are incorrectly classified as targeted samples B.} 
\emph{FMR$_{mc}$} is computed as the ratio of false monitored to the total number of targeted samples: $FMR_{mc} = \frac{FM_{mc}}{FM_{mc}+TP_{mc}+FN_{mc}}$.
%
% \markred{How are FM, TP, and FN defined? Define as in Section \ref{subsec:binary}.}

\figurename{ \ref{fig:multiclass_performance}} plots the \emph{FMR$_{mc}$}, \emph{FPR$_{mc}$} (fraction of non-targeted samples that are incorrectly classified), and \emph{TPR$_{mc}$} ($=\frac{TP_{mc}}{FM_{mc}+TP_{mc}+FN_{mc}}$, fraction of targeted samples that are fingerprinted correctly), for different number of non-targeted training samples (50k non-targeted testing samples).
We find:
% \vspace{-2mm}
\begin{itemize}
    \item \emph{FPR$_{mc}$} is low---even with just 10k non-targeted training samples, only 4\% (of 50k) non-targeted testing samples are misclassified with DuckDuckGo.
    
    \item \emph{FMR$_{mc}$} decreases when a larger number of non-targeted training samples are used---the trend is more obvious with Google, for which the \emph{FMR$_{mc}$} reduces to nearly 0 when 190k non-monitored training samples are used.
    
    \item Finally, \emph{TPR$_{mc}$} also decreases along with \emph{FMR$_{mc}$}, for larger number of non-targeted training samples---this implies that FN$_{mc}$ must be increasing, and a larger number of targeted samples get classified as non-targeted 
    (consistent with \figurename{ \ref{fig:multilevel_openset_performance_b}}).
\end{itemize}

\parlabel{Conclusion} Based on our multi-level and multi-class evaluations, we reach the following conclusions:
% \vspace{-2mm}
\begin{itemize}
    \item Determining which targeted keywords is searched for by a user, in the presence of large numbers of non-targeted samples, is challenging but not impossible. 
    For example, with DuckDuckGo, an attacker is able to correctly classify more than 80\% targeted keywords when training with 10k non-targeted samples.
    % (with an additional 10\% of targeted samples assigned an incorrect targeted label).
    \item Incorporating more non-targeted samples during training can help decrease the false positive rate, but also decreases the true positive rate and increases the false negative rate. 
    To achieve a different balance between these metrics, attackers can either adjust the decision threshold in binary classification or consider ``top-k" predictions in multi-class classification \cite{herrmann2009website,sirinam2018deep,oh2019p1};
    \item Consistent with closed-world evaluations, DuckDuckGo samples are more vulnerable than Google samples to be fingerprinted in the presence of non-targeted samples.

    \item A better performance is achievable when the attacker trains two classifiers using multi-level classification instead of one classifier with the ``one-step'' multi-class classification scheme, especially with Google samples. For example, when training with 190k non-targeted samples, an attacker is able to identify 90\% of targeted keywords and correctly classify them with 50\% accuracy in {\em multi-level} classification. However, the attacker is only able to correctly classify around 20\% of targeted samples with {\em multi-class} classification.
\end{itemize}
\section{Discussion}
\subsection{Regular Web page Visit vs. Search Query}
One challenge remaining is whether it is possible for attackers to distinguish search query from regular web page visit. Although Oh et al.\cite{oh2017fingerprinting} attempted to study the problem in Tor, the results are not conclusive due to the use of two datasets collected using different platforms at different times. Due to the access to per-connection headers and IP addresses, the problem in HTTPS are much easier.
We crawled the web pages in \emph{https://www.yahoo.com} on Jan. 17th, 2020 and obtained in total 73,605 urls served from \emph{yahoo} -- among them, 24,016 are search query result pages, including 16,175 web search, 3,211 image search, 2,272 video search and 2,353 other searches (e.g., shopping, recipes, answers, news and finance) while the remaining 49,589 are regular web page visits severed under \emph{yahoo.com} (e.g., \emph{https://www.yahoo.com/lifestyle/about-makers-183936026.html}). 
We visited all 73,605 urls during Jan.17th and 18th, 2020 with chrome and collected the associated traffic traces using the same methodology in Section \ref{subsec:dataset_summary}. 
The SNI field in TLS header reveals the server name for each TCP connection and in total we obtain 150 domains under \emph{yahoo.com}. 
By counting the number of TCP connections served from each of the yahoo domain, we create a feature matrix for training a Extra-Tree classifier to differentiate between web page visit and search query traffic. 
In the experiment, we randomly selected 10,000 web page visit and 10,000 search query results for training and another 10,000 web page visit and 10,000 search query results for testing. 
The experiment is repeated 50 times and on average the accuracy is $99.29 \pm 0.06$ \%, which indicates attackers' ability to easily separate web page visits from search query using server names obtained from SNI field.

\subsection{Effect of Cache}
In this section, we investigated the cache policy of each of the four search engines from its traffic traces. Below is a summary of cache policies adopted by search engines:
\begin{enumerate}
    \item When Google/Bing return the search results, the index.html is never cached (\emph{cache control = private, max-age = 0}). Furthermore, all images in Google and most in Bing that are displayed on the returned page are encoded in the \emph{index.html}, instead of sending GET request to fetch the object from another server. The only few contents being cached are general Google elements such as the googlemic figure, activityindictor gif and favicon.ico.
    \item For DuckDuckGo, the max-age for \emph{index.html} is set to 1 second (smaller than time to type a new keyword). DuckDuckGo results generally contain less images and are served from external-content.duckduckgo.com.
    \item For Yahoo, \emph{index.html} is not cached while almost all images are served from \emph{yimg.com} and cache is applied. But attackers can get rid of the caching effect by considering only TCP connections to \emph{yahoo.com} -- which has been shown (Section V-D) to improve the accuracy of Yahoo.
\end{enumerate}

% \subsection{Traffic Segmentation}
% \input{sections/related.tex}
\section{Limitation and Concluding Remarks}
\label{sec:conclusion}

In this paper, we study the vulnerability of keyword fingerprinting in HTTPS traffic with a large-scale dataset of nearly 4 million search samples, and study the impact of several factors.
Our evaluation methodology differs from \cite{oh2017fingerprinting} in the several ways: (i) we consider keyword fingerprinting in HTTPS traffic (versus Tor traffic); (ii) we study the impact of several factors on keyword fingerprinting---including client browser diversity, web search engine, time, as well as noisy traffic features; and (iii) our evaluation dataset is larger by several orders of magnitude.
% 
% collected using different search engines and browsers. 
% With a large scale dataset that contains up to 400,000 monitored and non-monitored samples, 
Our key findings include:
\begin{itemize}
    \item Search engines differ in the vulnerability of their users to keyword fingerprinting---Bing is the least vulnerable (up to 45\%), followed by Google (up to 80\%); while Yahoo and DuckDuckGo (up to 96\%) users are the most vulnerable. 
    
    \item An attacker can achieve high fingerprinting accuracy: (i) by ignoring traffic going to secondary domains, other than the search engine contacted by a user; (ii) by training on data collected using diverse client browser platforms; and (iii) by re-training their classifiers on data collected every 2-3 days.
    
    % \item The use of prominent countermeasures that help obfuscate packet size and ordering, drastically reduces the classification accuracy for fingerprinting keywords at the cost of significant bandwidth overhead.
    % To make it challenging for a search engine to be fingerprinted, it can either generate a lot of dynamic background traffic to obfuscate the actual search results, such as Yahoo or follow a specific pattern to make traffic generated when searching different queries indistinguishable, e.g., Bing. 
    % Although in the first case, it is possible for attackers to prune the traffic by considering only connections from specific domains to improve the classification accuracy (Section \ref{subsec:feature_set}). 
    
    % \item As suggested in \cite{alan2019client}, attackers should consider diverse sets of client platforms to guarantee the classification performance. 
    % Otherwise, the accuracy may decrease significantly when there is a mismatch between the training and testing client platforms.
    
    \item Search query fingerprinting is indeed a potential privacy concern even in open-world scenarios in which large scale unseen samples may be encountered---attackers are able to identify specific targeted search queries with 80\% \emph{recall} and 85\% \emph{precision} with 10k non-targeted training samples and 50k non-targeted testing samples from DuckDuckGo/Chrome (Section \ref{sec:open_world}).
    
\end{itemize}
We find our observations alarming about the possibility of fingerprinting search keywords being used by current Internet users. We believe this topic should receive significant and immediate attention from the research community. Some open issues that need to be addressed are:
\begin{itemize}
    \item \emph{Countermeasures}: Given the dominance of HTTPS in world-wide Internet traffic, our results urge for the study of efficient countermeasures against keyword fingerprinting. 
    In Appendix \ref{subsec:countermeasures}, we study the efficiency of two countermeasures designed for website fingerprinting with HTTPS traffic: PadToMTU and HTTPOS \cite{luo2011httpos}. Our preliminary results suggest that these two countermeasures can help decrease the classification accuracy, albeit, after incurring significant bandwidth overhead. 
    Thus, we consider an extensive study of the design and evaluation of efficient countermeasures (e.g.,  \cite{dyer2012peek,luo2011httpos, wang2017walkie}) as important future work.

    \item \emph{Mobile Web Browsers and Voice-based Searches:} An increasing number of users are relying on mobile devices, rather than PCs, to access the Internet.
    % than from desktop computers. 
    Meanwhile, voice services such as Amazon Alexa, Google Home, and Apple Siri has gained a lot of popularity in recent years. As part of future work, we plan to study the vulnerability of search query fingerprinting for users with mobile devices.
    % Both of them are not covered in the paper and remain to be an important future work.
    % Thus whether the difference between desktop and mobile devices will affect the search query traffic and its fingerprintability remain to be an important future work.
  
      \item \emph{Traffic Segmentation}: This paper assumes that a user's traffic has already been cleanly segmented on a per-search session basis---such an assumption has been used in nearly all prior work on website/webpage/keyword fingerprinting, both in HTTPS and Tor \cite{liberatore2006inferring,herrmann2009website,dyer2012peek,wang2013improved,wang2014effective,panchenko2016website,hayes2016k,rimmer2017automated,yan2018feature,oh2017fingerprinting,alan2019client}. In practice, traffic from a user is likely to contain overlapping connections from multiple tabs, multiple browsers, as well as traffic from several web requests multiplexed onto pipelined HTTPS connections---segmenting such a mix into traffic corresponding to individual web requests remains an important open problem in this field.
    % Given our results with addressbar searching, we may not need to identify all connections associated with a search, but just the ones to the primary domains serving the search results. \textcolor{blue}{However, when a user searches more than one queries within a short amount of time, the same TCP connection used for returning the auto-suggestion list and actual search results may be reused. 
    % Thus one problem is whether the classifier is able to identify that case and segment the traffic into different chunks associated with different queries. }
  
    \item \emph{Deep Learning}: The application of deep learning for website fingerprinting in Tor has recently been explored in \cite{abe2016fingerprinting,rimmer2017automated,Sirinam:2018:DFU:3243734.3243768}---this body of work has shown that deep learning is able to achieve comparable or even better performance compared to traditional machine learning classification frameworks, without the need for manual feature engineering. It is important for deep learning to also be explored for fingerprinting HTTPS traffic---as future work, we plan to do so for keyword fingerprinting.
    % However, the feasibility of applying deep learning models for HTTPS traffic has not been thoroughly studied yet and should be considered as a future work.
\end{itemize}

% \pagebreak

\bibliographystyle{IEEEtranS}
\bibliography{acmart.bib}

\appendix

\subsection{Related Work: Broader Discussion}
\label{subsec:related}

% \markred{Include discussions of all papers pointed out by the last reviewer.}

\parlabel{Keyword Fingerprinting}
The possibility of identifying users' search query has draw attentions from researchers. 
Chen et al. points out that sensitive information is being leaked out from several web applications, including search engines, despite the protection of HTTPS \cite{chen2010side}. 
They consider the auto-suggestion/auto-complete features implemented by search engines as the main reason for query word leakage since for different combination of letters, list of packet sizes responded by search engines for each type-in is unique. 
% Thus their methodology is to look at the responded packet size 
More specifically, Sharma et al. demonstrates that for every character typed in Google search box, the exchanged packet is followed by a fixed pattern---the request size is increased by one byte for every character typed in while the response size reveals the suggestion made by the search engine \cite{sharma2012implementing}. In their threat model, the attacker sends 26 requests for each character (a-z) and captures the size of suggestion for comparison after collecting the sequence of packet sizes from networking trace. Thus to find a string of size \emph{n}, the attacker needs to send $26^n$ automated search requests.
Later, Schaub et al. study how to use stochastic algorithm to deal with variable packet lengths considering that Google has supported variable packet lengths for a given query with payload randomization and \emph{Gzip} compression since 2012 \cite{schaub2014attacking}.
For a given length, they create a prefix tree to represent the set of all possible words based on a chosen dictionary and preform hierarchical matching based on the observed size of responded packets.
Oh et al. \cite{oh2017fingerprinting} first extend standard website fingerprinting attacks to fingerprint individual keywords that contains more than one word in Tor and further apply deep learning to the threat model \cite{oh2019p1}.
% They studied 300 monitored queries and 80,000 background queries with Google, Bing and DuckDuckGo using Tor, and obtained 80\% precision to determine whether a search query is monitored considering 10,000 background query traces and 48\% accuracy to identify a specific monitored query among a set of monitored queries.

\parlabel{Website/Webpage Fingerprinting} 
Website fingerprinting refers to the task of learning which website/webpage is being visited based on the information available from the TCP/IP headers in network traffic. 
Researchers have demonstrated the possibility to fingerprint websites/webpages in several communication scenarios including HTTPS (e.g., \cite{miller2014know,trevisan2016towards,macia2010isp,alan2019client}), encrypted tunnel (e.g., \cite{liberatore2006inferring,herrmann2009website,panchenko2011website,lu2010website,dyer2012peek}) and Tor (e.g., \cite{cai2012touching,yu2012attacking,wang2013improved,cai2014systematic,panchenko2016website,wang2016realistically}). We refer readers to \cite{yan2018feature} for a detailed discussion about website/webpage fingerprinting in each communication scenario.

More recently, Abe et al.\cite{abe2016fingerprinting}, Rimmer et al. \cite{rimmer2017automated} and Sirinam et al. \cite{sirinam2018deep} explore the application of deep learning to further boost the classification accuracy of website fingerprinting without manual feature selection. 
Jansen et al. \cite{jansen2018inside} explore traffic analysis attacks on Tor with middle relays rather than with relays from entry or exit positions. 

\parlabel{Other Learning-based Traffic Analysis Topics}
Other than website/webpage fingerprinting, machine learning has been applied in other fields of traffic analysis. 
Specifically, Wright et al. \cite{wright2010uncovering} shows the possibility to identify the phrase spoken within encrypted VoIP calls using knowledge of the phonetic pronunciation of words and Hidden Markov Model.
Coull et al. \cite{coull2014traffic} shows that information from instant messaging services such as users actions, the language of messages and even the length of the message can be learnt with more than 96\% accuracy using only the sizes of encrypted packets.
Alan et al. \cite{alan2016can} and Vincent et al. \cite{taylor2017robust} study the possibility of identifying applications installed on a smartphone using side-channel data such as packet size and direction. 
Schuster et al. \cite{schuster2017beauty} and Gu et al. \cite{gu2018walls} demonstrate the possibility of fingerprinting streaming videos by modeling their unique burst patterns using machine learning. 
Barradas et al. \cite{barradas2018effective} explores the application of semi-supervised and unsupervised machine learning techniques to identify multimedia protocol tunneling systems including Facet, CovertCast and Deltashaper.

\subsection{Distribution and Examples of search queries}
\label{appendix:example_non_monotored}

\begin{figure}[htbp]
\centering
\includegraphics[width=0.7\linewidth]{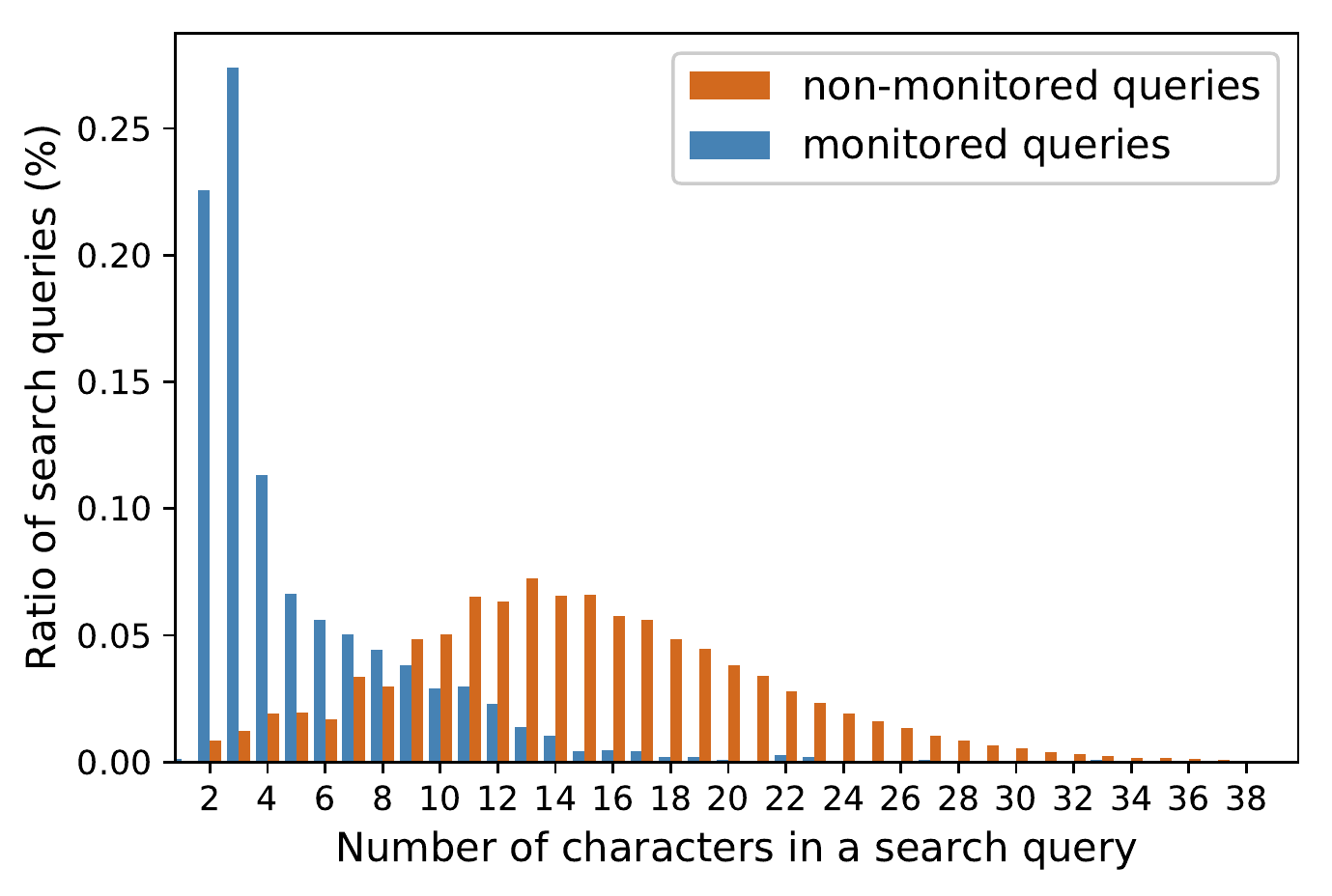}
\vspace{-2mm}
\caption{Number of Characters in Search Keywords.}
\vspace{-4mm}
\label{fig:length_of_queries}
\end{figure}

\figurename{ \ref{fig:length_of_queries}} plots the distribution of number of characters in each search keyword. As can be seen, most of keywords are composed of multiple characters.

% \vspace{-2mm}
% \vspace{-2mm}

\tablename{ \ref{tab:example_non_targeted}} shows 10 examples of the harvested non-targeted keywords. 
The non-targeted keyword list is representative since: (i) it is composed of more than 200K popular web search queries, (ii) it contains more than one language, and (iii) it covers different topics.

\vspace{-2mm}
\begin{table}[H]
    \centering
    \caption{10 Examples of non-targeted search queries.}
    \vspace{-2mm}
    \resizebox{0.7\linewidth}{!}{
    \begin{tabular}{llll}
    \hline
    1& iron man 3 lego 
    & 6 & gmail windows app \\
    2&\$5 pizza hut deals
    & 7 & cool things to buy
    \\
    3&$ (1-x^2)^3 $ 
    & 8 & python 
    \\ 
    4 & \begin{CJK*}{UTF8}{bsmi} 澳門教青局 \end{CJK*} &  9 & \begin{CJK}{UTF8}{}
     \CJKfamily{mj} 스크래치 \end{CJK}  
     \\
    5 & proceso de fecundación 
    & 10 & comcast specials \\
    % \multicolumn{4}{c}{\textbf{... ...} } 
    \\
    \hline
    \end{tabular}}
    \vspace{-2mm}
    \label{tab:example_non_targeted}
\end{table}

\begin{figure*}[htbp]
\centering
\begin{subfigure}[t]{0.95\linewidth}
    \centering\includegraphics[width=1\linewidth]{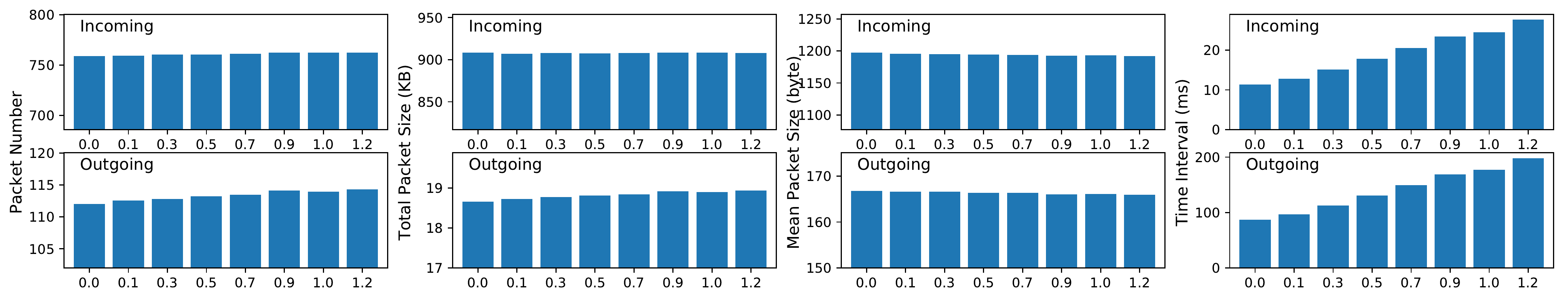}
    \vspace{-5mm}
    \caption{Statistical information for traffic traces generated with different typing speeds with Google/Chrome.}
    \label{fig:type_speed_stats}
  \end{subfigure}
 \begin{subfigure}[t]{0.9\linewidth}
    \centering\includegraphics[width=1\linewidth]{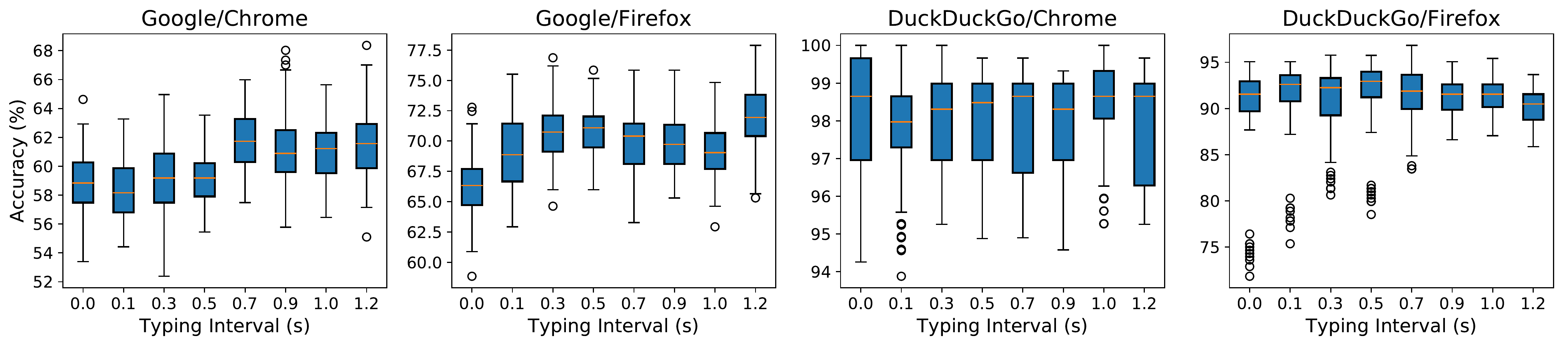}
    \vspace{-5mm}
    \caption{9-fold cross-validation accuracy with Google and DuckDuckGo samples.}
    \label{fig:clf_accuracy_type_interval}
  \end{subfigure}
\vspace{-3mm}
\caption{Impact of Typing Speeds}
% \vspace{-2mm}
\end{figure*}

\subsection{Impact of Typing speeds}
\label{subsec:type_speed}

In this section, we study how typing speeds of entering search queries affect the traffic trace and the keyword classification accuracy in closed-world scenario. We collect a monitored dataset with 4 search engines (i.e., Google, Bing, Yahoo, DuckDuckGo) and 2 browsers (Chrome, Firefox), and vary the intervals between typing of consecutive characters in the search query from 0 to 1.2 seconds (representing different typing speeds). In total, we collect around 315,000 samples and analyze the following three aspects of the traffic traces collected using different typing speeds:
\begin{enumerate}
    \item \emph{General statistical information such as the total incoming/outgoing packet number and bytes.} \figurename{ \ref{fig:type_speed_stats}} presents the results with Google/Chrome: the average inter-arrival time between subsequent incoming/outgoing packets does increase as typing speed decreases. However, there is no obvious variations in terms of packet number and bytes. The same trend is observed across other search engines and browsers.
    \item {\em Linear discriminant analysis to understand how separable are traces collected with different typing speeds in their feature space.} 
    We consider around 6,000 samples for each typing speed and use packet size count as features considering its good performance in prior experiments.
    \figurename{ \ref{fig:lda_type_interval}} shows the result when projecting each feature space onto the three-dimensional subspace. As can be seen, samples collected with different typing speeds are densely clustered together and can not be easily differentiated from each other, which indicates the similarity in their feature spaces.
    \begin{figure}[htbp]
\centering
\vspace{-2mm}
\includegraphics[width=0.75\linewidth]{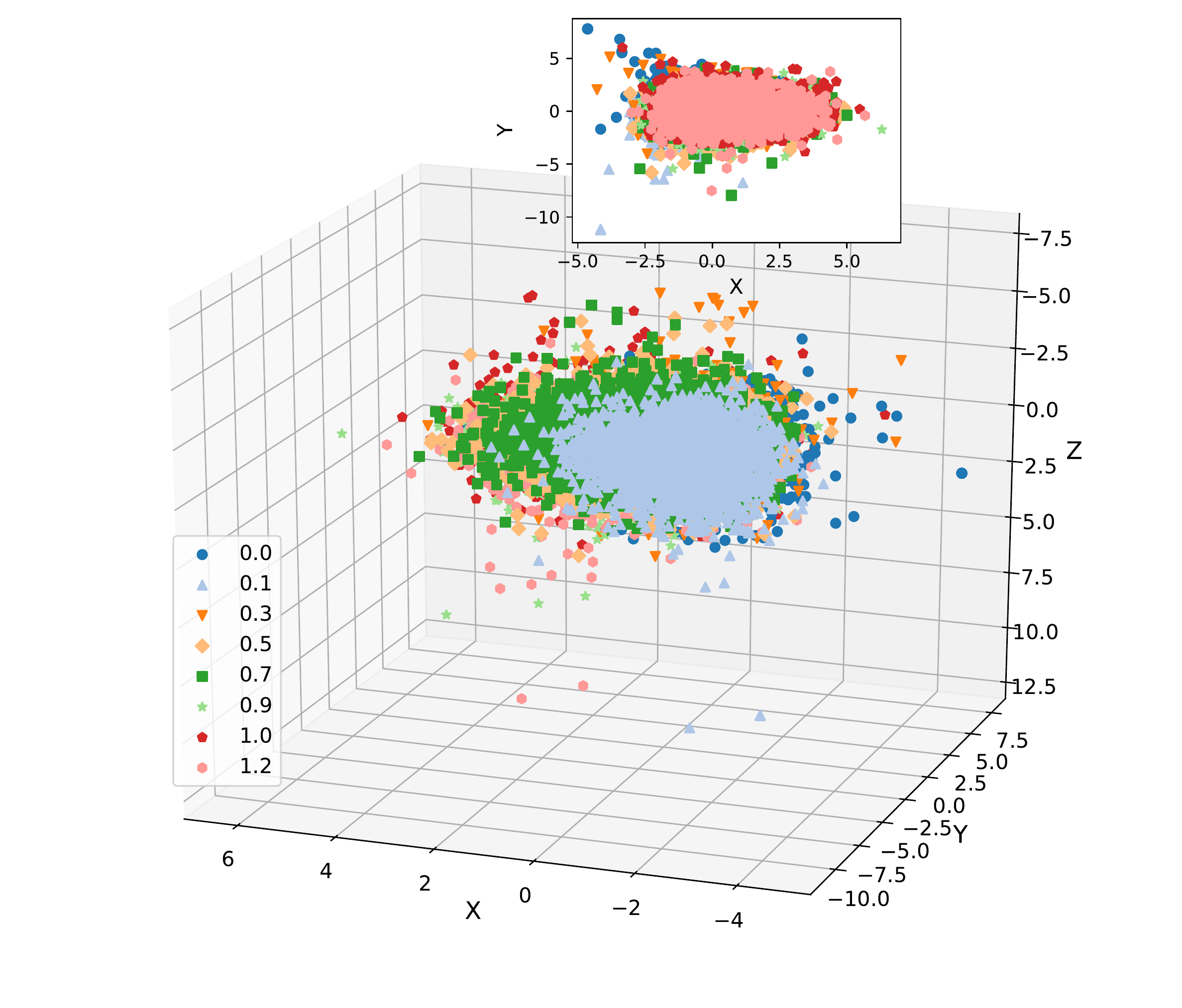}
\vspace{-4mm}
\caption{Linear discriminant analysis for traces with different typing speeds using \emph{PSC}.}
\vspace{-6mm}
\label{fig:lda_type_interval}
\end{figure}
    \item {\em Evaluating the vulnerability of traces collected with different typing speeds in face of keyword fingerprinting.} We measure the classification accuracy with cross-validation using \emph{PSC}.
    \figurename{ \ref{fig:clf_accuracy_type_interval}} shows the results obtained with 140 targeted keywords, each with 18 samples, for Google and DuckDuckGo traces collected with different typing speeds. 
    As can be seen, the accuracy does not consistently increase or decrease as the typing speed varies. 
\end{enumerate}

Based on the above investigations, we did not observe any obvious evidence to indicate that a users' typing speed impacts their vulnerability of being fingerprinted.

\subsection{Wfin++: Forward Feature Selection}
\label{appendix:wfin}

The \emph{Wfin} methodology for feature selection ranks features based on their importance for classification, and then groups the features that account for 99\% importance into semantically-relevant feature categories---all of these feature categories are then used to perform website fingerprinting \cite{yan2018feature}. 

\begin{figure}[htbp]
\centering
\begin{subfigure}{1\linewidth}
  \centering
  \includegraphics[width=1\linewidth]{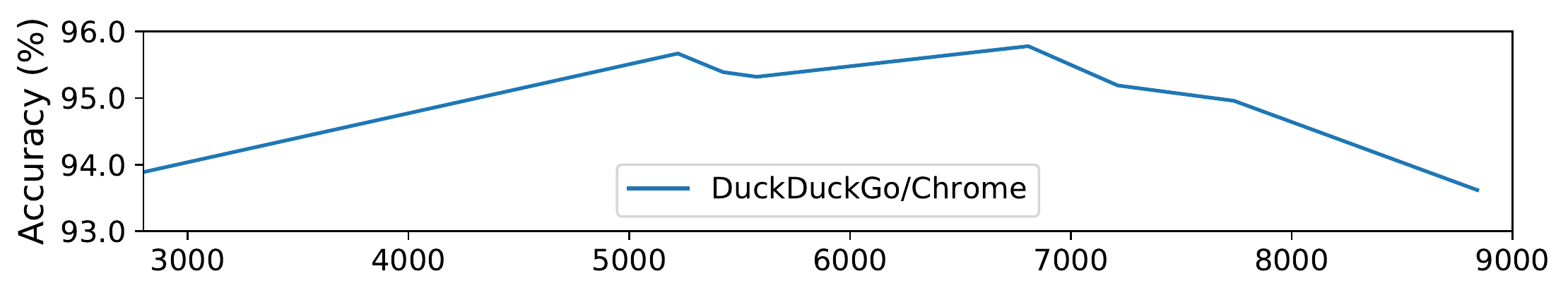}
\end{subfigure}%

\begin{subfigure}{1\linewidth}
  \centering
  \includegraphics[width=1\linewidth]{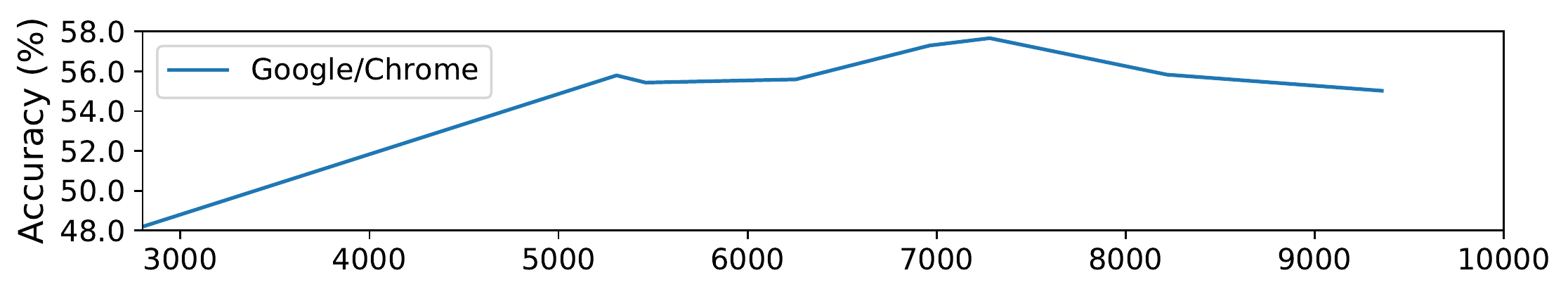}
\end{subfigure}

\begin{subfigure}{1\linewidth}
  \centering
  \includegraphics[width=1\linewidth]{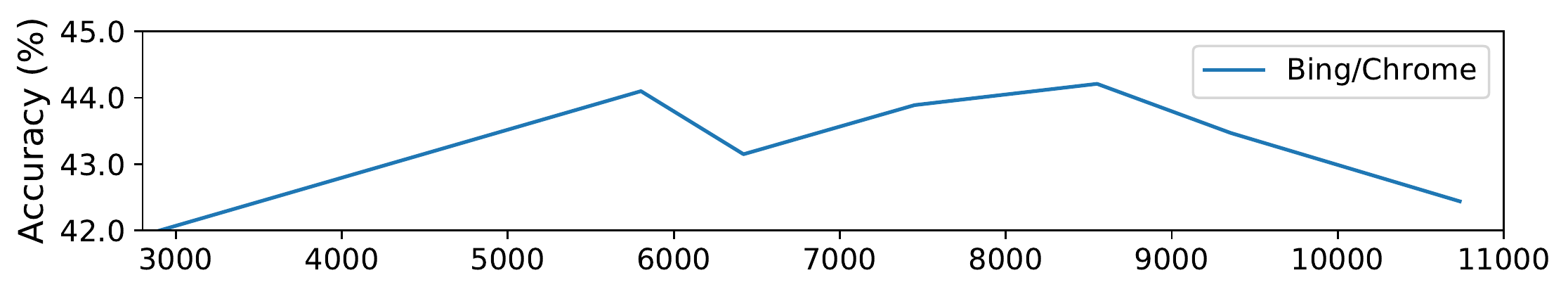}
\end{subfigure}%

% \vspace{-2mm}
\begin{subfigure}{1\linewidth}
  \centering
  \includegraphics[width=1\linewidth]{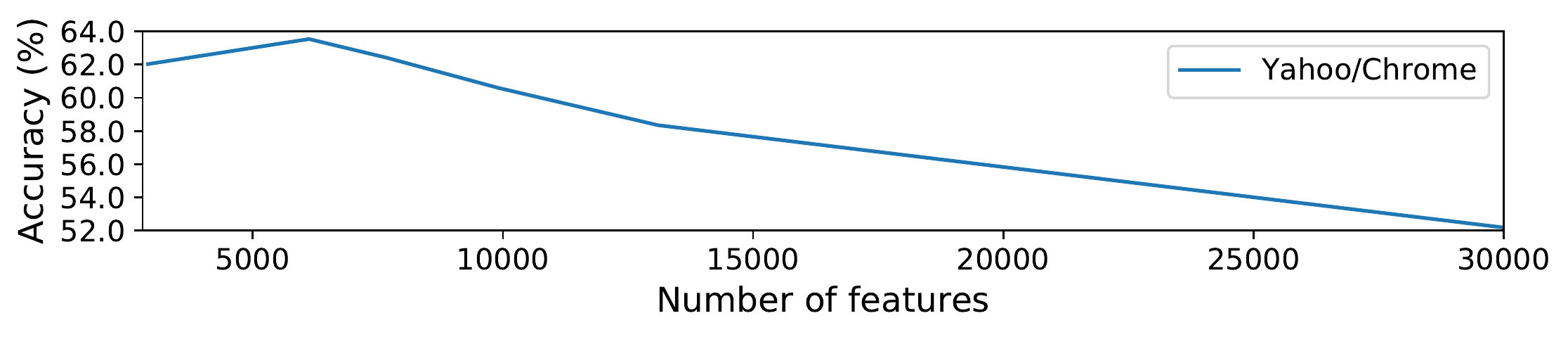}
\end{subfigure}%
\vspace{-3mm}
\caption{Validation Accuracy: top-\emph{N} features used for classification (Chrome, \emph{homepage searching} mode)}
\vspace{-2mm}
\label{fig:wfin_chrome}
\end{figure}

We believe that the above methodology does not consider the significant amount of noise that can be retained when a large number of features are used by a classifier---we believe that a better classification performance can be achieved by reducing features further in a manner similar to forward selection \cite{guyon2003introduction,das2001filters,jansen2018inside}. 
In \emph{Wfin++}, we consider the final ranked list of feature categories returned by \emph{Wfin}, and compute the validation accuracy by considering only the top-\emph{N} feature categories.
% s \markred{features, or feature categories?} \textcolor{blue}{feature categories}. 
The goal is to find the \emph{N} that yields the best validation accuracy. \figurename{ \ref{fig:wfin_chrome}} plots the validation accuracy versus \emph{N}, when the top-\emph{N} feature categories are used for keyword classification with the four search engines accessed using the Chrome browser (Extra-Trees classifier, number of trees: 700). 
% Although the absolute validation accuracy differs across search engines, 
A similar trend is observed with all search engines---after the initial increase in classification accuracy, adding more features causes the accuracy to drop.
% on Chrome, Firefox and Edge: the accuracy increases in the beginning and at some point start to keep decreasing.

Thus the final feature list in \emph{Wfin++} is selected as the one that achieves the highest validation accuracy---just like \emph{Wfin}, this feature list differs across the search engines. For instance, the final number of features selected for training with Yahoo/Chrome is around 5,820, while with Google/Chrome is around 7,200.

\vspace{-2mm}

\subsection{HTTPS vs. Tor}
\label{subsec:https_tor}

% \textcolor{blue}{
Tor traffic has attracted lots of attentions for traffic analysis in website/keyword fingerprinting \cite{oh2019p1, wang2013improved, juarez2014critical, wang2014effective, hayes2016k, panchenko2016website, wang2017walkie,oh2017fingerprinting, Li:2018:MIL:3243734.3243832,sirinam2018deep}.
Hence we compare the performance of Tor and HTTPS traffic in face of keyword fingerprinting to understand: (i) \emph{How vulnerable HTTPS traffic is compared with Tor?} (ii) \emph{Can classifiers designed for Tor be directly applied onto HTTPS traffic?}
% }

% \textcolor{blue}{
To minimize the impact of irrelevant variations (e.g., data collection platform, visiting time and targeted keywords etc.) on classification accuracy, we collect another dataset with 2 search engines (Google and DuckDuckGo) and 2 browsers (Chrome and Tor (version 8.5.5)) in a round-robin way using the same 100 targeted keywords as Oh et al. \cite{OhGit}.
Each keyword contains at least 100 samples, with the first 100 samples used for training/validation and the last 10 for testing.
\tablename{ \ref{tab:https_tor}} shows the testing accuracy with the \emph{PSC}, \emph{k-FP} and \emph{EtResp/SvmResp}.
% feature sets in Section \ref{sec:feature-sets}. 
Besides, we train with \emph{SvmResp} using the code provided by Oh et al. \cite{OhGit} to compare the performance of Extra-Trees and SVM. \begin{table}[htbp]
    \centering
    \caption{Classification accuracy with 100 targeted keywords under Chrome vs. Tor.}
    % \vspace{-3mm}
    \resizebox{0.95\linewidth}{!}{
    \begin{tabular}{|l|c|c|c|c|}
    \hline
    \multirow{2}{*}{} & \multicolumn{2}{c|}{DuckDuckGo} & \multicolumn{2}{c|}{Google}  \\ \cline{2-5}
    & Tor & Chrome & Tor & Chrome \\ \hline
    \emph{PSC} & 11.43 $\pm$ 0.65 & 100.00 & 4.60 $\pm$ 0.56 & 95.75 $\pm$ 0.21\\ 
    % \emph{Wfin++}  & & & & \\ 
    \emph{k-FP}  & 42.65 $\pm$ 0.57 & 98.56 $\pm$ 0.04 & 23.66 $\pm$ 0.33 & 89.74 $\pm$ 0.25 \\ 
    \emph{EtResp}  & 37.01 $\pm$ 0.39 & 97.25 $\pm$ 0.12  & 14.27 $\pm$ 0.21 & 41.96 $\pm$ 0.30 \\
    \hdashline
    \emph{SvmResp}  & 31.10 & 93.54 & 13.00 & 25.10 \\
    \hline
    \end{tabular}
    }
    \vspace{-3mm}
    \label{tab:https_tor}
\end{table}
We find that:
\begin{itemize}
    \item With the same set of features, Extra-Trees  achieves comparable and slightly better performance compared with SVM. Besides, the same trend is also observed when we re-run \emph{SvmResp} and \emph{EtResp} with the closed-world dataset provided by Oh et al. \cite{OhGit}. For 100 Google keywords, the precision achieved by \emph{EtResp} is 100\% and  82\% by \emph{SvmResp}, when for each keyword 80 samples are for training and 30 for testing.
    \item HTTPS traffic are indeed more vulnerable than Tor traffic, which makes keyword fingerprinting under HTTPS a more severe concern considering its vulnerability and the scale of potential victims.
    However, classifiers that achieve high performance in one communication scenario do not always obtain an equal high performance in other strict/relaxed communication scenarios \cite{yan2018feature}. 
   For example, \emph{PSC} achieves up to 95\% accuracy under HPPTS but only around 4\% to 11\% in Tor. The same trend is also observed with \emph{k-FP} and \emph{EtResp/SvmResp}.
    % \item Finally, although it is recommended to compare performance across different datasets collected at different time , the classification gap between two Tor datasets (ours vs. \cite{oh2017fingerprinting}) cautious us about the  generalization of feature-based classifiers. 
    % It further demonstrates the importance of the scale of the dataset and feature selection.
\end{itemize}
% }

\subsection{Vulnerability of Search Engines with Edge}
\label{appendix:edge_search_engine}

\tablename{ \ref{tab:search_engine_acc_edge}} shows the classification accuracy achieved using four different search engines with the Edge browser. The targeted dataset is split into three disjoint subsets for training, validation, and testing, with the ratio of \emph{8:1:1}: for every 10 consecutive visits of a search query, the first 8 samples are used for training, the 9th for validation, and the 10th for testing. 
For each targeted keyword, we consider 32 training samples, 4 validation samples and 4 testing samples during evaluation. 
To obtain the test accuracy shown in \tablename{ \ref{tab:search_engine_acc_edge}}, we use 36 samples for training (including both samples from training and validation dataset) and 4 for testing.
% \textcolor{blue}{Based on the results in \tablename{ \ref{tab:search_engine_acc_edge}}, we have the following observations:
We observe that:
\begin{itemize}
    \item \emph{Wfin++} and \emph{PSC} consistently achieve better accuracy than \emph{EtResp} and \emph{k-FP}.
    
    \item Among all four search engines, DuckDuckGo makes the user the most vulnerable to keyword fingerprinting,  while Bing offers the least vulnerability in \emph{homepage searching} mode.
    
    \item After filtering out homepage traffic in \emph{address searching} mode, classification accuracy of Yahoo increased by about 10\%---which makes Yahoo traffic the most vulnerable.
    The classification accuracy of DuckDuckGo and Bing did not increase much, while that of Google decreased by about 4\%.
    % However, the accuracy is decreased by around 4\% for Google. 
\end{itemize}

\vspace{-1mm}
\begin{table}[htbp]
    \centering
    \caption{Classification Accuracy Achieved: Edge.}
    \vspace{-3mm}
    \resizebox{1\linewidth}{!}{
    \begin{tabular}{|l|cccc|}
    
    \multicolumn{5}{c}{\textbf{Homepage Searching Mode}} \\ 
    
    \hline
        \textbf{Edge} & Bing & Yahoo & Google & DuckDuckGo  \\ \hline
         \emph{Wfin++} & \fbox{35.10 $\pm$ 0.12} &\fbox{77.52 $\pm$ 0.15} & \fbox{64.09 $\pm$ 0.25}
 & \fbox{78.75 $\pm$ 0.08}\\
         \emph{PSC} & 28.81 $\pm$ 0.36 &72.22 $\pm$ 0.13 & 59.31 $\pm$ 0.32& 76.15 $\pm$ 0.19 \\
         \emph{EtResp} & 5.57 $\pm$ 0.10
         & 0.49 $\pm$ 0.03 & 10.60 $\pm$ 0.09&  8.17 $\pm$ 0.08\\
         \emph{k-FP} & 9.53 $\pm$ 0.07& 1.11 $\pm$ 0.07& 13.85 $\pm$ 0.14& 11.56 $\pm$ 0.11 \\
    \hline
    
    % \multicolumn{5}{l}{} \\ 
    \multicolumn{5}{c}{\textbf{Addressbar Searching Mode}} \\   
    \hline
        \textbf{Edge} &  Bing &Yahoo & Google & DuckDuckGo  \\ \hline
         \emph{Wfin++} & \fbox{35.27 $\pm$ 0.22} & \fbox{87.96 $\pm$ 0.09}& 59.54 $\pm$ 0.28 &\fbox{78.82 $\pm$ 0.18} \\
         \emph{PSC} & 30.02	$\pm$ 0.17 & 84.90 $\pm$ 0.11
         & \fbox{60.34 $\pm$ 0.06} & 76.30 $\pm$ 0.1 \\
         \emph{EtResp} & 5.71 $\pm$ 0.02 & 0.98 $\pm$ 0.01
         &  11.60 $\pm$ 0.09& 10.53 $\pm$ 0.12 \\
         \emph{k-FP} & 8.67	$\pm$ 0.07 & 2.31 $\pm$ 0.07 & 15.64 $\pm$ 0.19 & 13.19 $\pm$ 0.23\\
    \hline
    
    \end{tabular}
    }
    \vspace{-2mm}
    \label{tab:search_engine_acc_edge}
\end{table}

\subsection{Informative Features Selected by Wfin++}
\label{appendix:top_features_wfin_ddg}

\tablename{ \ref{tab:top_features_wfin_ddg}} lists the top 20 informative feature categories selected by \emph{Wfin++} with DuckDuckGo (Chrome) in \emph{addressbar searching} mode.

\begin{table}[htbp]
    \centering
    \caption{Top 20 informative feature categories with Wfin (DuckDuckGo)}
    \vspace{-2mm}
    \resizebox{0.75\linewidth}{!}{
    \begin{tabular}{llr}
% \toprule
\hline

0  &                                  unique packet size &  24.397 \\
1  &                            initial 30 outgoing packets &  10.767 \\
2  &                                     packet size count &   6.336 \\
3  &              first 300 incoming packets preposition &   3.824 \\
4  &                  first 300 outgoing packets position &   3.703 \\
5  &                first 300 outgoing packets preposition &   3.362 \\
6  &                first 300 incoming packets position &   2.971 \\
7  &                        initial outgoing bursts &   2.864 \\
8  &                average outgoing inter-arrival time &   2.742 \\
9  &                              initial 30 packets &   2.338 \\
10 &                           initial 30 incoming packets &   2.248 \\
11 &                                    unique burst size &   1.993 \\
12 &        first 20 largest outgoing bytes per TCP conn. &   1.827 \\
13 &                         initial incoming bursts &   1.708 \\
14 &               ratio of incoming bytes per TCP conn. &   1.566 \\
15 &              initial 30 outgoing in first TCP conn. &   1.379 \\
16 &                                    burst size count &   1.279 \\
17 &       first 20 largest outgoing bytes per hostname &   1.264 \\
18 &                        outgoing bytes per TCP conn. &   1.071 \\
19 &       outgoing bytes per TCP conn. w.r.t. Port 443/80 &   0.974 \\
\hline
\end{tabular}}
\vspace{-2mm}
    % \caption{Caption}
    \label{tab:top_features_wfin_ddg}
\end{table}{}

\subsection{Parameter tuning in Extra-Trees classifier}
\label{appendix:parameter_tuning}

We use the implementation of Extra-Trees classifier from sklearn \cite{sklearn} in python3 \cite{python3} and tune two parameters (\emph{criterion} and \emph{n\_estimators}) based on the validation accuracy achieved with the monitored dataset. For other parameters, we use the default value in sklearn.

\subsubsection{Gini Index  vs.  Information Gain}

\begin{figure}[htbp]
\centering
\includegraphics[width=0.7\linewidth]{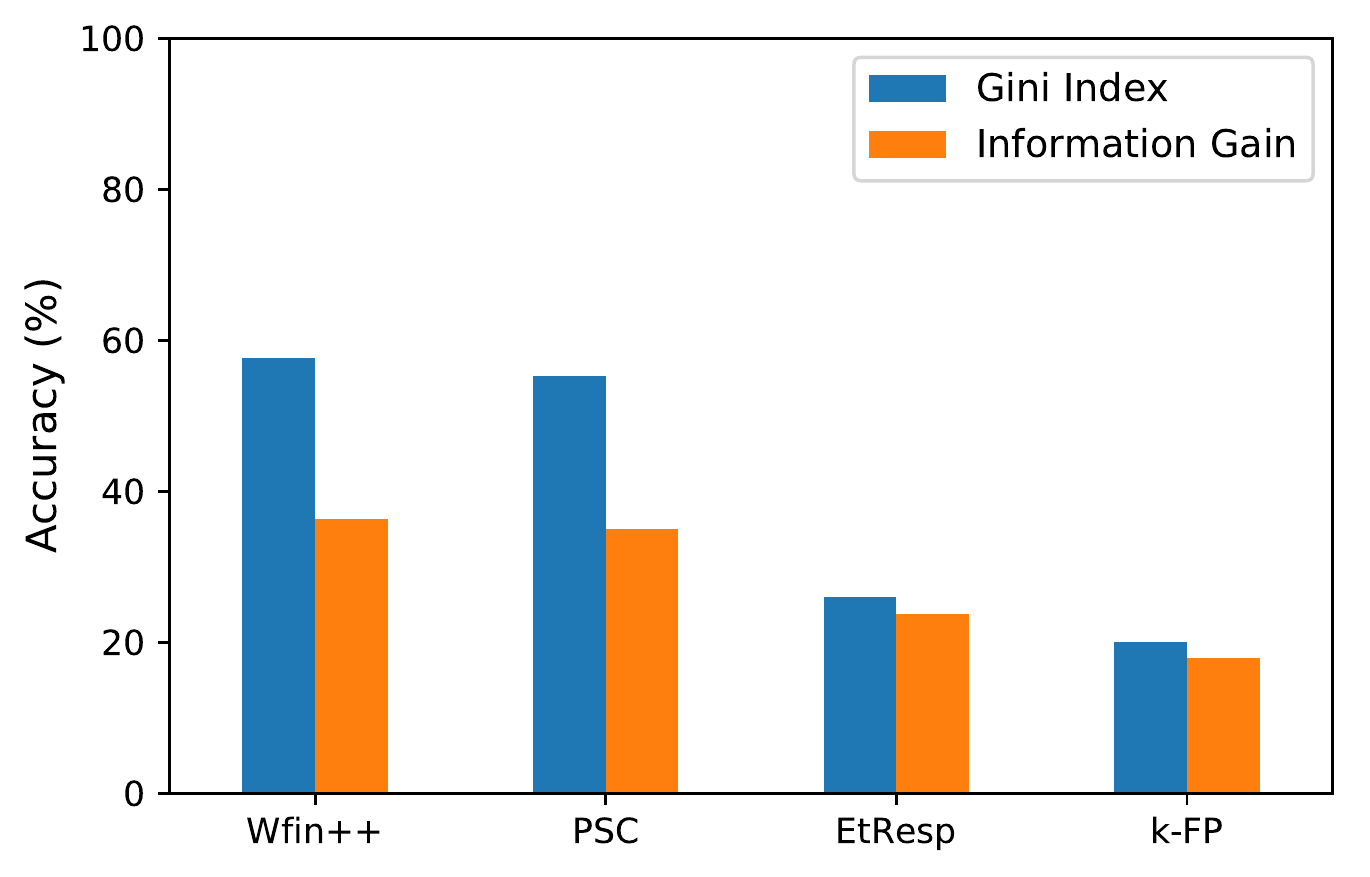}
\vspace{-3mm}
\caption{Classification Accuracy: Different Split Criteria}
\vspace{-2mm}
\label{fig:gini_mi_accuracy}
\end{figure}

The selection of features at each node of the tree to split the data (split criterion) directly affects the performance with decision tree-based ensemble methods. Two widely used split criterion is Gini Index and Information Gain. A lot of research was dedicated to understand which of them produce the best decision tree for a given dataset \cite{mingers1989empirical,lim2000comparison, stoffel2001selecting, raileanu2004theoretical}. Although most of empirical studies concluded that there is no significant differences between those two criteria and the disagreement is generally no higher than 2\% of all cases, we show the validation accuracy achieved with different feature sets when using different criteria with Extra-Trees classifier (n\_estimators = 700) in \figurename{ \ref{fig:gini_mi_accuracy}} for Google/Chrome (the overall trend is consistent across different search engines and browsers).

Based on the outcome in  \figurename{ \ref{fig:gini_mi_accuracy}}, Gini Index achieves better accuracy compared with Information Gain with the monitored dataset and the gap ranges from around 2\% to 21\% across different feature sets with Google/Chrome. 
Thus we choose Gini Index as the split criterion in Extra-Trees classifier.

% \subsection{Bootstrap Enabled vs. Disabled}
% Bootstrap aggregation, or Bagging for short, draws a new training set with replacement from the original training when building each tree. 
% Breiman suggests to use bootstrap in tandem with random feature selection to enhance accuracy and to obtain ongoing estimates of the generalization error and correlations with out-of-bag estimate~\cite{breiman2001random}.
% Although Extra-Trees is designed to use the whole training samples to grow each tree \cite{geurts2006extremely}, sklearn supports bootstrap in its implementation. 
% Thus we experiment with bootstrap enabled and disabled with the monitored dataset and present the accuracy achieved with four feature sets with Extra-Trees classifier (n\_estimators = 1,000) in \tablename{ \ref{tab:bootstrap_accuracy}} with DuckDuckGo/Chrome.

% \input{tables/bootstrap_appendix_a.tex}

\subsubsection{Number of Trees}
\begin{figure}[htbp]
\centering
\includegraphics[width=0.75\linewidth]{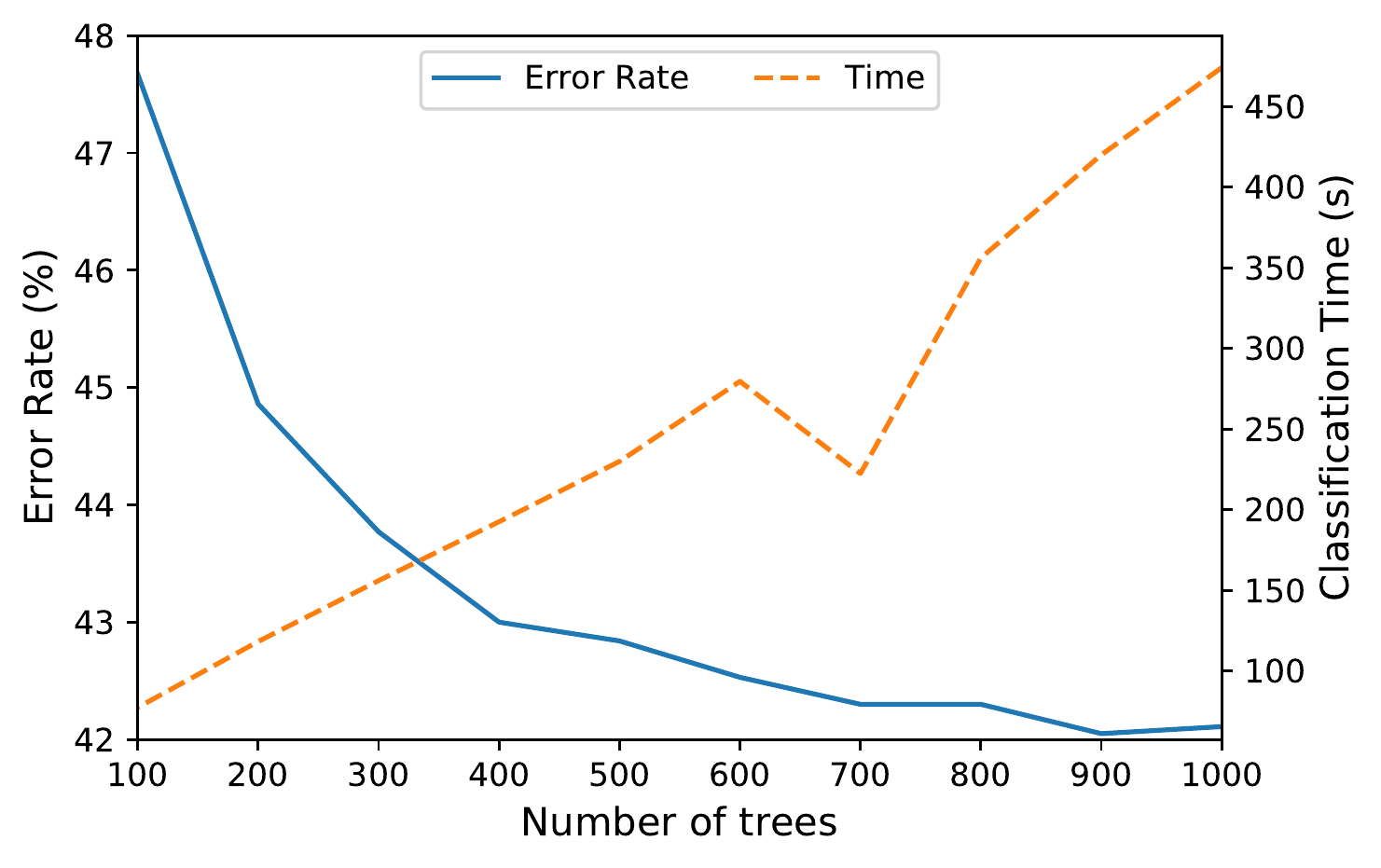}
\vspace{-3mm}
\caption{Error Rate and Classification Time (\emph{Wfin++}, Google/Chrome): impact of number of trees in Extra-Trees}
\vspace{-2mm}
\label{fig:n_trees_impact}
\end{figure}

As Breiman stated in \cite{breiman2001random}, the behavior of prediction error for randomization methods is a monotonically decreasing function of number of trees in the ensemble. Thus the more the trees, the better the accuracy and the higher the computational overhead. In \figurename{ \ref{fig:n_trees_impact}}, we show the average error rate obtained by \emph{Wfin++} and the classification time when ranging the number of trees from 100 to 1,000 with Google/Chrome. 
As the results suggested, the error rate tends to stabilize and the classification time increase significantly as we keep increasing the number of trees -- the error rate decrease around 0.1\% while the classification time is increased by nearly 2/3 from 700 to 800.
The same trend is observed with different search engines and browsers.
In both closed-world and open-world experiments, therefore, we set the number of trees to 700.

\vspace{-2mm}

% \section{Discussion}

% \textcolor{blue}{\textbf{collect a Tor dataset with the same keyword as Oh and will update the results}}

\subsection{Impact of Number of Targeted Keywords}
\label{subsec:num_of_keywords}

\begin{figure}[htbp]
\centering
% \vspace{-2mm}
\includegraphics[width=0.75\linewidth]{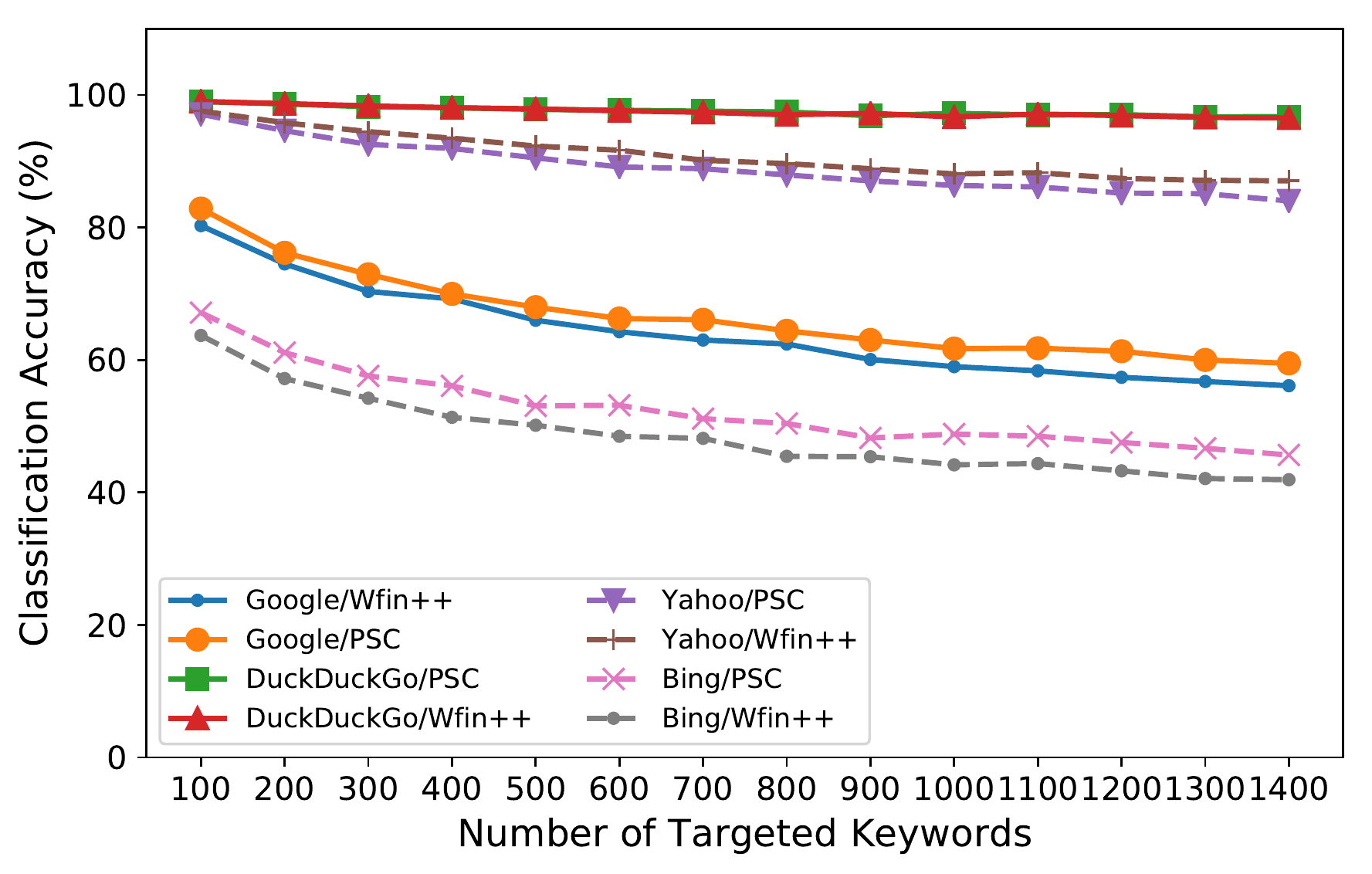}

\caption{Classification accuracy achieved by \emph{Wfin++} and \emph{PSC} with different number of targeted keywords (Chrome browser).}
\vspace{-6mm}
\label{fig:varying_size_acc}
\end{figure}

In this section, we study how the accuracy changes as more targeted keywords are considered for classification by varying the number of keywords from 100 to 1400. 
For each keyword, 45 samples are used for training and 9 for testing as in Section \ref{subsec:vul_search_engine}. \figurename{ \ref{fig:varying_size_acc}} shows the accuracy with four search engines using \emph{Wfin++} and \emph{PSC} with Chrome in \emph{addressbar searching} mode. As the results indicate, the performance keeps decreasing as more targeted keywords are considered although the rate of decline differs among search engines. For example, when number of targeted keywords is increased from 100 to 1,440, the accuracy drops significantly from around 67\% to 45\% with Bing/Chrome, but slightly from 99\% to 96\% with DuckDuckGo/Chrome.\footnote{A consistent trend is also observed with Firefox samples but the results are omitted due to space constraints.} 
Thus when the number of targeted keyword keeps growing, more advanced machine learning techniques such as deep learning may help boost the performance \cite{oh2019p1}.

\begin{figure*}[htbp]
\centering
\includegraphics[width=1.0\textwidth]{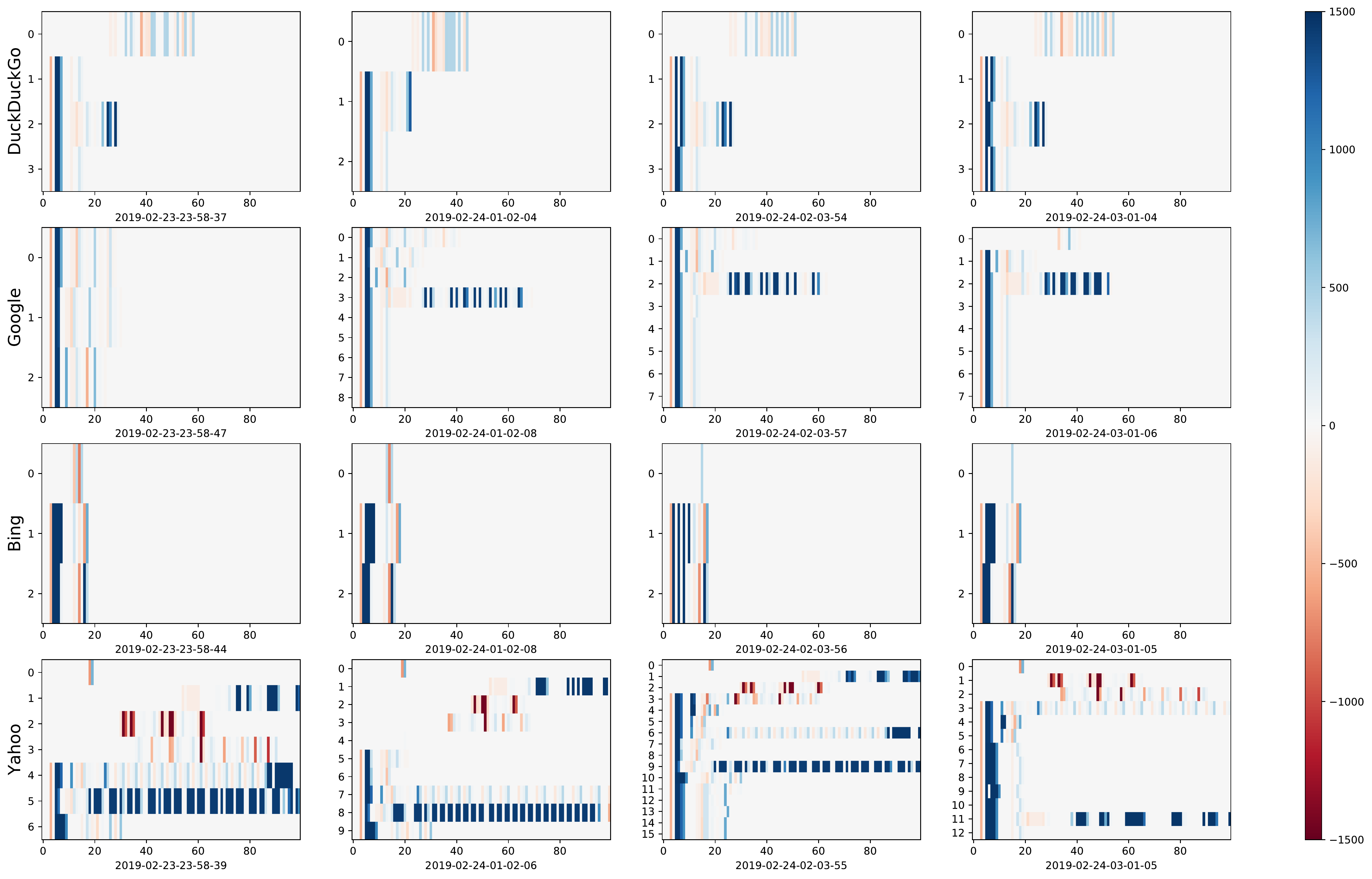} 
% \caption{Comparison of network traffic generated by four search engines with Chrome for a specific search query searched approximately the same time after filtering out homepage traffic. The corresponding search engine from top to bottom is DuckDuckGo, Google, Bing and Yahoo. The caption in each sub-figure indicates the searching time following the format \emph{Year-Month-Day-Hour-Minute-Second}.}
\vspace{-7mm}
\caption{Traffic generated using Chrome: \emph{Addressbar Searching} (caption below each sub-figure indicates time of search)}
\vspace{-3mm}
\label{fig:search_engine_no_homepage}
\end{figure*}

\subsection{Addressbar Searching Mode: Generated Traffic Pattern}
\label{appendix:traffic_pattern_addressbar}

\figurename{ \ref{fig:search_engine_no_homepage}} displays the traffic pattern generated by four different search engines for the same search query in \figurename{ \ref{fig:search_engine}} in \emph{addressbar searching} mode. \tablename{ \ref{tab:hostname_yahoo_no_homapge}} shows the second-level server name contained in the traffic trace in  \figurename{ \ref{fig:search_engine_no_homepage}} with Yahoo (Chrome).

\vspace{-2mm}
\begin{table}[htbp]
    \centering
    \caption{Second-level Sever Names: Yahoo with Chrome in \emph{addressbar searching} mode}
    \vspace{-2mm}
    \resizebox{0.65\linewidth}{!}{%
    \begin{tabular}{lr|lr}
    \hline
    \multicolumn{2}{l|}{2019-02-25-16-48-03} & 
    \multicolumn{2}{l}{2019-02-25-17-39-38} \\ \hline
    yahoo.com & 4 & yahoo.com & 5 \\
    yimg.com & 1 & yimg.com & 1 \\
    bing.com & 1 & addthis.com & 1 \\
    &  & google.com & 1 \\
    &  & bing.com & 1 \\ \hline
    % \vspace{-5mm} \\
    \multicolumn{4}{c}{} \\ \hline
    \multicolumn{2}{l|}{2019-02-25-18-30-39} &
    \multicolumn{2}{l}{2019-02-25-19-20-47} \\ \hline
     yahoo.com & 5 & bing.net & 6 \\
     yahoodns.net & 5 & yahoo.com & 5 \\
     yimg.com & 1 & krxd.net & 1 \\
     nexac.com & 1 &  &  \\
     addthis.com & 1 &  &  \\
     google.com & 1 &  &  \\
     bing.com & 1 &  &  \\ \hline
    \end{tabular}%
    }
    \vspace{-2mm}
    \label{tab:hostname_yahoo_no_homapge}
\end{table}

\subsection{Safari and Chrome in homepage searching mode}

\tablename{ \ref{tab:cross_browser_homepage}} shows the performance of \emph{Wfin++} and \emph{PSC} in cross-browser scenario with Google and Firefox samples. Similar trend is observed as in \emph{addressbar searching} mode--the accuracy degrades drastically when the training and testing browser do not match.

\vspace{-2mm}
\begin{table}[htbp]
    \centering
    \caption{Classification Accuracy: Cross-browser Attacks in \emph{Homepage Searching}}
    \vspace{-2mm}
    \resizebox{1\linewidth}{!}{
    \begin{tabular}{|l|cc|cc|}

    \hline
    \multirow{2}{*}{\textbf{DDG}} & \multicolumn{2}{c|}{Firefox} & \multicolumn{2}{c|}{Chrome} \\ \cline{2-5}
    & \emph{Wfin++} & \emph{PSC}  & \emph{Wfin++} & \emph{PSC} \\ \hline
    Firefox & 91.95 $\pm$ 0.10 & \fbox{92.23 $\pm$ 0.02} & 1.22 $\pm$ 0.09 & 1.82 $\pm$ 0.05 \\ 
    Chrome & 0.22 $\pm$ 0.01&0.97 $\pm$ 0.05 & 96.15 $\pm$ 0.01 & \fbox{96.33 $\pm$ 0.04} \\
    Fire/Chr & 91.40	$\pm$ 0.08& \fbox{92.14 $\pm$ 0.06}& \fbox{96.30 $\pm$ 0.04} & 96.04 $\pm$ 0.04 \\ 
    
    \hline
    % \multicolumn{5}{c}{} \\ 
    \hline
    
    \multirow{2}{*}{\textbf{Google}} & \multicolumn{2}{c|}{Firefox} & \multicolumn{2}{c|}{Chrome} \\ \cline{2-5}
    & \emph{Wfin++} & \emph{PSC}  & \emph{Wfin++} & \emph{PSC} \\ \hline
    Firefox & 75.56 $\pm$ 0.07 & \fbox{76.75 $\pm$ 0.02} & 0.38 $\pm$ 0.07 &1.42 $\pm$ 0.04  \\ 
    Chrome &0.07& 0.98 $\pm$ 0.13 & \fbox{60.32 $\pm$ 0.07} & 57.72 $\pm$ 0.05 \\
    Fire/Chr & 75.17 $\pm$ 0.13& \fbox{75.88 $\pm$ 0.13} & \fbox{61.91 $\pm$ 0.31} &57.12 $\pm$ 0.14\\ 
    \hline  
    \hline
    \multirow{2}{*}{\textbf{Yahoo}} & \multicolumn{2}{c|}{Firefox} & \multicolumn{2}{c|}{Chrome} \\ \cline{2-5}
    & \emph{Wfin++} & \emph{PSC}  & \emph{Wfin++} & \emph{PSC} \\ \hline
    Firefox & \fbox{58.06 $\pm$ 0.25} & 49.87 $\pm$ 0.13 & 15.61 $\pm$ 0.35 & 12.42 $\pm$ 0.21 \\ 
    Chrome & 4.01 $\pm$ 0.39& 2.47 $\pm$ 0.04& \fbox{64.60 $\pm$ 0.21} & 57.85 $\pm$ 0.26\\
    Fire/Chr  & \fbox{57.56 $\pm$ 0.12}& 48.73 $\pm$ 0.18 & \fbox{64.22 $\pm$ 0.22} & 56.23 $\pm$ 0.05 \\ 
    
    \hline
    
    \end{tabular}
    }
    \vspace{-5mm}
    \label{tab:cross_browser_homepage}
\end{table}

\subsection{Time Effect in homepage searching mode}
\label{appendix:time_effect_addressbar}

\tablename{ \ref{tab:time_effect_close_world}} summarizes the impact of time gap between training and test samples, in closed-world evaluations with the homepage searching mode (Chrome browser).

\vspace{-2mm}
\begin{table}[htbp]
    \centering
    \caption{Impact of Time (Chrome, \emph{homepage searching} mode)}
    \vspace{-3mm}
    \resizebox{0.75\linewidth}{!}{
    \begin{tabular}{l|ccccc}
    
    \hline
     \textbf{DDG} & \emph{test-3}  & \emph{test-4}  & \emph{test-8} & \emph{test-10} & \emph{test-14}\\ \hline
        \emph{Wfin++} &92.20 & 90.56 & 81.22 & 71.79 & 71.70\\
        \emph{PSC} & 92.47 & 90.84 & 80.2 & 69.74 &  69.39 \\ 
        \emph{EtResp} & 26.65 & 22.04 & 11.92 & 7.61 & 6.58 \\
        \emph{k-FP} & 28.82 & 27.07 & 20.98 & 14.62 & 14.00 \\
        \hline
    % \multicolumn{6}{c}{} \\ 
    \hline
        
       \textbf{Google} & \emph{test-3}  & \emph{test-4}  & \emph{test-8} & \emph{test-10} & \emph{test-14}\\ \hline 
        \emph{Wfin++} & 54.29 & 44.54 & 23.44 & 18.92 & 18.16\\
        \emph{PSC} &47.33 & 35.39 & 9.92 & 9.83 & 14.69\\
        \emph{EtResp} & 24.78  &  22.05 & 14.06 & 6.89 & 6.29 \\
        \emph{k-FP}  & 18.96 & 19.76 & 15.26 & 12.35 & 10.38\\
        \hline
       
    % \multicolumn{6}{c}{} \\
    % \multicolumn{6}{l}{\textbf{\emph{Addressbar Searching Mode}}} \\ \hline
    
    % DuckDuckGo & \emph{test-3}  & \emph{test-4}  & \emph{test-8} & \emph{test-10} & \emph{test-14}\\ \hline
    %     \emph{Wfin++} & 92.20 & 90.86 & 81.84 & 72.90 & 72.46\\
    %     \emph{PSC} & 92.85 & 91.09 & 80.34 & 69.65 & 69.15\\ 
    %     \emph{EtResp} & 31.22 & 22.09 & 11.07 & 7.94 & 7.29\\
    %     \emph{k-FP} & 37.35 & 34.78 & 27.99 & 20.30 & 19.72\\
    %     \hline
        
    % \multicolumn{6}{c}{} \\ \hline
        
    %   Google & \emph{test-3}  & \emph{test-4}  & \emph{test-8} & \emph{test-10} & \emph{test-14}\\ \hline 
    %     \emph{Wfin++} & 55.14 &43.34 & 22.53 & 19.99 & 22.16\\
    %     \emph{PSC} & 47.74 & 34.67 & 9.75 & 10.14 & 14.72\\
    %     \emph{EtResp} & 26.92 &24.05 & 18.78 & 12.91 & 12.34\\
    %     \emph{k-FP}  & 23.96 & 24.05 & 22.19 & 17.29 & 15.54\\
    %     \hline
        
    \end{tabular}}
    \vspace{-2mm}
    \label{tab:time_effect_close_world}
\end{table}

\subsection{Cross-browser attack with Edge and Safari}
\label{appendix:cross_browser_edge}

\tablename{ \ref{tab:cross_browser_edge_safari}} shows the accuracy achieved when using 36 Edge samples for training and 4 Firefox samples for testing in both \emph{homepage searching} mode and \emph{addressbar searching} mode. 
Consistent with the results obtained in Section \ref{subsec:client_platform}, the classification accuracy degrades severely when using samples from different browser/device for training and testing. For instance, when testing with Safari samples the accuracy is dropped to less than 0.2\% compared with around 78\% with Edge samples.

\begin{table}[htbp]
    \centering
    \vspace{-2mm}
    \caption{Classification accuracy (\%) in cross-browser attack with Edge and Safari. (DDG: DuckDuckGo)}
    \vspace{-2mm}
    \resizebox{0.9\linewidth}{!}{%
    \begin{tabular}{|l|cc|cc|}
    \multicolumn{5}{c}{\textbf{Homepage Searching Mode}} \\ 
    \hline
    \multirow{2}{*}{\textbf{DDG}} & \multicolumn{2}{c|}{Edge} & \multicolumn{2}{c|}{Safari} \\ \cline{2-5}
    & \emph{Wfin++} & \emph{PSC}  & \emph{Wfin++} & \emph{PSC} \\ \hline
    Edge &  78.75 $\pm$ 0.08 & 76.15 $\pm$ 0.19  & 0.19 $\pm$ 0.04 & 0.15 $\pm$ 0.02 \\ 
    \hline
    \hline
    
    \multirow{2}{*}{\textbf{Google}} & \multicolumn{2}{c|}{Edge} & \multicolumn{2}{c|}{Safari} \\ \cline{2-5}
    & \emph{Wfin++} & \emph{PSC}  & \emph{Wfin++} & \emph{PSC} \\ \hline
    Edge &  64.09 $\pm$ 0.25 & 59.31 $\pm$ 0.32  & 0.09 $\pm$ 0.02 & 0.07 \\ \hline
    \multicolumn{5}{c}{\textbf{Addressbar Searching Mode}} \\ 
    \hline
    \multirow{2}{*}{\textbf{DDG}} & \multicolumn{2}{c|}{Edge} & \multicolumn{2}{c|}{Safari} \\ \cline{2-5}
    & \emph{Wfin++} & \emph{PSC}  & \emph{Wfin++} & \emph{PSC} \\ \hline
    Edge &  78.82 $\pm$ 0.18 & 76.30 $\pm$ 0.1 & 0.19 $\pm$ 0.05 & 0.12 $\pm$ 0.02 \\ 
    \hline
    \hline
    
    \multirow{2}{*}{\textbf{Google}} & \multicolumn{2}{c|}{Edge} & \multicolumn{2}{c|}{Safari} \\ \cline{2-5}
    & \emph{Wfin++} & \emph{PSC}  & \emph{Wfin++} & \emph{PSC} \\ \hline
    Edge & 59.54 $\pm$ 0.28 & 60.33 $\pm$ 0.07 & 0.46 $\pm$ 0.06 & 0.07 \\ 
    \hline
    \end{tabular}
    }
    \vspace{-2mm}
    \label{tab:cross_browser_edge_safari}
\end{table}

\begin{figure*}[htbp]
\centering

\begin{subfigure}{0.3\textwidth}
    \centering
    \smallskip
    \includegraphics[width=1\textwidth]{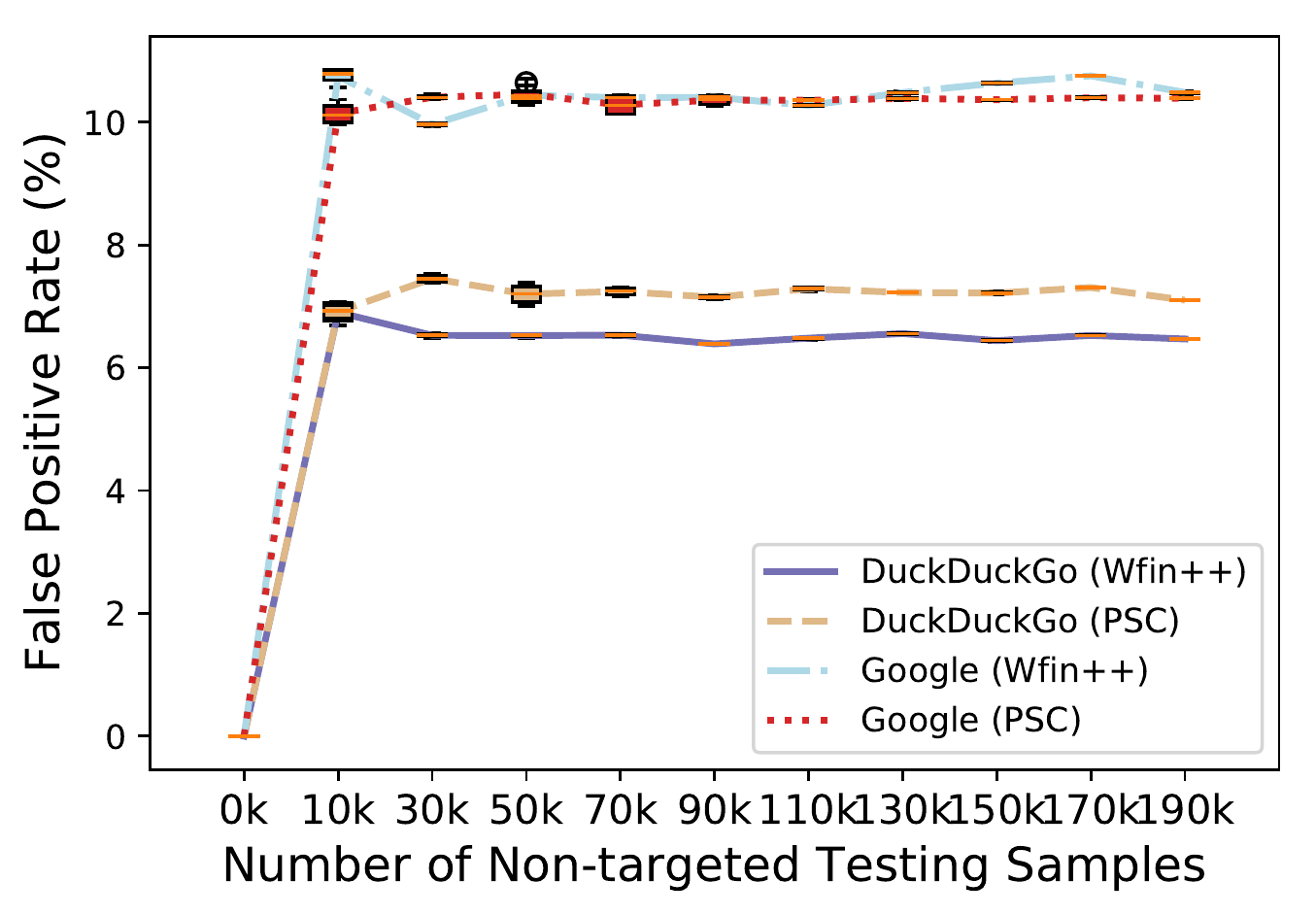}
    \vspace{-6mm}
    \caption{False Positive Rate} %{Light Unit}
    \label{fig:multilevel_vary_test_a}
\end{subfigure}
\begin{subfigure}{0.3\textwidth}
    \centering
    \smallskip
    \includegraphics[width=1\textwidth]{figures/fnr_vary_test_multilevel.pdf}
    \vspace{-6mm}
    \caption{False Negative Rate} %{Light Unit}
    \label{fig:multilevel_vary_test_b}
\end{subfigure}
\begin{subfigure}{0.3\textwidth}
    \centering
    \smallskip
    \includegraphics[width=1\textwidth]{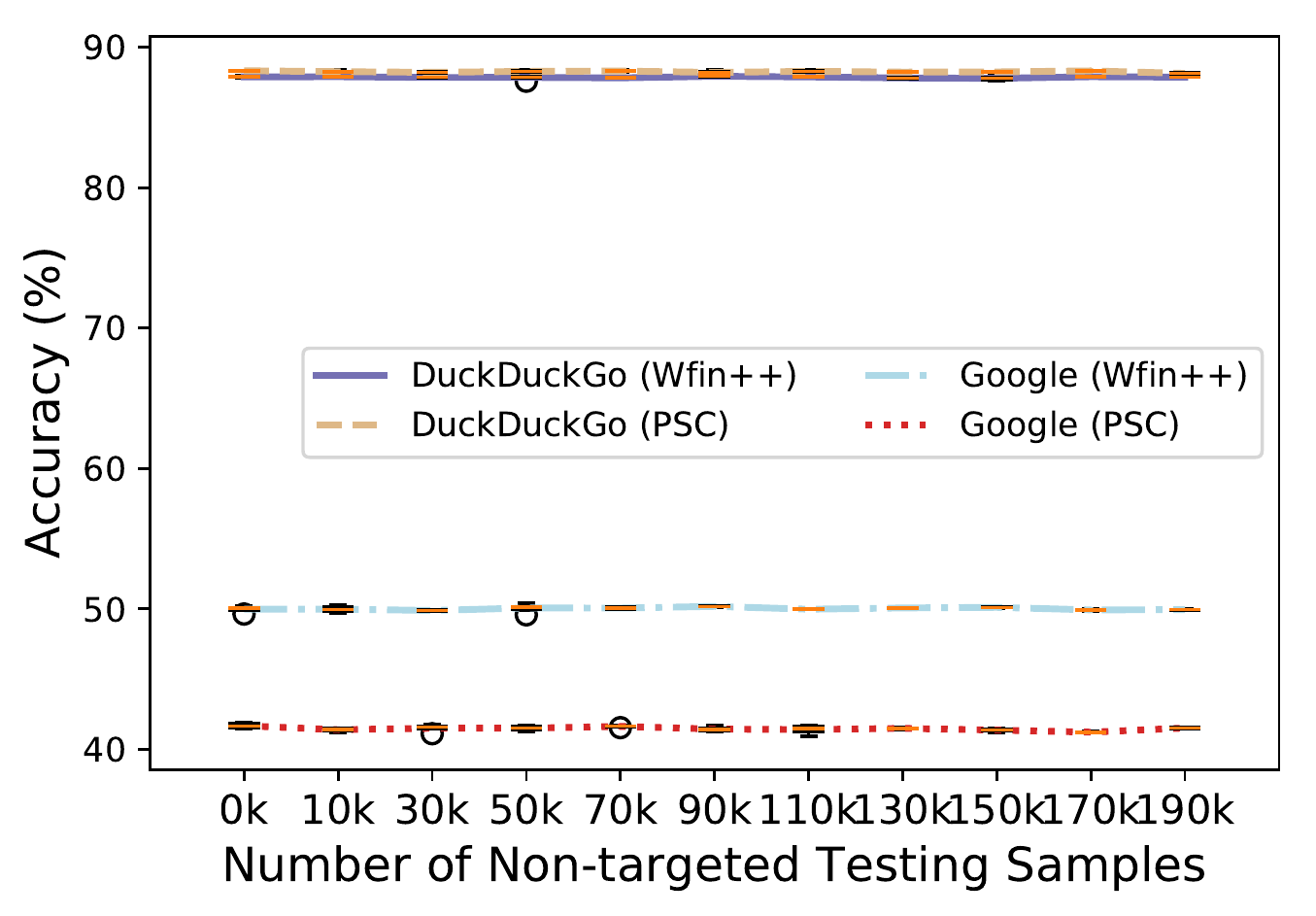}
    \vspace{-6mm}
    \caption{Accuracy} %{Light Unit}
    \label{fig:multilevel_vary_test_c}
\end{subfigure}
\vspace{-3mm}
\caption{Multi-level Classification: Impact of Number of Non-targeted Test Samples (50k Non-targeted Training Samples)}
\vspace{-6mm}
\label{fig:multilevel_vary_test}
\end{figure*}

\subsection{Eliminating Noise from ``Other'' Domains}
\label{appendix:top_domains_others}

\tablename{ \ref{tab:top_domains_others}} summarizes the classification accuracy when connections to only top-level domains are considered by the attacker.\\
\vspace{-2mm}
\begin{table}[htbp]
    \centering
    \caption{Classification Accuracy when only considering top domains in \emph{addressbar searching} mode.}
    \vspace{-3mm}
    \resizebox{1\linewidth}{!}{
    \begin{tabular}{|l|cc|cc|}

    \hline
    \multirow{2}{*}{\textbf{DDG}} & \multicolumn{2}{c|}{Firefox} & \multicolumn{2}{c|}{Chrome} \\ \cline{2-5}
    & \emph{Wfin++} & \emph{PSC}  & \emph{Wfin++} & \emph{PSC} \\ \hline
    Firefox &  93.50 $\pm$ 0.06 & 93.42 $\pm$ 0.06 & 1.00 $\pm$ 0.14 & 0.96 $\pm$ 0.09\\ 
    Chrome & 1.79 $\pm$ 0.20 & 1.99 $\pm$ 0.14 & 96.63 $\pm$ 0.03 & 96.62 $\pm$ 0.04\\
    \hline
    % \multicolumn{5}{c}{} \\ 
    \hline
    
    \multirow{2}{*}{\textbf{Google}} & \multicolumn{2}{c|}{Firefox} & \multicolumn{2}{c|}{Chrome} \\ \cline{2-5}
    & \emph{Wfin++} & \emph{PSC}  & \emph{Wfin++} & \emph{PSC} \\ \hline
    Firefox & 78.20 $\pm$ 0.10 & 80.53 $\pm$ 0.16 & 2.21 $\pm$ 0.06 & 1.80 $\pm$ 0.05 \\ 
    Chrome & 5.13 $\pm$ 0.11 & 1.30 $\pm$ 0.17
 & 62.39 $\pm$ 0.10 & 59.16 $\pm$ 0.08\\
    \hline
    % \multicolumn{5}{c}{} \\ 
    \hline
    % \multirow{2}{*}{\textbf{Bing}} & \multicolumn{2}{c|}{Firefox} & \multicolumn{2}{c|}{Chrome} \\ \cline{2-5}
    % & \emph{Wfin++} & \emph{PSC}  & \emph{Wfin++} & \emph{PSC} \\ \hline
    
  \multirow{2}{*}{\textbf{Bing}} & \multicolumn{2}{c|}{Firefox} & \multicolumn{2}{c|}{Chrome} \\ \cline{2-5}
    & \emph{Wfin++} & \emph{PSC}  & \emph{Wfin++} & \emph{PSC} \\ \hline
    Firefox &  46.37 $\pm$ 0.13 & 46.23 $\pm$	0.20 & 2.94 $\pm$ 0.16  &3.04 $\pm$ 0.21\\ 
    Chrome & 6.77 $\pm$ 0.14 & 2.72 $\pm$ 0.24 & 45.87 $\pm$ 0.14 & 45.77 $\pm$ 0.16 \\
    \hline
  
    \end{tabular}
    }
    \vspace{-2mm}
    \label{tab:top_domains_others}
\end{table}

\subsection{Multi-level Classification: Impact of Number of Non-targeted Test Samples}
\label{appendix:multi_level_vary_test}

\figurename{ \ref{fig:multilevel_vary_test}} shows the performance of \emph{Wfin++} and \emph{PSC} in multi-level classification, for different number of non-targeted testing samples. We observe that:
\begin{itemize}
    \item More non-targeted samples are incorrectly classified as targeted ones (\emph{FPR}), as the number of non-targeted testing samples increases.
    
    \item The likelihood of incorrectly classifying a targeted sample (\emph{FNR}) is not affected by the number of non-targeted testing samples. 
    % Although it keeps increasing as we increase the number of non-monitored testing samples as shown in Section \ref{subsec:multilevel}.
\end{itemize}

\subsection{Countermeasures}
\label{subsec:countermeasures}

 In this section, we evaluate keyword fingerprinting in the presence of countermeasures in closed-world scenarios.
 % \markred{both closed-world and open-world scenarios}. 
 By examining the most-informative features yielded by \emph{Wfin++},\footnote{Top 30 informative features categories selected by \emph{Wfin++} in \emph{addressbar searching} mode with DuckDuckGo (Chrome) are listed in Appendix \ref{appendix:top_features_wfin_ddg}.} we find that most of these are extracted from packet sizes (e.g., \emph{unique packet size} and \emph{packet size count}) and packet ordering (e.g., \emph{initial 30 outgoing packets} and \emph{first 300 incoming/outgoing packets preposition}).
 We next consider two prominent HTTPS countermeasures that obfuscate actual packet size or packet ordering---PadToMTU \cite{dyer2012peek} and HTTPOS \cite{luo2011httpos}.\footnote{Advanced countermeasures designed for encrypted tunnels or Tor (e.g., \cite{cai2014systematic,wang2017walkie}) are not considered since they are not directly applicable in HTTPS.}
 PadToMTU aims at hiding actual packet sizes, which is one of the most informative features for website fingerprinting in HTTPS \cite{liberatore2006inferring,herrmann2009website, panchenko2016website,dyer2012peek,yan2018feature,alan2019client}, by padding each packet to MTU bytes. 
 HTTPOS is a browser-side defense to obfuscate traffic by exploiting several TCP and HTTP features---including, MSS negotiation, advertised window, HTTP Range, and HTTP Pipelining \cite{luo2011httpos}. 
 Specifically, it injects dummy requests within the user traffic in order to obfuscate actual traffic patterns, and modifies advertised window size to make servers pack responses into blocks of MSS-byte packets (and hide actual packet size).
 Luo et al. \cite{luo2011httpos} shows that HTTPOS is able to effectively prevent attackers from inferring information from 1,000 Google search queries over HTTPS. 

% \input{tables/pesudo_code_httpos.tex}

% \markred{How do you implement these two in your experiments?} 
% Add text.\footnote{The MTU-byte is set to 1,500 bytes. In HTTPOS, we set the MSS-size to 1,000 bytes as described in \cite{luo2011httpos} and sample the size of dummy requests and responses from a uniform distribution between [0, MTU] and [0, 3*MTU] respectively.}

% \textcolor{blue}{(Should we move this paragraph to footnote to avoid attention?)}
In order to implement PadToMTU, we replace the sizes of all incoming and outgoing packets in our traces with MTU bytes. In order to implement HTTPOS, we pad the packet size of each outgoing packet (request) (\emph{p\_size}) to a value chosen randomly from the discrete uniform distribution [\emph{p\_size}, MTU]; for each incoming packet (response), we pad the packet size to a value chosen randomly from the discrete uniform distribution  [\emph{p\_size}, 3*MSS], and segment into multiple packets with MSS bytes in each. 
Specifically, we set MTU to 1,500 and MSS to 1,000, as described in \cite{luo2011httpos}.\footnote{The integration of HTTPOS into different browsers to perform automated large scale data collection with Selenium is considered as a future work.}

% \vspace{-2mm}

\setlength\dashlinedash{0.2pt}
\setlength\dashlinegap{2.0pt}
\setlength\arrayrulewidth{0.3pt}

\begin{table}[htbp]
    \centering
    \caption{Classification Accuracy Against Countermeasures: \emph{Addressbar Searching}}
    % \vspace{-5mm}
    \resizebox{0.95\linewidth}{!}{%
    \begin{tabular}{|l|cccc|}
    \hline
        \multicolumn{5}{c}{\textbf{{PadToMTU}}} \\
        \hline
        \textbf{Chrome} & Bing & Yahoo  & Google & DuckDuckGo  \\ \hline
        \emph{Wfin++} 
         & 9.26 $\pm$ 0.17&2.34 $\pm$ 0.06 & 27.95 $\pm$ 0.05& 44.22 $\pm$ 0.01\\
         \emph{PSC} & 
         0.90 $\pm$ 0.03& 0.11 $\pm$ 0.02& 1.45 $\pm$ 0.06&
         3.18 $\pm$ 0.04\\
                  \hdashline
         BW overhead & 209.6\% & 292.6\% & 193.9\% & 199.4\%\\
         
         \hline
         \hline
        \textbf{Firefox} & Bing & Yahoo  & Google & DuckDuckGo  \\ \hline
         \emph{Wfin++} 
         & 7.86 $\pm$ 0.03 &4.54 $\pm$ 0.01 & 27.72 $\pm$ 0.07& 40.72 $\pm$ 0.04\\
         
         \emph{PSC} & 0.85 $\pm$ 0.01 & 0.11 $\pm$ 0.03& 1.78 $\pm$ 0.05& 2.43 $\pm$ 0.05\\
                  \hdashline
         BW overhead & 220.2\% &330.1\% & 207.4\% & 218.7\% \\

        \hline
        % \vspace{-3mm}
        \multicolumn{5}{c}{} \\ 
         
        \multicolumn{5}{c}{\textbf{{HTTPOS}}} \\
         \hline
        \textbf{Chrome} & Bing & Yahoo & Google & DuckDuckGo  \\ \hline
         \emph{Wfin++} 
         & 7.42 $\pm$ 0.12& 2.11 $\pm$ 0.08 &25.53 $\pm$ 0.02 &36.23 $\pm$ 0.05  \\
         \emph{PSC} & 0.29 $\pm$ 0.07
& 0.11 $\pm$ 0.00
& 0.25 $\pm$ 0.01
& 0.32 $\pm$ 0.02
 \\
                  \hdashline
         BW overhead & 403.3\%
 & 514.5\% & 387.5\% & 387.4\% \\
         \hline
         \hline
         \textbf{Firefox} & Bing & Yahoo  & Google & DuckDuckGo  \\ \hline
         \emph{Wfin++} 
         &6.54 $\pm$ 0.02 & 4.05 $\pm$ 0.06& 24.23 $\pm$ 0.07& 33.96 $\pm$ 0.13\\
         \emph{PSC}
         & 0.26 $\pm$ 0.00 & 0.08 $\pm$ 0.03 & 0.28 $\pm$ 0.02 & 0.26 $\pm$ 0.04\\
         \hdashline
         BW overhead & 420.5\% & 562.9\% & 405.1\% & 414.3\% \\ \hline
    \end{tabular}%
    }
    \vspace{-2mm}
    \label{tab:cm_close_world_padtomtu}
    
\end{table}

\tablename{ \ref{tab:cm_close_world_padtomtu}} shows the accuracy of \emph{Wfin++} and \emph{PSC} in the presence of PadToMTU and HTTPOS for fingerprinting 1,440 targeted keywords---when only connections to top domains are used (Section \ref{subsec:feature_set}) in \emph{addressbar searching} mode.
The training, validation, and testing dataset are the same as in Section \ref{subsec:vul_search_engine}---for each query, 36 samples are used for training, 9 for validation, and 9 for testing. The accuracies of \emph{PSC} and \emph{Wfin++} are significantly lower in the presence of PadToMTU and HTTPOS (compared to Tables \ref{tab:top_domains_yahoo} and \ref{tab:top_domains_others})---for instance, for samples collected with DuckDuckGo/Chrome, the accuracy achieved by \emph{Wfin++} decreases from around 96\% to 44\% or 36\%, while the accuracy achieved by \emph{PSC} decreases from 96\% to 3\% or 0.3\%, respectively, in the presence of PadToMTU and HTTPOS. 
% Besides, the performance is further decreased to around 25\% and 0.2\% respectively when injecting dummy requests/responses with HTTPOS.

However, both PadToMTU and HTTPOS incur significant bandwidth overhead---for instance, with Google/Chrome the overhead is as high as 193\% for PadToMTU, and 387\% for HTTPOS.
The PadToMTU overhead is due to padding each packet to MTU bytes; the HTTPOS overhead is due to insertion of dummy requests/responses. 
We leave the design and evaluation of additional countermeasures against keyword fingerprinting as important future work.
% The obvious decrease of performance of \emph{Wfin++} and \emph{PSC} in face of PadToMTU and HTTPOS shows the efficiency of countermeasures against keyword fingerprinting and further indicates the importance of packet sizes for building unique fingerprints for each search query. 
% Indeed, it is more significant 
% Thus how to further reduce the bandwidth overhead while at the meantime degrade the performance of keyword fingerprinting is considered as an important future work.


% \section*{Acknowledgment}



%\bibliographystyle{IEEEtranS}

\end{document}